\documentclass[11pt,a4paper]{article}
\usepackage[colorlinks=true, linkcolor=black!50!blue, urlcolor=blue, citecolor=blue, anchorcolor=blue]{hyperref}
\usepackage[font=small,labelfont=bf,margin=0mm,labelsep=period,tableposition=top]{caption}
\usepackage[a4paper,top=3cm,bottom=2.5cm,left=2.5cm,right=2.5cm,bindingoffset=0mm]{geometry}

\usepackage{graphicx,placeins}
\usepackage{float}
\usepackage{afterpage}
\usepackage{epsfig,cite}
\usepackage{amssymb}
\usepackage{amsmath}
\usepackage{dsfont}
\usepackage{multirow}
\usepackage{url}
\usepackage{xcolor,colortbl}
\usepackage{float}
\usepackage{afterpage}
\usepackage{url}
\usepackage{hyperref}
\usepackage{booktabs}
\usepackage{mathrsfs}

\usepackage{tikz}
\usepackage{tikz-3dplot}
\usepackage[compat=1.0.0]{tikz-feynman}

\usepackage{enumitem}
\usepackage{hyperref}
\usepackage{cite}

\usepackage{pifont}

\usetikzlibrary{shapes, arrows}
\usetikzlibrary{decorations.pathreplacing}
\usetikzlibrary{positioning, calc}
\tikzstyle{fitted} = [rectangle, minimum width=5cm, minimum height=1cm, text centered, draw=black, fill=red!30]
\tikzstyle{operations} = [rectangle, rounded corners, minimum width=2cm,text centered, draw=black, fill=red!30]
\tikzstyle{roundtext} = [rectangle, rounded corners, minimum width=2cm, minimum height=0.8cm, text centered, draw=black, fill=red!30]
\tikzstyle{n3py} = [rectangle, rounded corners, minimum width=3cm, minimum height=1cm, text centered, draw=black, fill=green!30]
\tikzstyle{myarrow} = [thick,->,>=stealth]
\tikzstyle{line} =[draw, -latex']
\tikzstyle{decision} = [diamond, draw, fill=red!20, text width=7.5em, text centered,  inner sep=0pt, minimum height=2em, aspect=4]
\tikzstyle{cloud} = [draw, ellipse,fill=green!20, minimum height=2em]
\tikzstyle{inout} = [rectangle, draw, fill=green!20, text width=9.5em, text centered, rounded corners, minimum height=2em, minimum width=10em]
\tikzstyle{block}=[rectangle, draw, fill=blue!20, text width=9.5em, 
                   text centered, rounded corners, minimum height=2em, 
                   minimum width=10em]

\definecolor{darkgreen}{rgb}{0.0, 0.5, 0.13}

\newcommand{\be}{\begin{equation}}
\newcommand{\ee}{\end{equation}}
\newcommand{\bea}{\begin{eqnarray}}
\newcommand{\eea}{\end{eqnarray}}
\newcommand{\bi}{\begin{itemize}}
\newcommand{\ei}{\end{itemize}}
\newcommand{\ben}{\begin{enumerate}}
\newcommand{\een}{\end{enumerate}}

\newcommand{\lc}{\left[}
\newcommand{\rc}{\right]}
\newcommand{\lp}{\left(}
\newcommand{\rp}{\right)}

\def\gsim{\mathrel{\rlap{\lower4pt\hbox{\hskip1pt$\sim$}}
    \raise1pt\hbox{$>$}}}         %greater than or approx. symbol
\def\lsim{\mathrel{\rlap{\lower4pt\hbox{\hskip1pt$\sim$}}
    \raise1pt\hbox{$<$}}}         %less than or approx. symbol

\newcommand{\draft}[1]{}

\def\beq{\begin{equation}}
\def\eeq{\end{equation}}

\def\lapprox{\lower .7ex\hbox{$\;\stackrel{\textstyle <}{\sim}\;$}}
\def\gapprox{\lower .7ex\hbox{$\;\stackrel{\textstyle >}{\sim}\;$}}

\def\TeV{{\rm TeV}}

\def\d{{\rm d}}

\def\barq{\bar{q}}

\numberwithin{equation}{section}
\numberwithin{figure}{section}
\numberwithin{table}{section}

\usepackage{tabularx}
\newcolumntype{C}[1]{>{\centering\arraybackslash}p{#1}}

\usepackage{amsmath}
\usepackage{amsfonts}
\usepackage{amssymb}
\usepackage{dsfont}
\usepackage{pifont}
\usepackage{booktabs}
\usepackage{graphicx}
\usepackage{epstopdf}
\usepackage{epsfig}
\usepackage{framed}
\usepackage{makeidx}
\usepackage{siunitx}
\usepackage[capitalise]{cleveref}
\usepackage{hyperref}
\usepackage{placeins}
\usepackage[font=small,labelfont=bf]{caption}
\newcommand{\llb}{\ell\bar{\ell}}
\newcommand{\mll}{m_{\llb}}
\newcommand{\yll}{y_{\llb}}
\newcommand{\sign}{\operatorname{sign}}

\begin{document}
\newgeometry{top=1.5cm,bottom=1.5cm,left=1.5cm,right=1.5cm,bindingoffset=0mm}

\vspace{-2.0cm}
\begin{flushright}
Edinburgh 2022/14 \\
Nikhef-2022-013\\
TIF-UNIMI-2022-13\\
\end{flushright}
\vspace{0.3cm}

\begin{center}
  {\Large \bf Parton Distributions and New Physics Searches: \\[0.3cm]
   the Drell-Yan Forward-Backward Asymmetry as a Case Study}
  \vspace{1.1cm}

   Richard D. Ball,$^{1}$
  Alessandro Candido,$^{2}$
Stefano Forte,$^{2}$
Felix Hekhorn,$^{2}$ \\[0.2cm]
Emanuele R. Nocera,$^{3}$ 
Juan Rojo$^{4,5}$, and
Christopher Schwan$^{6}$\\

 \vspace{0.7cm}
 
 {\it \small

 ~$^1$The Higgs Centre for Theoretical Physics, University of Edinburgh,\\
   JCMB, KB, Mayfield Rd, Edinburgh EH9 3JZ, Scotland\\[0.1cm]
 ~$^2$Tif Lab, Dipartimento di Fisica, Universit\`a di Milano and\\
   INFN, Sezione di Milano, Via Celoria 16, I-20133 Milano, Italy\\[0.1cm]
 ~$^3$ Dipartimento di Fisica, Universit\`a degli Studi di Torino and\\
   INFN, Sezione di Torino, Via Pietro Giuria 1, I-10125 Torino, Italy\\[0.1cm]
 ~$^4$Department of Physics and Astronomy, Vrije Universiteit, NL-1081 HV Amsterdam\\[0.1cm]
 ~$^5$Nikhef Theory Group, Science Park 105, 1098 XG Amsterdam, The Netherlands\\[0.1cm]
 ~$^6$Universit\"at W\"urzburg, Institut f\"ur Theoretische Physik und Astrophysik, 97074 W\"urzburg, Germany\\[0.1cm]

   }

 \vspace{0.7cm}

{\bf \large Abstract}

\end{center}
We discuss the sensitivity of theoretical predictions of observables used in
searches for new physics to parton
distributions (PDFs) at large momentum fraction $x$.
Specifically, we consider the neutral-current Drell-Yan production of
gauge bosons with invariant masses in the TeV range, for which    
the forward-backward asymmetry of charged leptons
from the decay of the gauge boson in its rest frame is a traditional
probe of new physics. We show that the qualitative  behaviour of the asymmetry 
depends strongly on the assumptions made in determining the underlying PDFs.
 We discuss and compare the large-$x$
 behaviour of various different PDF sets, and find that they 
 differ significantly.
 Consequently, the shape of the asymmetry observed at lower
 dilepton invariant masses, where all PDF sets are in reasonable agreement 
because of  the presence of experimental constraints,
 is not necessarily reproduced at large masses where the
 PDFs are mostly unconstrained by data.
 It follows that the shape 
of the asymmetry at high masses may depend on 
assumptions made in the PDF parametrization, 
and thus deviations from the traditionally expected behaviour cannot be taken as a reliable 
indication of new physics.
We demonstrate that forward-backward asymmetry measurements
could help in constraining PDFs at large $x$ and discuss the accuracy that would be required to
disentangle the effects of new physics from uncertainties in the PDFs in this region.

\clearpage

\tableofcontents

\section{Introduction}

An important direction for ongoing and future studies of 
new physics beyond the Standard Model (BSM) at
the Large Hadron Collider (LHC) is the search for novel heavy resonances.
The LHC is uniquely suited to direct searches for these resonances,
thanks to its unparalleled center of mass energy, $\sqrt{s}=13.6$ TeV in the
recently started 
Run III, and the  high statistics to be accumulated in the coming years,
especially in the high-luminosity (HL)
phase.
For instance, considering representative
benchmark BSM scenarios, the HL-LHC is sensitive~\cite{CidVidal:2018eel} to 
searches for sequential Standard Model (SM) 
$W'$ gauge bosons 
up to $m_{W'}=7.8$ TeV, $E_6$ model $Z'$ gauge bosons up to $m_{Z'}=5.7$ TeV, and
Kaluza-Klein resonances decaying into a $t\bar{t}$ pair up to
$m_{KK}=6.6$ TeV.

The production of such high-mass states proceeds via partonic
scattering that involves large 
values of the momentum fractions $x_1$ and $x_2$ of the colliding
partons, because the center of mass energy of the partonic collision
is $\hat s= x_1 x_2 s$.
For instance, the on-shell production  of a state
with invariant mass $m_{X}=8$ TeV requires
$x_1x_2 \gsim 0.3 $, hence for central production at leading order
$x_1=x_2\approx 0.6$. This is problematic because parton distribution functions
(PDFs)~\cite{Gao:2017yyd,Kovarik:2019xvh} are poorly known for $x\gsim
0.4$, as there is limited data
included in current PDF determinations to constrain this kinematic region.
Indeed, in the past, claims of possible BSM signals~\cite{CDF:1996yow} 
were subsequently traced to poor modeling of the PDFs in the large-$x$
region~\cite{Lai:1996mg}.
The impact of lack of knowledge of the PDFs
on BSM searches is thus a delicate issue~\cite{Beenakker:2015rna}.

Here we wish to further investigate this by specifically considering
neutral-current (NC) Drell-Yan (DY) dilepton production and associated
observables, frequently  used for BSM searches at the LHC.
NC Drell-Yan production
is one of the cleanest processes in the search for  both narrow and
broad heavy resonances decaying into dileptons,
$pp \to X \to \ell^+\ell^-$, since 
the two charged leptons can be detected with excellent energy and
angular resolution.
This also enables the search for
smooth, non-resonant distortions with respect to the SM
backgrounds, such as those arising in the context of contact interactions 
or, more generally,
induced by Effective Field Theory
(EFT) higher-dimensional operators that lead to 
direct couplings between quarks and leptons~\cite{Ethier:2021bye,Dawson:2018dxp,Ellis:2020unq,Greljo:2021kvv}.
Indeed, both ATLAS
and CMS have extensively explored this channel in their BSM search
program~\cite{ATLAS:2014gys,ATLAS:2020yat,ATLAS:2019erb,CMS:2021ctt,ATLAS:2021mla,CMS:2018nlk}.
To this purpose, it is mandatory to have a detailed understanding of
the dominant SM background, namely dilepton
production from quark-antiquark annihilation mediated by
a virtual electroweak (EW) boson, $q\bar{q} \to  \gamma^*/Z \to \ell^+\ell^-$,
with subleading processes involving the quark-gluon
and photon-photon initial states.

Drell-Yan production
is one of the SM processes which is known to highest perturbative
accuracy: indeed, both N$^3$LO QCD results~\cite{Duhr:2021vwj}
and the full mixed QCD-EW corrections at NNLO~\cite{Buccioni:2020cfi,Buccioni:2022kgy,Bonciani:2020tvf,Bonciani:2021zzf,Armadillo:2022bgm}
have become available recently.
Therefore, the main uncertainty on
theoretical predictions for this process is mostly due to the
PDFs, which, as mentioned, are poorly known at large $x$.
Experimentally, uncertainties are minimized when considering
observables in which several systematics cancel in part or entirely.
An example relevant for the DY process is the forward-backward asymmetry $A_{\rm fb}$ of the
angular distribution of the dilepton pair in the center-of-mass frame
of the partonic collision, i.e.\ the asymmetry in the
so-called Collins-Soper angle $\theta^*$, recently
measured from the Run II dataset by ATLAS~\cite{ATLAS:2017rue} and CMS~\cite{CMS:2022uul}.
The sensitivity of this observable to both PDFs and BSM signals has
been emphasized
recently~\cite{Fiaschi:2021sin,Fiaschi:2021okg,Accomando:2019vqt,Accomando:2018nig,Fiaschi:2022wgl},
as well as its relevance to extractions of the weak mixing angle
$\sin^2\theta_W$ at the LHC~\cite{CMS:2018ktx}.
These studies  are mostly restricted to the vicinity of the $Z$-boson
peak, $m_{\ell\bar{\ell}} \sim m_Z$ with $m_{\ell\bar{\ell}}$ being the dilepton mass, though in
 a recent study by CMS~\cite{CMS:2022uul} the forward-backward asymmetry has been used
to obtain a lower mass limit
(of 4.4~TeV) on a hypothetical $Z'$ heavy gauge boson.

In this work, we assess to which extent  different assumptions on the large-$x$ behavior of PDFs, as well as different
estimates of the PDF uncertainty in this region, may affect BSM searches,
by specifically studying neutral-current Drell-Yan production, and
the forward-backward asymmetry in particular. To this purpose, we 
explain the dependence of the general
qualitative features of the asymmetry   on the
behavior of PDFs, based on an understanding of the analytic dependence
of the asymmetry on the partonic luminosities.
We then present
detailed computations of the forward-backward asymmetry at the LHC, with realistic experimental
cuts, using a variety of PDF sets.

We find that first, the large-$x$ PDF shape and uncertainty can differ
considerably between different PDF sets, with
NNPDF4.0~\cite{Ball:2021leu} generally displaying a more flexible shape
and a wider uncertainty.
And second, that all PDF sets except NNPDF4.0 lead to a qualitative
behavior
of the asymmetry which in the large-mass
multi-TeV region reproduces the shape found around the $Z$-peak
region, even though there is no fundamental reason why this should be
the case
We will then trace the observed behavior of the asymmetry to that of the
underlying PDFs.

The structure of the paper is the following.
First  in Sect.~\ref{sec:HMDY} we review the leading-order (LO) expressions
for the Drell-Yan differential distributions and forward-backward asymmetry, in order to 
explain  how the leading qualitative behavior of the
asymmetry  --- specifically the reason for an asymmetry, and
its sign --- is related to the underlying parton luminosities. We 
will also show that this LO picture is not qualitatively modified 
by higher-order perturbative corrections. 
Then in Sect.~\ref{sec:largexpdfs} we investigate the way the
shape of the asymmetry (and specifically its sign) is determined
by the large-$x$ behavior of the PDFs. After discussing this in a 
toy model, we examine current PDF sets:
ABMP16~\cite{Alekhin:2017kpj},
CT18~\cite{Hou:2019efy}, MSHT20~\cite{Bailey:2020ooq}, and
NNPDF4.0.
Specifically, we  compare the
behavior of the PDFs and the asymmetry as the
final-state dilepton invariant mass is varied.
Finally, in Sect.~\ref{sec:afb} we present predictions 
for high-mass DY production, specifically 
the forward-backward asymmetry, at the LHC with realistic experimental
cuts, and accounting for NLO QCD and electroweak corrections.
For completeness, we present in App.~\ref{app:nnpdf31} a comparison to results
obtained using the previous, widely used  NNPDF3.1 PDF set.

\section{Anatomy of Drell-Yan production}
\label{sec:HMDY}

The aim of this  section is
to scrutinize the PDF dependence of the neutral-current
Drell-Yan differential cross-section and of the associated forward-backward asymmetry by
reviewing the LO kinematics, determining LO analytic expressions, and finally
comparing these analytical calculations to the results of LO and NLO  numerical simulations
obtained  using {\sc\small MadGraph5\_aMC@NLO}~\cite{Alwall:2014hca}
interfaced to {\sc\small PineAPPL}~\cite{Carrazza:2020gss,christopher_schwan_2022_7023438}.
Specifically, we will relate the behavior of the differential
distribution and asymmetry to the relevant parton luminosities.

\subsection{Drell-Yan kinematics and cross-sections at LO}
\label{sec:dylo}

\begin{figure}[t]
\centering

\begin{tikzpicture}
\begin{feynman}
  \tikzfeynmanset{large}

  \vertex (b);
  \vertex [above left=of b] (a) {\(q\)};
  \vertex [below left=of b] (f1) {\(\bar{q}\)};
  \vertex [right=of b] (c);
  \vertex [above right=of c] (f2) {\(\ell\)};
  \vertex [below right=of c] (f3) {\(\bar{\ell}\)};

  \diagram* {
  (a) -- [fermion, momentum=\(\hat{k}_1\)] (b) -- [fermion, rmomentum=\(\hat{k}_2\)] (f1),
  (b) -- [boson, edge label'=\(\gamma / Z\), momentum=\(q\)] (c),
  (c) -- [anti fermion, momentum=\(p_{\ell}\)] (f2),
  (c) -- [fermion, momentum=\(p_{\bar{\ell}}\)] (f3),
  };

\end{feynman}
\end{tikzpicture}

\caption{Neutral-current Drell-Yan production at LO in the quark-antiquark channel.}
\label{fig:lo-dy}
\end{figure}
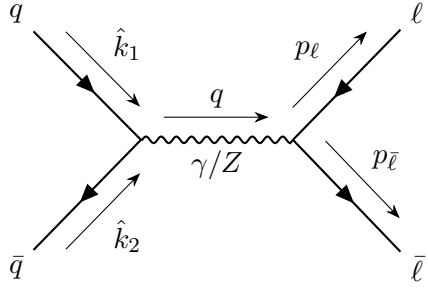

We consider dilepton production via the exchange of an electroweak
neutral gauge boson $Z/\gamma^*$
in proton-proton collisions:
\begin{equation}
  \label{eq:DYprocess}
  \mathrm{p}(k_1) + \mathrm{p}(k_2) \to Z/\gamma^*(q) \to \ell(p_{\ell}) + \bar{\ell}(p_{\bar{\ell}}) + X \text{.}
\end{equation}
The hadronic differential cross-section $\d\sigma^{\mathrm{p}\mathrm{p}\to\llb}$  is factorized in
terms of PDFs $f_i$ and the partonic cross sections
$\d\hat\sigma_{ij}$ for incoming partons of species $i,\,j$ as
\begin{equation}
  \d\sigma^{\mathrm{p}\mathrm{p}\to\llb} = \sum_{ij} \int\limits_0^1\!\d x_1 \d
  x_2 f_i(x_1,\mu_F^2) f_j(x_2,\mu_F^2) \d\hat\sigma_{ij}(\hat k_1 = x_1
  k_1, \hat k_2 = x_2 k_2).
  \label{eq:factorization}
\end{equation}
In the sequel we will set the  factorization scale $\mu_F$ to the
invariant mass of the gauge boson, i.e.\ the dilepton
invariant mass, so $\mu^2_F = \mll^2=(p_\ell + p_{\bar{\ell}})^2$.
The kinematics and Feynman diagram of the LO partonic process
in the quark-antiquark channel are shown in \cref{fig:lo-dy}.
We do not consider photon-initiated processes, as they do not affect
the qualitative features of our discussion.

%----------------------

%----------------------

At LO, the momentum fractions of the two incoming partons are fully
fixed by knowledge of the invariant mass and rapidity of the gauge
boson, i.e.\ of the dilepton pair  $\yll = (y_\ell + y_{\bar{\ell}})/2$: 
\begin{equation}
  \label{eq:x_fractions}
  x_1 = \frac{ \mll}{\sqrt{s}}\exp(\yll) \, ,\quad x_2 = \frac{
  \mll}{\sqrt{s}}\exp(-\yll) \, ,
\end{equation}
where the center of mass energy of the hadronic collision is
$s=(k_1+k_2)^2$ and at LO
$\mll^2 = \hat s = x_1 x_2 s$. The absolute dilepton
rapidity thus lies in the range $|\yll|\le \ln (\sqrt{s}/\mll)$.
Beyond LO there might be extra radiation in the final state, so the LO
kinematics provides a lower bound on the momentum fractions of the
incoming partons, and all values of the momentum
fractions such that $x_{1,2}\ge \mll/\sqrt s$ are allowed.

It is useful to define the so-called  Collins-Soper
angle $\theta^*$~\cite{Collins:1977iv}, which in the hadronic center-of-mass
(CoM) frame is defined as
\begin{equation}
\begin{split}
  \cos\theta^* &= \sign (\yll) \cos\theta \, ,\\
  \cos\theta &\equiv\frac{p_\ell^+ p_{\bar{\ell}}^- - p_\ell^- p_{\bar{\ell}}^+}{\mll \sqrt{\mll^2 + p_{\mathrm{T},\ell\bar{\ell}}^2}} \text{,} \quad p^\pm = p^0 \pm p^3 \text{.}
  \label{eq:cosine-cs-angle}
\end{split}
\end{equation}
It is easy to show that the Collins-Soper angle $\theta^*$ coincides with the
scattering angle of the lepton in the partonic CoM frame, $\bar\theta$.
The latter is  
defined in terms of the lepton momentum as 
\begin{equation}
 \cos\bar\theta \equiv \frac{p^z_\ell}{\mll} \,, \label{eq:coscm}
\end{equation}
where the $z$ axis is along the direction of the incoming
quark-antiquark pair.
In the partonic CoM frame, of course,
$p^z_\ell=-p^z_{\bar \ell}$ and $\yll =0$, so
\begin{equation}
p^\pm_\ell=p^{\mp}_{\bar{\ell}}=  \mll \left( 1\pm \cos{\bar\theta} \right)\, ,
\end{equation}
and substituting in
\cref{eq:cosine-cs-angle} it immediately follows that, taking
the convention 
$\sign (\yll)=\sign(0)=+1$, 
$\cos\theta^*=\cos\theta=\cos{\bar\theta}$.
The expression of $\cos\theta$ in \cref{eq:cosine-cs-angle} is
manifestly invariant upon boosts along the $z$ axis, so the
identification of $\theta$ with the CoM scattering angle $\bar\theta$
remains true in any reference frame.

Note that the definition
\cref{eq:coscm} requires a choice for the positive direction of
the $z$ axis, which is usually taken along the direction of the
incoming fermion (quark).
This direction is  not experimentally accessible in proton-proton collisions,
so the Collins-Soper angle is defined by always
taking the positive $z$ axis in the direction of the boosted dilepton
pair, i.e., at LO, along the direction of the incoming quark with
largest momentum fraction, i.e.\ by supplementing in the definition
a factor $\sign(\yll)$.
Hence $\cos\theta^*=\cos{\bar\theta}$ ($\cos\theta^*=-\cos{\bar\theta}$)  if
the momentum fraction of the incoming quark (antiquark) is the largest.

The hard scattering matrix elements that enter the partonic cross-section
in \cref{eq:factorization} are the sum of a pure photon-exchange
contribution, a photon-$Z$ interference term, and a pure $Z$-exchange
contribution.
Of course, in the region
$\mll \gtrsim m_Z$ these contributions are all of the same order.
Standard arguments~\cite{Peskin:1995ev} then imply that, because in the
Standard Model the photon
coupling to leptons is vector  while the $Z$ coupling is chiral,
the pure photon and pure $Z$ contributions to the cross-section are
necessarily  even in $\cos\theta^*$ while the interference term is
odd.

\begin{figure}
  \centering
  \includegraphics[width=0.49\linewidth]{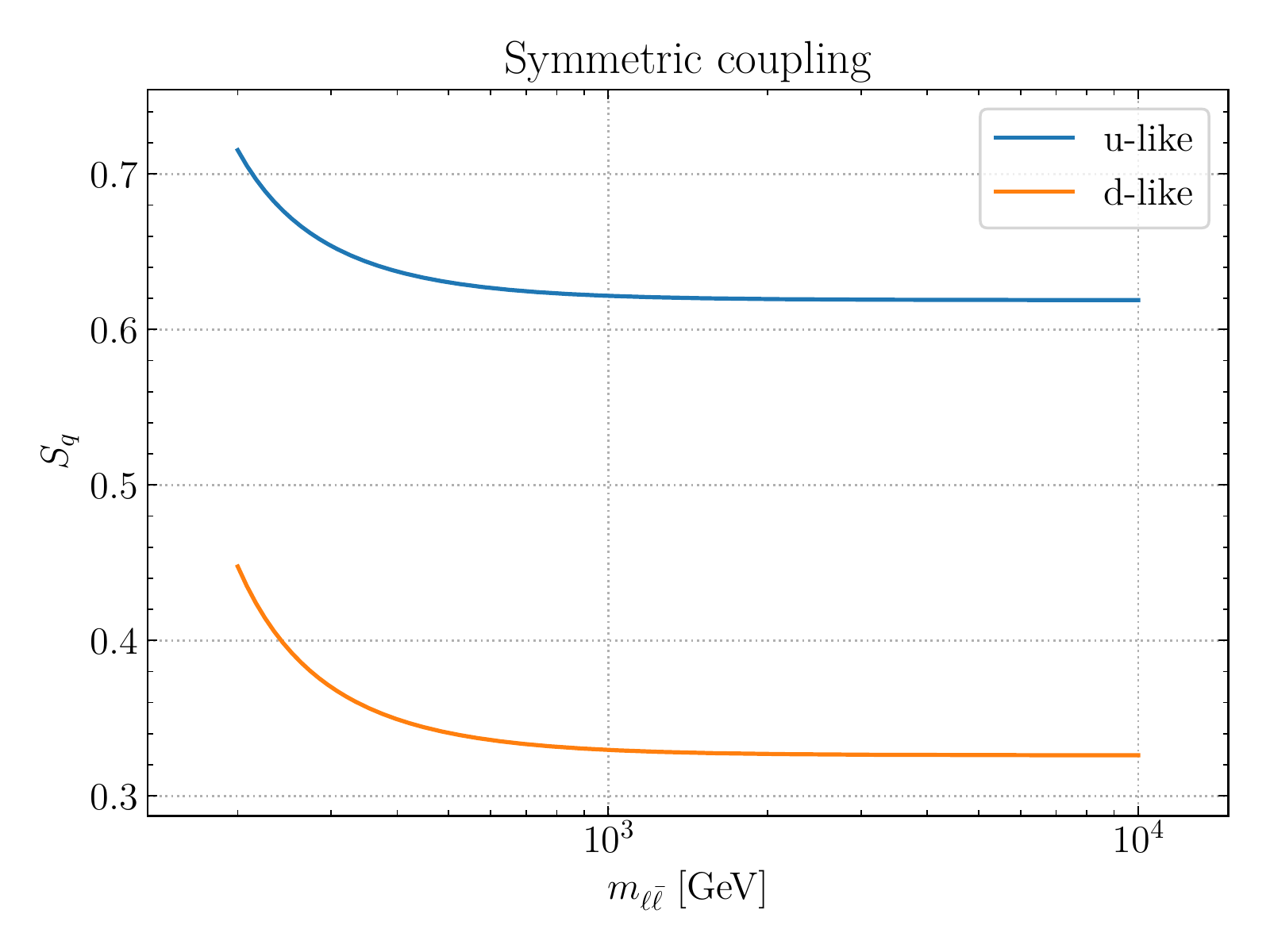}
  \includegraphics[width=0.49\linewidth]{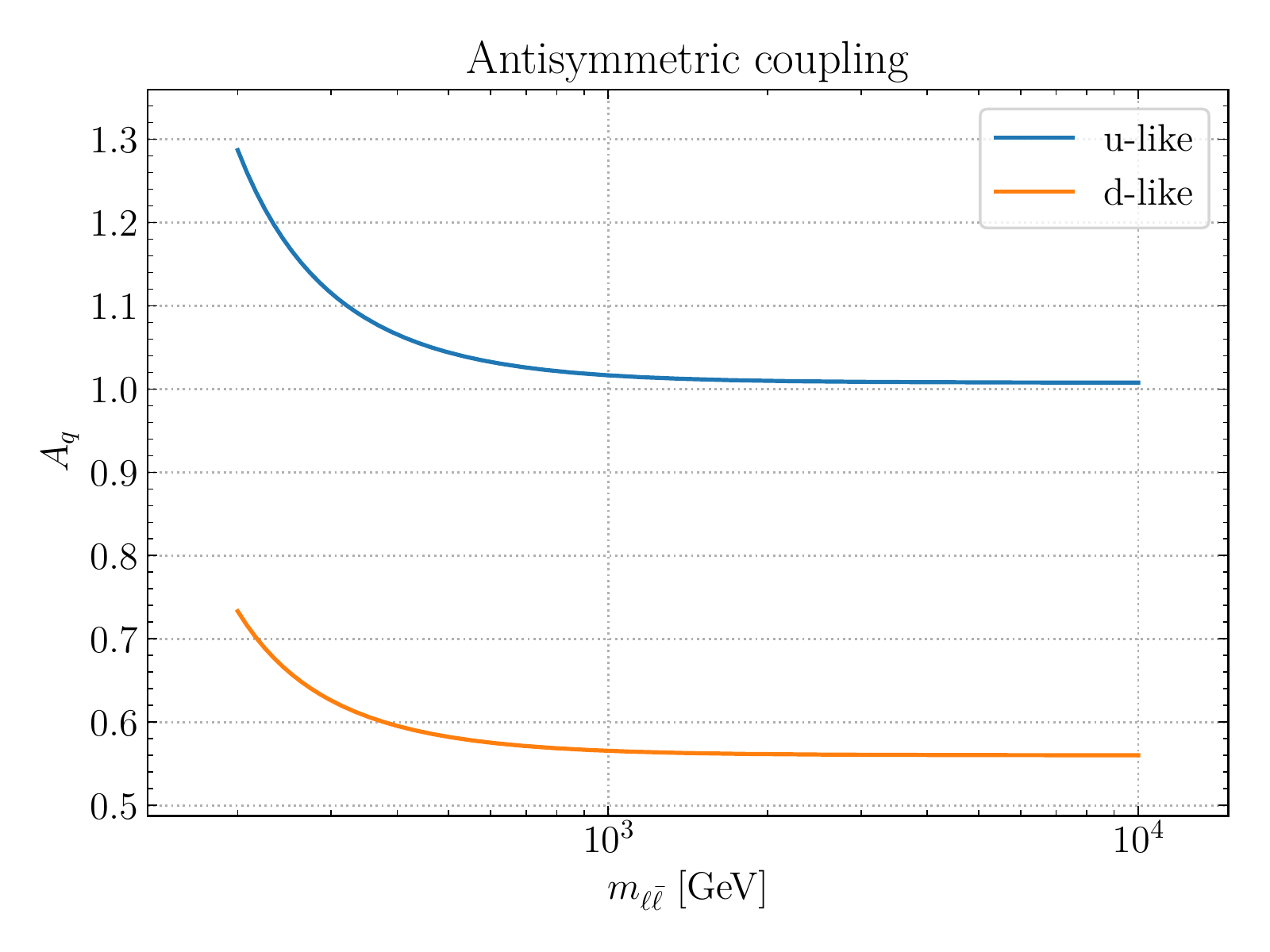}
  \caption{The symmetric $S_q$ (left) and antisymmetric $A_q$ (right)
    couplings, \cref{eq:coup}, for up-like and
    down-like quarks, as a function of 
 the dilepton invariant mass $\mll$.
  }
  \label{fig:lo-couplings}
\end{figure}

Specifically, at LO the fully differential hadronic cross-section can
be obtained from the well-known result~\cite{Peskin:1995ev} for
$e^+e^-\to\mu^+\mu^-$ by replacing the incoming lepton charges  with
those of the quarks, and accounting for the PDFs, with
the result
\begin{align}
    \frac{\d^3 \sigma}{\d \mll \, \d \yll \, \d\cos\theta^*} &= \frac{\pi \alpha^2}{3 \mll s} \left((1+\cos^2({\theta^*})) \sum_q S_q \left[f_q(x_1,\mll^2) f_{\barq}(x_2,\mll^2) + f_q(x_2,\mll^2) f_{\barq}(x_1,\mll^2) \right] \right. \nonumber\\
    &\hspace*{15pt} + \left. \cos\theta^* \sum_q A_q \sign (\yll) \left[ f_q(x_1,\mll^2) f_{\barq}(x_2,\mll^2) - f_q(x_2,\mll^2) f_{\barq}(x_1,\mll^2)\right] \right) \, ,
    \label{eq:lo-triple-diff}
\end{align}
where  $\alpha$ is the QED coupling and the even (symmetric) and
odd  (antisymmetric) couplings are given by
\begin{align}
  \label{eq:coup}
    S_q &= e_l^2 e_q^2 + P_{\gamma Z} \cdot  e_l v_l e_q v_q + P_{ZZ} \cdot  (v_l^2+a_l^2)(v_q^2+a_q^2) \nonumber \\
    A_q &= P_{\gamma Z} \cdot 2 e_l a_l e_q a_q  + P_{ZZ} \cdot 8 v_l a_l  v_q a_q \, ,
\end{align}
in terms of the electric charges  $e_l$, $e_q$ and the vector and
axial couplings $v_l$, $v_q$ and $a_l$, $a_q$  of the leptons and
quarks, and the propagator factors
\begin{align}\label{eq:propgz}
    P_{\gamma Z}(\mll) &= \frac{2\mll^2 (\mll^2  - m_Z^2)}{\sin^2(\theta_W) \cos^2(\theta_W)\left[(\mll^2 - m_Z^2)^2 + \Gamma_Z^2 m_Z^2\right]}\\
\label{eq:propzz}
    P_{ZZ}(\mll) &= \frac{\mll^4}{\sin^4(\theta_W) \cos^4(\theta_W)\left[(\mll^2 - m_Z^2)^2 + \Gamma_Z^2 m_Z^2\right]},
\end{align}
with $m_Z$  and $\Gamma_Z$ respectively the $Z$ mass and width and $\theta_W$ the weak mixing angle.
In \cref{fig:lo-couplings} we display the
symmetric $S_q$ (left) and antisymmetric $A_q$ (right)
couplings, \cref{eq:coup}, for up-like and
down-like quarks, as a function of 
the dilepton invariant mass $\mll$.
Both couplings are around a factor 2 larger for
up-like quarks than for down-like quarks, and
become $\mll$-independent for $\mll \gsim 1$ TeV, where they take
the asymptotic values $\bar S_q$, $\bar A_q$ obtained by
substituting in \cref{eq:coup} the large-mass expressions of
the propagator factors
\begin{equation}\label{eq:propasympt}
    \bar P_{\gamma Z} = \frac{2}{\sin^2(\theta_W)\cos^2(\theta_W)},
    \qquad 
    \bar P_{ZZ} = \frac{1}{\sin^4(\theta_W) \cos^4(\theta_W)},
\end{equation}
to which $P_{\gamma Z}$ and $P_{ZZ}$ respectively reduce up to $O(m^2_Z/\mll^2)$ corrections.

%%%%%%%%%%%%%%%%%%%%%%%%%%%%%%%%%%%%%%%%%%%%%%%%%%%%
%%%%%%%%%%%%$$$$$$$$$$$$$$$$$$$$$%%%%%%%%%%%%%

The interference term proportional to
$A_q$ is odd in the Collins-Soper angle $\cos\theta^*$, leading to a forward-backward
scattering asymmetry.
In a proton-proton collision the initial state is completely
symmetric, so the quark and antiquark contributions to the
cross-section \cref{eq:lo-triple-diff} are necessarily symmetric
upon the interchange of the incoming quark and antiquark, with the
corresponding momentum fractions fixed at LO by
\cref{eq:x_fractions}.
However, as mentioned, there
is a sign change in the relation between $\cos\theta^*$ and
$\cos\theta$ according to whether the incoming parton with largest
momentum fraction is a quark or an antiquark, i.e.,
when interchanging
$x_1$ with $x_2$ in the argument of the quark and antiquark PDFs,
thereby leading to the result of  \cref{eq:lo-triple-diff}.
This leads
to a forward-backward asymmetry whenever the quark and antiquark
PDFs have different $x$ dependence.

In order to understand the relation of this forward-backward
asymmetry in terms of the
behavior of the PDFs, it is convenient
to rewrite the PDF combinations that contribute to the differential
cross-section \cref{eq:lo-triple-diff} in terms of symmetric and
antisymmetric parton luminosities, defined as
\begin{align}
  \mathcal{L}_{S,q}(\mll, \yll) &\equiv f_q(x_1,\mll^2) f_{\barq}(x_2,\mll^2) + f_q(x_2,\mll^2) f_{\barq}(x_1,\mll^2) \, ,
  \nonumber\\
  \mathcal{L}_{A,q}(\mll, \yll) &\equiv \sign (\yll) \left[ f_q(x_1,\mll^2) f_{\barq}(x_2,\mll^2) - f_q(x_2,\mll^2) f_{\barq}(x_1,\mll^2)\right] \, , \label{eq:symm_asymm_lumis}
\end{align}
where the momentum fractions $x_1$ and $x_2$ are given in terms of $\mll$, $\yll$,
and $\sqrt{s}$ in \cref{eq:x_fractions}.
Note that both parton luminosities are invariant under
the interchange $x_1\leftrightarrow x_2$, upon which $\yll \to -\yll$.
In terms of these luminosities, the triple differential cross-section \cref{eq:lo-triple-diff}
takes the compact form
\begin{equation}
  \label{eq:lo-triple-diff-lumis}
  \frac{\d^3 \sigma}{\d \mll \, \d \yll \, \d\cos\theta^*} =
  \frac{\pi \alpha^2}{3 \mll s} \left( (1+\cos^2({\theta^*})) \sum_q S_q \mathcal{L}_{S,q}(\mll, \yll)
  + \cos\theta^* \sum_q A_q \mathcal{L}_{A,q}(\mll, \yll)  \right) \, ,
\end{equation}
which explicitly displays
its symmetry properties upon the transformation $\cos\theta^* \to -\cos\theta^*$,
equivalent to a charge conjugation transformation 
$q\leftrightarrow \bar q$ and $\ell \leftrightarrow \bar{\ell} $.

The symmetric and antisymmetric parton luminosities \cref{eq:symm_asymm_lumis} can also be expressed
in terms of the sum and difference of quark and antiquark PDFs,
\begin{equation}
  \label{eq:fqpm}
  f_{q}^\pm \left( x, Q\right) = f_{q} \left( x, Q\right) \pm f_{\bar{q}} \left( x, Q\right) \, ,
\end{equation}
where $f_{q}^-$ is usually called the valence PDF combination,
and $f_{q}^+$  the total quark PDF\@. Note that at LO, and more
generally in factorization schemes in which PDFs are
positive, such as $\overline{\rm MS}$~\cite{Candido:2020yat}, $f^+_q$
is positive while $f_q^-$
in general is not, and $f_{q}^+>|f_{q}^-|$.
We can write the symmetric and antisymmetric parton luminosities in
\cref{eq:symm_asymm_lumis} as
\begin{align}
  \mathcal{L}_{S,q}(\mll, \yll) &= \frac {1} 2 \left( f_q^+(x_1,\mll^2) f_{q}^+(x_2,\mll^2) - f_q^-(x_2,\mll^2) f_{q}^-(x_1,\mll^2)  \right) \, \label{eq:lumiss_qpm}\\
  \mathcal{L}_{A,q}(\mll, \yll) &= \frac {\sign (\yll)} 2 \left( f_q^-(x_1,\mll^2) f_{q}^+(x_2,\mll^2) - f_q^-(x_2,\mll^2) f_{q}^+(x_1,\mll^2)  \, \right) \,. \label{eq:lumisa_qpm}
\end{align}

The symmetric luminosity $\mathcal{L}_{S,q}$ is of course positive,
and it is
dominated by the $f_q^+(x_1,\mll^2) f_{q}^+(x_2,\mll^2)$ term, which
is always larger than the valence contribution 
$f_q^-(x_2,\mll^2) f_{q}^-(x_1,\mll^2)$.
The sign of the antisymmetric combination, that in turn
drives the sign of the
forward-backward asymmetry, is in general not determined
uniquely.
If $x_1$ is in the region of the valence peak, and $x_2$ in
the small $x$ region, then $f^-(x_1,\mll^2)\gg f^-(x_2,\mll^2)$, and the
antisymmetric luminosity is positive provided only that the valence PDF is positive.
As we will discuss in Sect.~\ref{sec:largexpdfs},
while this is indeed the case in
the $Z$-peak region, it is actually not necessarily the case in the
high dilepton mass region relevant for BSM searches.

\subsection{Single-differential distributions and the forward-backward asymmetry}
\label{sec:numlo}
Starting from the triple differential cross section,
\cref{eq:lo-triple-diff-lumis}, one can define 
single differential distributions by integrating the other two kinematic variables
over the available phase space.
In particular, the single-differential distribution in the
Collins-Soper angle $\theta^*$ is given by
\begin{equation}
  \label{eq:dsigma-dcos}
  \frac{\d\sigma}{\d\cos\theta^*} = \int\limits_{\mll^{{\rm min}}}^{\sqrt s}\!\d\mll\!\!\int\limits_{\ln(\mll/\sqrt s)}^{\ln(\sqrt s/\mll)}\!\!\d\yll\, \frac{\d^3 \sigma}{\d \mll \, \d \yll \, \d\cos\theta^*} \, ,
\end{equation}
where $\mll^{{\rm min}}$ is a lower kinematic cut in the dilepton invariant mass.
Since \cref{eq:lo-triple-diff-lumis}
falls off steeply with $\mll$, the region with $\mll \gsim \mll^{{\rm min}}$
will dominate the integral.
Given that the dependence of the fully differential cross-section
\cref{eq:lo-triple-diff-lumis}
on  the Collins-Soper angle factorizes with respect to the PDF
dependence, the integration over rapidity and invariant mass does not
affect the  $\cos\theta^*$ dependence, and the single-differential
cross section \cref{eq:dsigma-dcos} takes the simple form 
\begin{equation}
  \label{eq:dsigma-dcos-v2}
  \frac{\d\sigma}{\d\cos\theta^*} = (1+\cos^2\theta^*)\sum_q g_{S,q} + \cos\theta^*\sum_q g_{A,q} \, ,
\end{equation}
where the symmetric and antisymmetric coefficients $g_{S,q}$ and $g_{A,q}$ depend on the quark flavor
and on the invariant mass cut $\mll^{{\rm min}}$, but not on the
Collins-Soper angle itself.
The contributions relevant for the forward-backward asymmetry, $g_{A,q}$,
are given at LO by
\begin{equation}
\label{eq:gAq_integrated_1}
g_{A,q} =\frac{\pi \alpha^2}{3 s} \int\limits_{\mll^{{\rm min}}}^{\sqrt s}\frac{\d\mll}{\mll}A_{q}(\mll)\!\!\int\limits_{\ln(\mll/\sqrt s)}^{\ln(\sqrt s/\mll)}\!\!\d\yll\, \mathcal{L}_{A,q}(\mll,\yll) \, ,
\end{equation}
which in the large-$\mll$ region,
 expressing the longitudinal momentum integration in terms of
$x_1$ (assuming $x_1~\ge x_2$), becomes
\begin{equation}
\label{eq:gAq_integrated}
g_{A,q} = \frac{\pi \alpha^2\bar A_q}{3 s} \int\limits_{\mll^{{\rm min}}}^{\sqrt s}\frac{\d\mll}{\mll} \!\!\int\limits_{\mll/\sqrt s}^{1}\!\!\frac{{\rm d}x_1}{x_1}\, \mathcal{L}_{A,q}(\mll,x_1)+\mathcal{O} \left(\frac{m_Z^2}{\mll^2}\right) \, ,
\end{equation}
where the  $\mll$-independent effective couplings $\bar A_q$  are
given substituting in \cref{eq:coup} the expressions for
the asymptotic propagator factors \cref{eq:propasympt}.

Upon integration over the Collins-Soper angle, the
antisymmetric contribution vanishes: so for instance the
rapidity distribution
\begin{equation}
  \label{eq:dsigma-dyll}
  \frac{\d\sigma}{\d\yll} = \int\limits_{\mll^{{\rm min}}}^{\sqrt s}\!\d\mll \int\limits_{-1}^{1}\!\!\d\cos\theta^*\, \frac{\d^3 \sigma}{\d \mll \, \d \yll \, \d\cos\theta^*} \, ,
\end{equation}
does not depend on terms proportional to $A_q$.
Hence, for BSM searches in which one is
interested in the interference terms, as well as for PDF studies in which one is
interested in the valence-sea separation, the forward-backward
asymmetry is especially relevant.
This observable is
defined at the differential level as
\begin{equation}
  A_{\rm fb}(\cos\theta^*) \equiv \frac{ \frac{\d\sigma}{\d\cos\theta^*}(\cos\theta^*)
  - \frac{\d\sigma}{\d\cos\theta^*}(-\cos\theta^*)}{ \frac{\d\sigma}{\d\cos\theta^*}(\cos\theta^*)
  + \frac{\d\sigma}{\d\cos\theta^*}(-\cos\theta^*) } \, ,\quad \cos\theta^*>0 \, ,
  \label{eq:forward-backward-asymmetry}
\end{equation}
which in terms of the coefficients introduced in
\cref{eq:dsigma-dcos-v2} is given at LO by
\begin{equation}
  \label{eq:afb_lo}
  A_{\rm fb}(\cos\theta^*)   = \frac{\cos\theta^*}{(1+\cos^2(\theta^*))}\frac{\sum_q g_{A,q} }{\sum_{q'} g_{S,q'}} \, ,\quad \cos\theta^*>0 \,.
\end{equation}
This  shows that the dependence on $\cos\theta^*$ factorizes
and the PDF dependence only appears as an overall normalization factor
depending on the ratio of $\sum_q g_{A,q}$
and $\sum_q g_{S,q}$, which in turn depend on the antisymmetric and symmetric
partonic luminosities $ \mathcal{L}_{A,q}$ and $ \mathcal{L}_{S,q}$ respectively.
Note that the overall sign of $A_{\rm fb}$ remains in general undetermined.

%%%%%%%%%%%%%%%%%%%%%%%%%%%%%%%%%%%%%%%
\begin{figure}[t]
  \centering
  \includegraphics[width=0.49\linewidth]{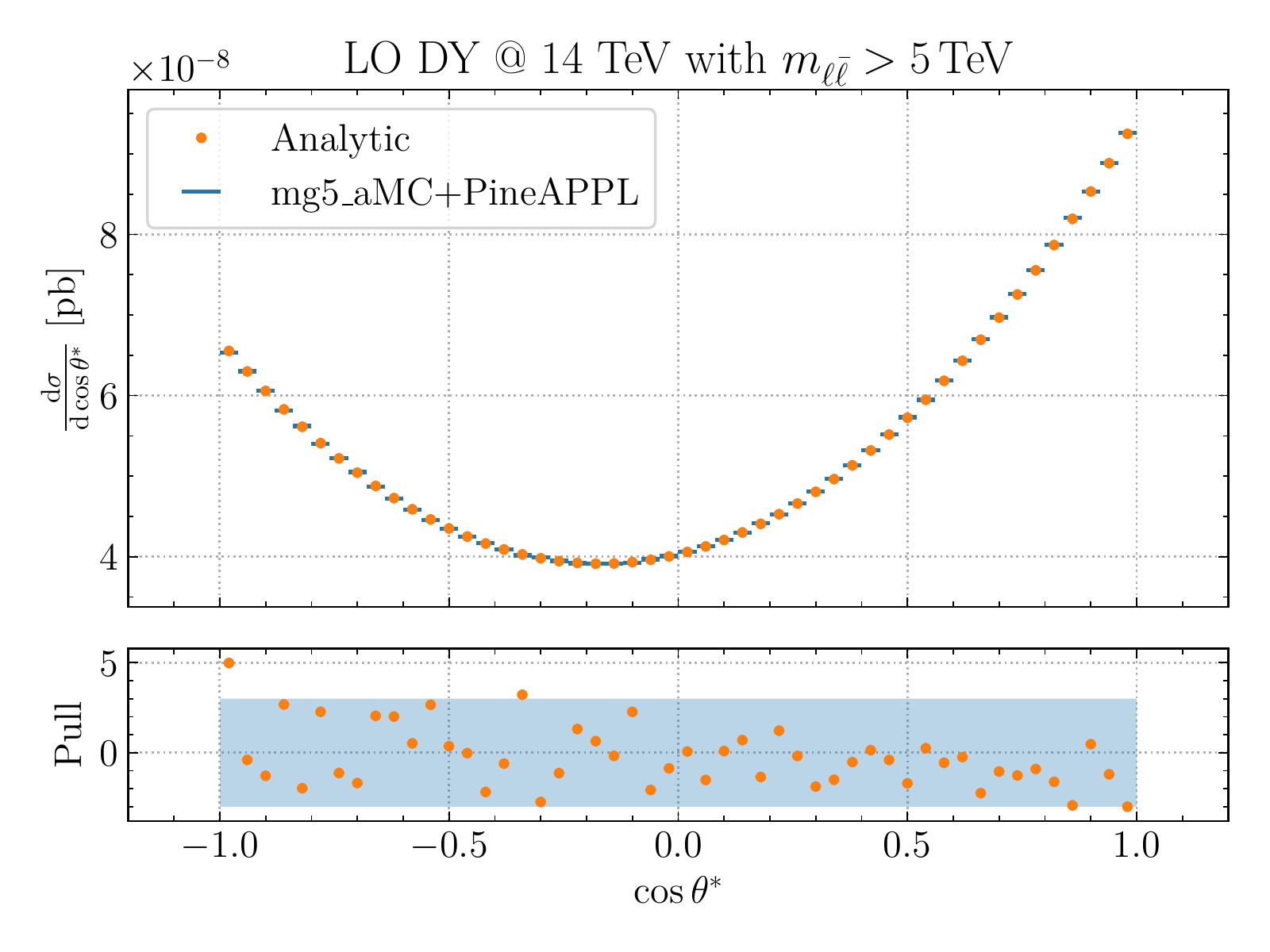}
  \includegraphics[width=0.49\linewidth]{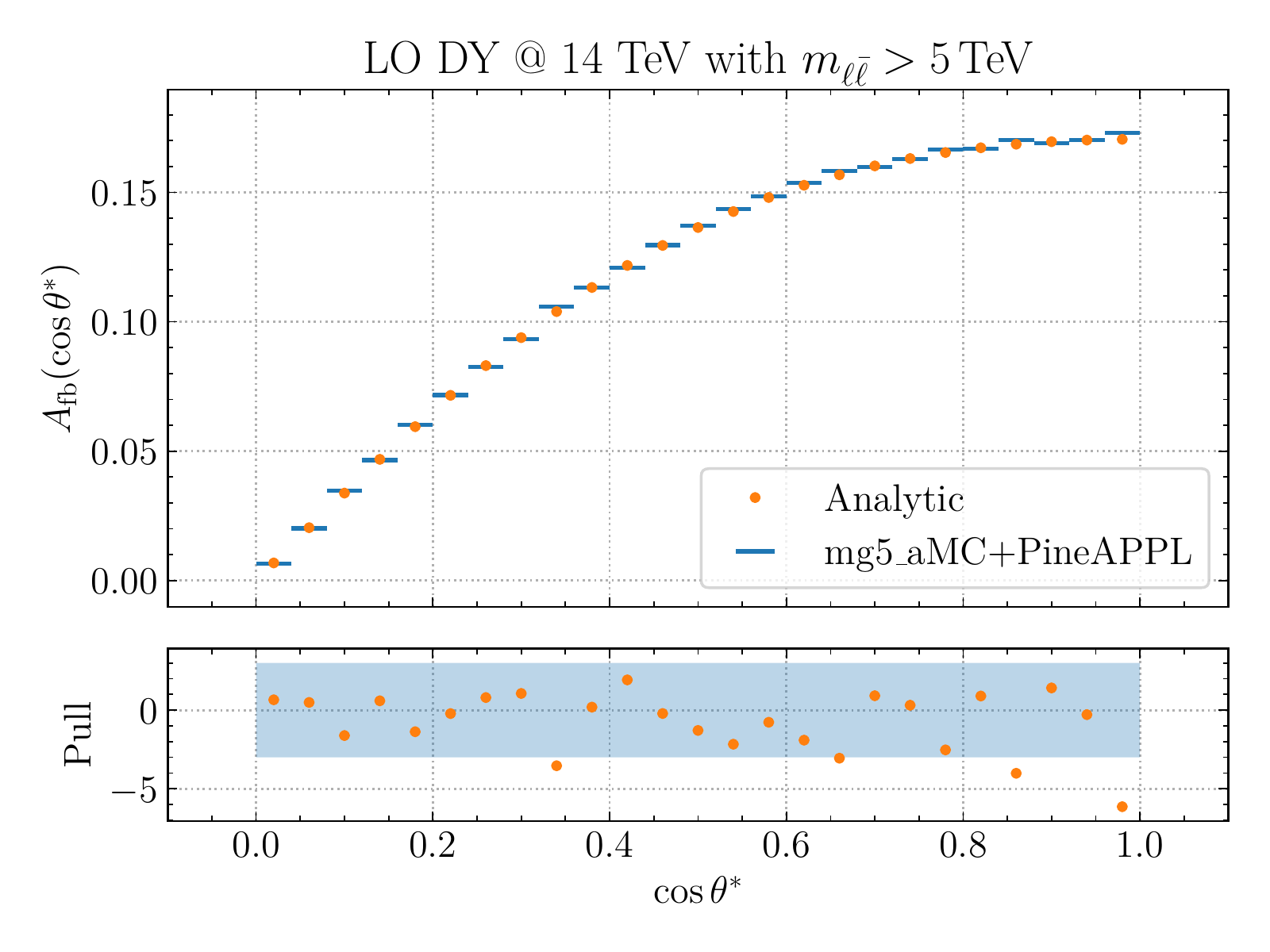}
  \caption{The single-inclusive differential distribution in
    the Collins-Soper angle $\cos\theta^*$,
    \cref{eq:dsigma-dcos},
    and the corresponding forward-backward asymmetry computed at LO,
    where the analytic calculation  \cref{eq:forward-backward-asymmetry}
    is compared with the numerical simulation based on 
    {\sc\small MadGraph5\_aMC@NLO}
    interfaced to {\sc\small PineAPPL}.
    The bottom panels display the difference between the analytic and
    numerical calculations relative to the Monte Carlo integration uncertainty.
    The blue band indicates the $3\sigma$ uncertainty interval.
    One of the replicas of the NNPDF4.0 NNLO PDF set is used as input
    to the calculation.
  }    
  \label{fig:lo-diff-cos}
\end{figure}
%%%%%%%%%%%%%%%%%%%%%%%%%%%%%%%

\begin{figure}[t]
  \centering
  \includegraphics[width=0.49\linewidth]{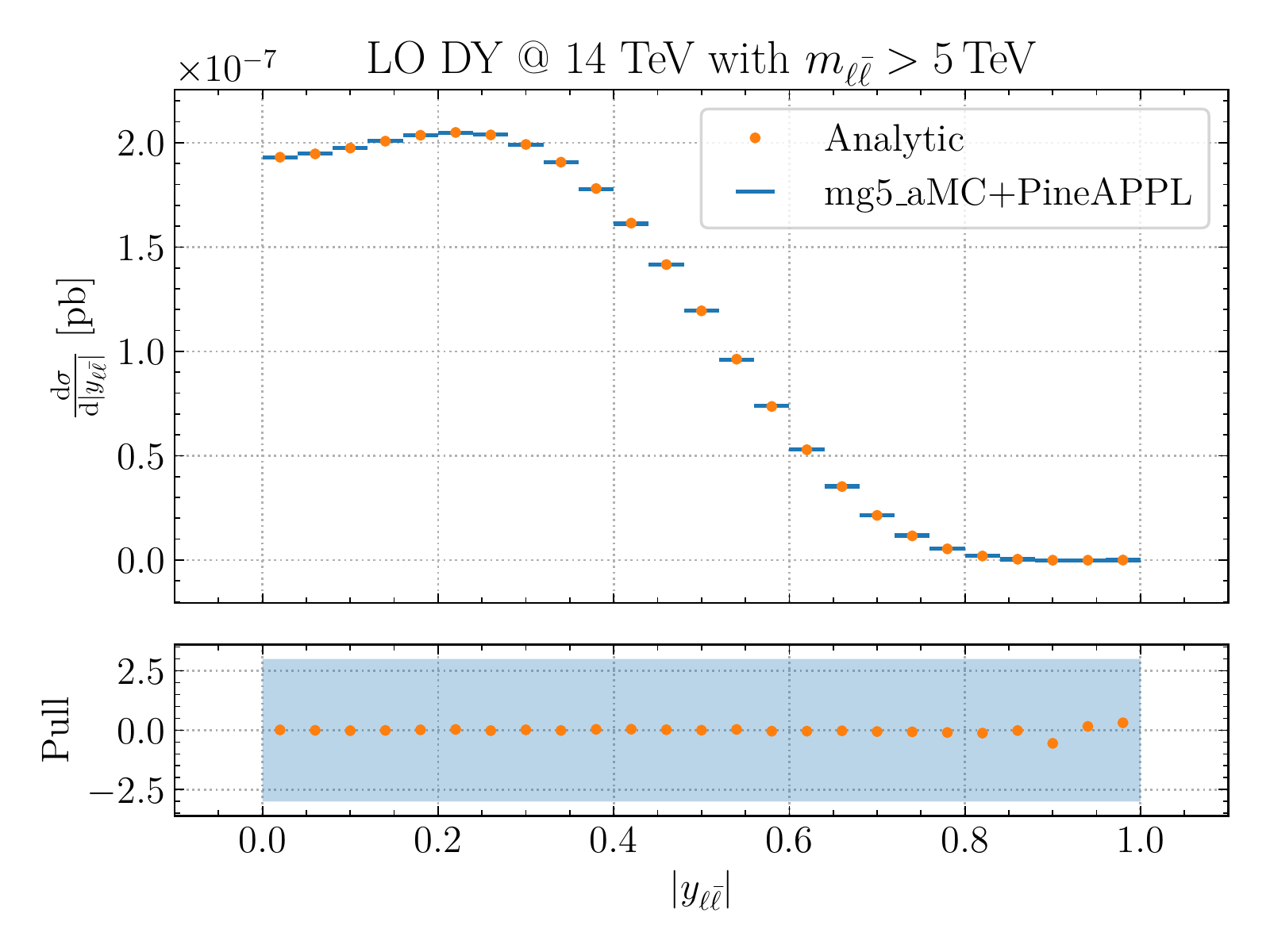}
  \caption{Same as \cref{fig:lo-diff-cos} but now for the absolute dilepton rapidity distribution $|\yll|$}
  \label{fig:lo-diff-yll}
\end{figure}

In order to illustrate concretely these results,  
in \cref{fig:lo-diff-cos} we display the
single-inclusive differential distribution in $\cos\theta^*$,
\cref{eq:dsigma-dcos},
and the corresponding forward-backward asymmetry,
\cref{eq:forward-backward-asymmetry} evaluated at LO
for $\mll^{\rm min}=\SI{5}{\TeV}$. The single-differential rapidity
distribution Eq.~(\ref{eq:dsigma-dyll}) is also shown for reference in Fig.~\ref{fig:lo-diff-yll}.
We display both a numerical  evaluation based on
 {\sc\small MadGraph5\_aMC@NLO}
interfaced to {\sc\small
  PineAPPL},
as well as analytic results found using the form \cref{eq:lo-triple-diff-lumis} of the triple
differential luminosity, with all the values of the parameters
entering \crefrange{eq:coup}{eq:propzz} set to the values used in
the {\sc\small MadGraph5\_aMC@NLO} runcard, and performing  numerically
the integrals in \cref{eq:dsigma-dcos,eq:dsigma-dyll}.
For validation purposes, no kinematic cuts are applied to the
rapidities and transverse momenta of final-state leptons.
The PDF input is taken to be given, for illustrative purposes, by
one of the replicas of the NNPDF4.0 NNLO set.
The relative difference between the analytic and
numerical calculation is shown in the bottom panels of
\cref{fig:lo-diff-cos} and demonstrates perfect agreement. 

\begin{figure}[t]
  \centering
  \includegraphics[width=0.49\linewidth]{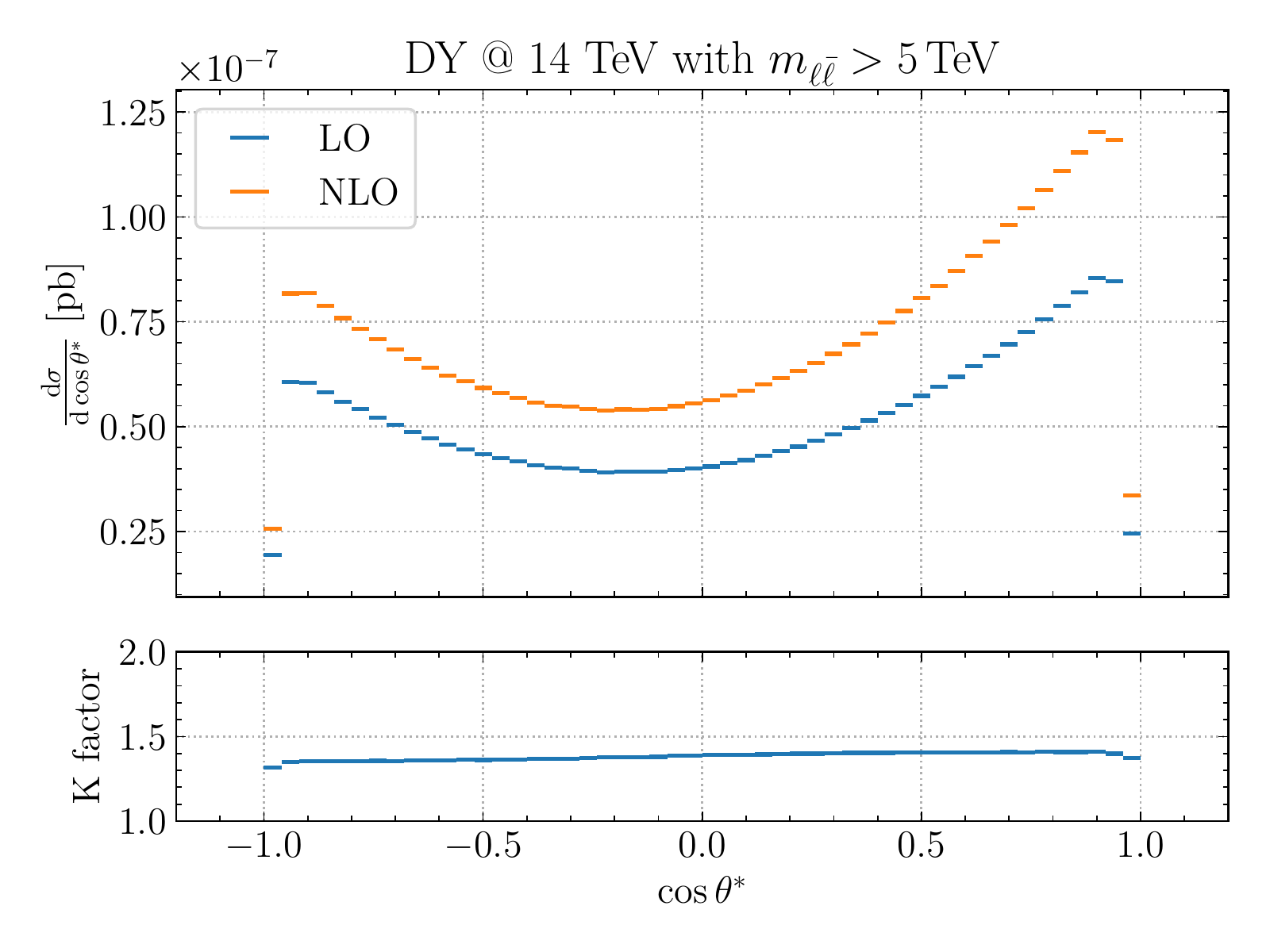}
  \includegraphics[width=0.49\linewidth]{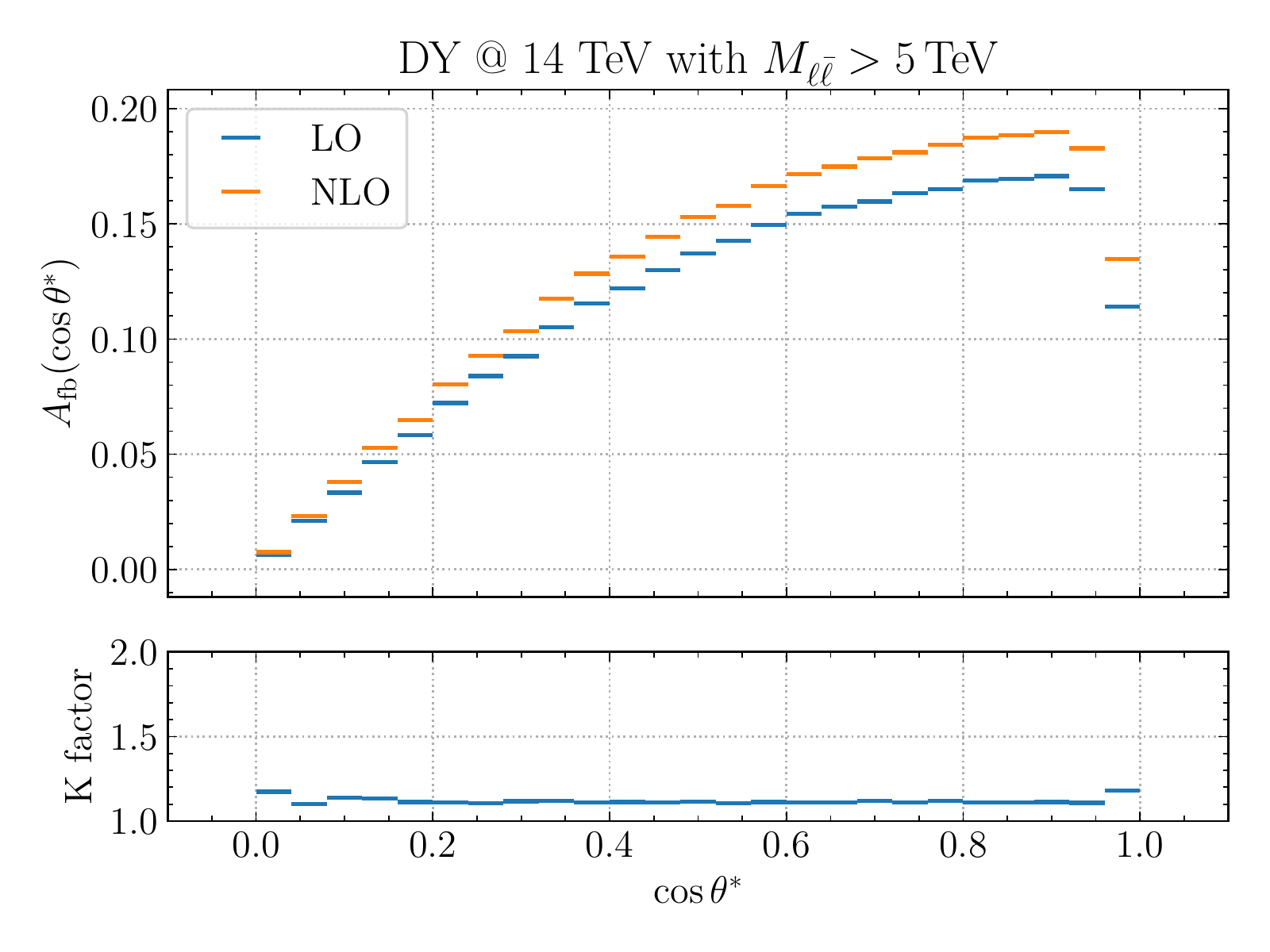}
  \caption{Same as \cref{fig:lo-diff-cos} now comparing the LO result
    to the NLO QCD result obtained using
    {\sc\small MadGraph5\_aMC@NLO}.
    The $K$-factor is shown in the lower panel.
  }
  \label{fig:lo-kfact}
\end{figure}

While the discussion so far has been presented  at LO, its
qualitative features are unaffected by 
higher-order corrections.
To illustrate this, in \cref{fig:lo-kfact} we compare
the LO result 
from \cref{fig:lo-diff-cos} to the corresponding NLO QCD result.
The bottom panels display the NLO $K$-factor
for the $\cos\theta^*$ distribution
and the forward-backward asymmetry.
Whereas the NLO $K$-factor in the
$\cos\theta^*$ distribution is quite large (around 40\%)
it exhibits only a mild dependence on
the Collins-Soper angle.
For $A_{\rm fb}$, the $K$-factor is at the 10\% level
and essentially independent of the value of $\cos\theta^*$.

%%%%%%%%%%%%%%%%%%%%%%%%%%%%%%%%%%%%
%%%%%%%%%%%%%%%%%%%%%%%%%%%%%%%%%%%%%%%%%%
 
%%%%%%%%%%%%%%%%%%%%%%%%%%
%%%%%%%%%%%%%%%%%%%%%%%%%

\section{The forward-backward asymmetry and the large-\texorpdfstring{$x$}{x} PDFs}
\label{sec:largexpdfs}

After our general discussion of the Drell-Yan process,
we now investigate
 proton structure at large-$x$, focusing on its
impact on the forward-backward asymmetry $A_{\rm fb}\lp \cos\theta^*\rp$
at large invariant masses.
First, we discuss the dependence of the
qualitative features of the asymmetry,
and specifically its sign, on the behavior of the underlying PDFs: we
illustrate this in a toy model, and compare results to a simple and 
commonly used approximation.
Subsequently,
we study the large-$x$ behavior of the PDFs from several
recent PDF sets: we compare PDFs, luminosities and the LO asymmetry
$A_{\rm fb}$ as a function of the dilepton invariant mass $\mll$.

\subsection{Qualitative features of \texorpdfstring{$A_{\rm fb}$}{Afb}}
\label{sec:afb_toy}

In order to understand the main qualitative features of  the $\cos\theta^*$
distribution and of the asymmetry $A_{\rm fb}$ and their dependence on the 
properties of the underlying
PDFs, it is instructive to evaluate predictions based on the
same computational setup adopted in Sect.~\ref{sec:HMDY}, namely
 LO matrix elements without kinematic cuts, using toy PDFs as input.
We consider toy quark and antiquark PDF with  the form
\begin{equation}
  \label{eq:toypdf}
  xf_q(x) = A_qx^{-a_q}(1-x)^{b_q} \, , \quad xf_{\bar{q}}(x) = A_{\bar{q}}x^{-a_{\bar{q}}}(1-x)^{b_{\bar{q}}} \, ,
\end{equation}
where $A_q$ and $A_{\bar{q}}$ are  normalization constants, irrelevant
for this discussion.
For simplicity we neglect the scale dependence of
the PDFs.
We then compute the single-differential distribution Eq.~(\ref{eq:dsigma-dcos}) and
the asymmetry Eq.~(\ref{eq:forward-backward-asymmetry}) with different assumptions on the
large $x$-behavior of these toy PDFs, i.e.\ different values of the large-$x$
exponents $b_q$, $b_{\bar{q}}$.

Since the overall normalization does not affect the shape
of the distribution, we set $A_q=A_{\bar{q}}=1$.
Furthermore, since we are not interested in the small-$x$ behavior,
we set $a_q=a_{\bar{q}}=1$. 
Hence, we consider simple scenarios in which 
\begin{align}\label{eq:toyp}
xf_q^+(x; b_q,b_{\bar{q}}) &= xf_q(x)+xf_{\bar{q}}(x) = x^{-1}\lc (1-x)^{b_q} +(1-x)^{b_{\bar{q}}}  \rc  \,, \\\label{eq:toym}
xf_q^-(x; b_q,b_{\bar{q}}) &= xf_q(x)-xf_{\bar{q}}(x) = x^{-1}\lc (1-x)^{b_q} -(1-x)^{b_{\bar{q}}}  \rc  \,, 
\end{align}
with different choices of the parameters  $b_q$ and $b_{\bar{q}}$.
Specifically, we consider a scenario with $b_q < b_{\bar{q}} $, in particular
$(b_q,b_{\bar{q}})=(3,5)$;  a  scenario with
$(b_q,b_{\bar{q}})=(3,3)$; and a third scenario
in which the quark PDFs at large-$x$ fall off more rapidly than the antiquarks,
$(b_q,b_{\bar{q}})=(5,3)$.

%-----------------------------------------------------------------
\begin{figure}[!t]
 \centering
 \includegraphics[width=0.49\linewidth]{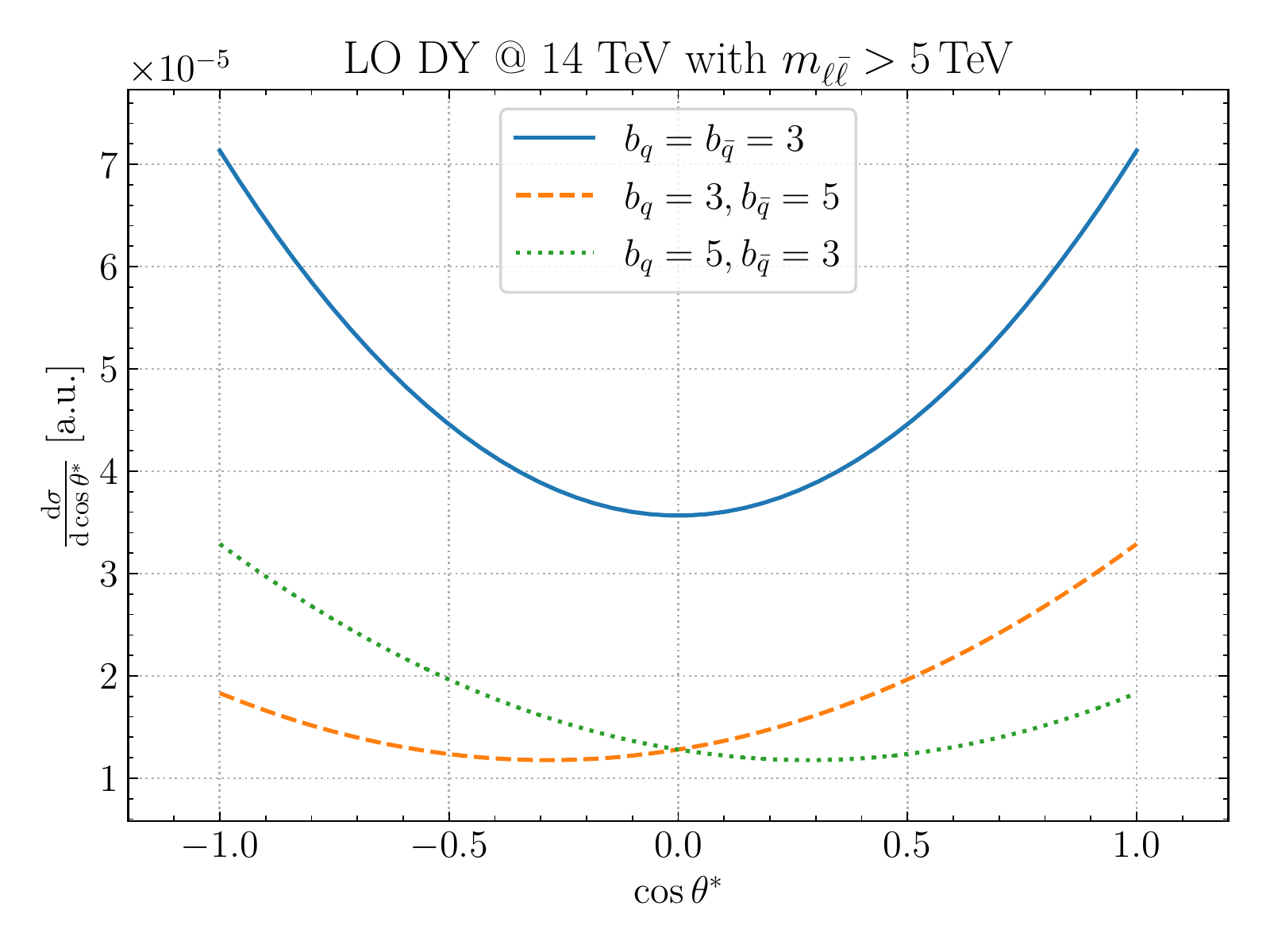}
 \includegraphics[width=0.49\linewidth]{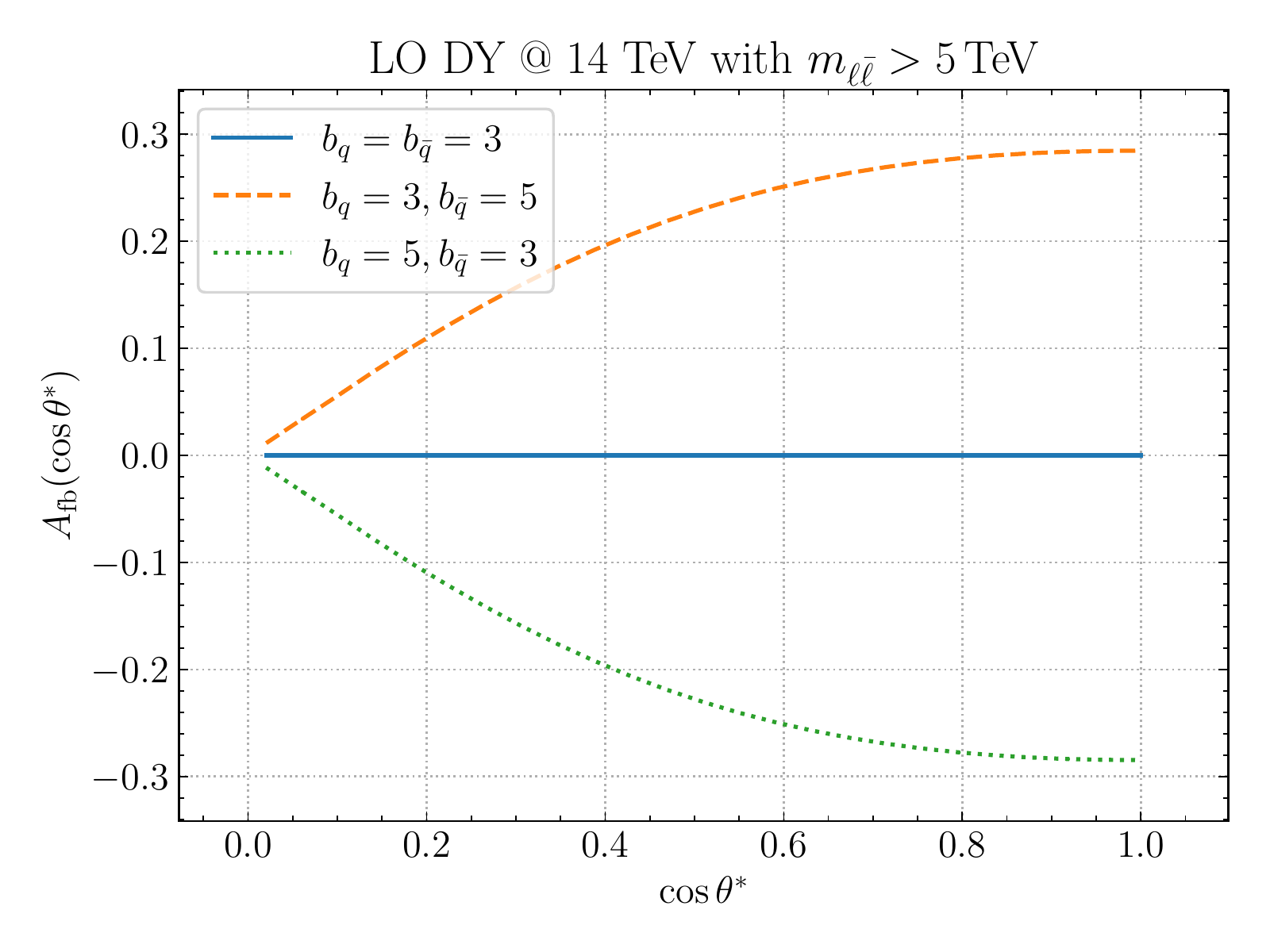}
 \caption{The single-inclusive $\cos\theta^*$ distribution
   Eq.~(\ref{eq:dsigma-dcos})  (left)
   and the corresponding forward-backward asymmetry
   (right panel) Eq.~(\ref{eq:forward-backward-asymmetry}) evaluated using 
    the toy PDFs of Eq.~(\ref{eq:toypdf}).
   No  kinematic cuts are applied except for $\mll^{\rm min}=5$ TeV.
 }    
 \label{fig:sigma_toy}
\end{figure}
%-----------------------------------------------------------------

In \cref{fig:sigma_toy} we display both the $\cos\theta^*$
single-inclusive distribution Eq.~(\ref{eq:dsigma-dcos}) and
the asymmetry Eq.~(\ref{eq:forward-backward-asymmetry}).
It is apparent that if the  antiquark PDFs fall off at large-$x$ faster than
the quarks, i.e.\ when $b_q < b_{\bar{q}}$ the forward-backward
asymmetry is positive, while if the converse is true it is
negative. Of course if the quark and antiquark PDFs behave in the same
way there is no asymmetry.
Indeed, the condition for a negative asymmetry is
(assuming $x_1>x_2$)
\begin{equation}\label{eq:signas}
   \sign\left[\mathcal{L}_{A,q}\right]=\sign\left[\frac{ f_q^+(x_2)}{
       f_q^+(x_1)}-\frac{
       f_q^-(x_2)}{f_q^-(x_1)}\right]=\sign\left[\frac{ f_q(x_2)}{
       f_q(x_1)}-\frac{f_{\barq}(x_2)}{f_{\barq}(x_1)}\right] \, , \quad x_1>x_2 \,.
\end{equation}
Namely, what determines the sign of the antisymmetric luminosity, and thus
of the forward-backward asymmetry, is the relative rate of decrease of
the quark and antiquark, or valence and total quark PDFs.

In the simple model that we discussed, this rate of decrease is
controlled by the values of the exponents  $b_q$ and $b_{\bar{q}}$.
The simple model has unphysical features, in that
a negative asymmetry corresponds to a negative
valence distribution, which conflicts with sum rules. It is easy to
construct a more contrived model, in which the valence drops faster
than the total quark PDF, yet it remains positive.
Also one could argue that
Brodsky-Farrar counting rules~\cite{Brodsky:1973kr,Brodsky:1974vy}
imply that  $b_q<  b_{\bar{q}}$ as $x\to1$, so a faster dropping
antiquark is favored.
However,
counting rules are supposed to only hold asymptotically, so whether
they apply in any given region of $x$ is a priori unclear.
Again, it  is easy
to construct more contrived models in which the value of the exponent
or effective exponent is $x$-dependent.
However, our purpose is to highlight which features determine the sign
of the asymmetry, and not to construct an explicit PDF model. In fact,
in Sect.~\ref{sec:subsec-largexPDFs} we will explicitly exhibit PDFs that do
lead to a negative asymmetry, while being consistent with sum rules
and not leading to contradiction with asymptotic counting.

It is interesting to note that different conclusions on the asymmetry
could be  reached by  using
an approximation to the asymmetry which is quite accurate  in the $Z$
peak region.
This approximation however turns out to fail at high
invariant mass.
Indeed, the expression Eq.~(\ref{eq:lumisa_qpm}) of the antisymmetric
luminosity in terms of the valence 
and total PDF combinations $f_q^+$ and $f_q^-$ PDF combinations
suggests an approximation based on the expectation
that the valence is dominant at large $x$ and the sea is dominant at
small $x$. Assuming $x_1> x_2$, one then expects that
\be
\mathcal{L}_{A,u}(\yll,\mll) \approx\frac {1}{2} f_u^-(x_1,\mll^2)
f_{u}^+(x_2,\mll^2)   \,  \, ,\quad x_1>x_2.
\label{eq:app_asym_lumi_2}
\ee
This is clearly true  in the $Z$-peak region, which  motivates the
suggestion to use the measurement of $A_{\rm fb}$ as a means to
 constrain the valence quark combinations~\cite{Accomando:2019vqt}.

%-------------------------------------------------------------------------------
\begin{figure}[!t]
 \centering
 \includegraphics[width=0.90\linewidth]{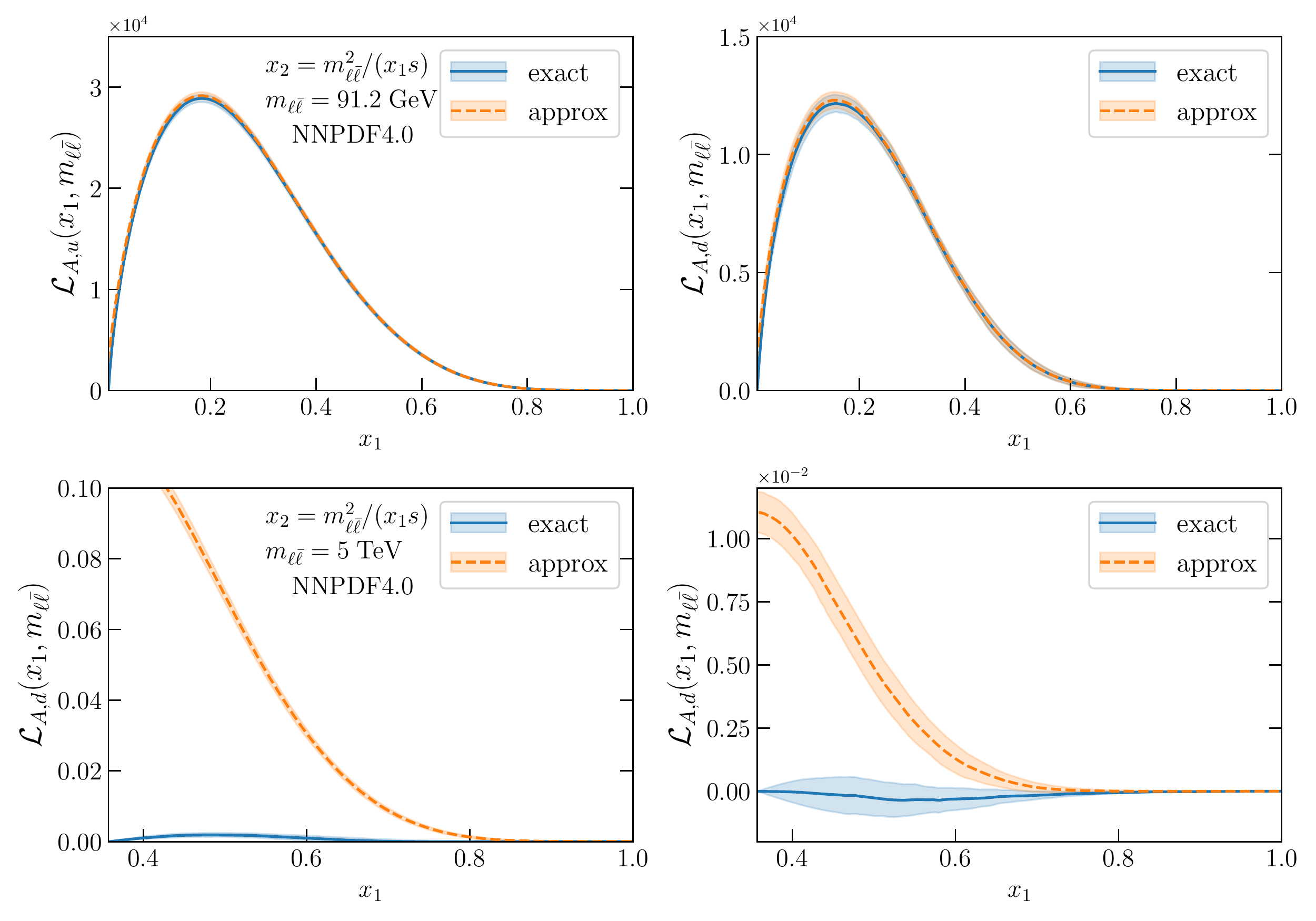}
 \caption{The  antisymmetric partonic luminosity $\mathcal{L}_{A,q}$, Eq.~(\ref{eq:lumisa_qpm}),
for the up and down quarks 
compared to the approximation 
Eq.~(\ref{eq:app_asym_lumi_2}) in the case of NNPDF4.0
at $\mll=m_Z$ (top)
and $\mll=5$ TeV (bottom panels).
 }    
 \label{fig:pdfplot-abs-DYlumis-minus-validation-lowQ-nnpdf40}
\end{figure}
%---------------------------------------------------------------------------

However, while Eq.~(\ref{eq:app_asym_lumi_2}) provides
a satisfactory approximation in the  $Z$-peak region,
it fails  at larger $\mll$ values. Indeed, for on-shell $Z$
production, with $\sqrt{s}=14$ TeV,
for a dilepton rapidity with $y_{\ell\bar{\ell}}\sim 2.5$, the limit of the
acceptance region
of ATLAS and CMS, the colliding partons have
$x_1=0.09$ and $x_2=6\times 10^{-4}$. So indeed the contribution in
which the valence PDF is evaluated at the smallest $x$ value is highly suppressed.
But for $\mll=5$~TeV, the smallest value of $x_2$, attained when
$x_1=1$, is $x_2=0.35$: so both momentum fractions are large and in fact
to the right of the valence peak.
In such case, there 
is no obvious hierarchy between
the different terms that contribute to to antisymmetric
luminosity $\mathcal{L}_{A,q}$.

This is illustrated in
Fig.~\ref{fig:pdfplot-abs-DYlumis-minus-validation-lowQ-nnpdf40},
where we compare the antisymmetric luminosity $\mathcal{L}_{A,q}$
for the up and down quarks 
to the approximation
Eq.~(\ref{eq:app_asym_lumi_2}), evaluated with NNPDF4.0 NNLO,
in the $Z$-peak region $\mll=m_Z$ 
and at $\mll=5$ TeV.
While indeed for $\mll=m_Z$ Eq.~(\ref{eq:app_asym_lumi_2}) reproduces
the exact luminosity, this is not the case for $\mll\gg m_Z$: both
the magnitude and the shape of the luminosity are  very different.
This qualitative behavior is common to all PDF sets: the approximation
fails equally badly regardless of the PDF set.

We conclude that there
is no simple relation between the sign of the asymmetry and that of
the valence PDF, and that the
behavior of the asymmetry must be determined by studying the large-$x$
behavior of the quark and antiquark PDFs.

\subsection{Parton distributions}
\label{sec:subsec-largexPDFs}

We assess now the large-$x$ behavior of
the quark and antiquark PDFs in different recent PDF
determinations: specifically, we compare
 ABMP16,
 CT18,  NNPDF4.0,
 and MSHT20.
 For
completeness, in App.~\ref{app:nnpdf31} we also present results
obtained with the widely used NNPDF3.1~\cite{Ball:2017nwa} set.

%-------------------------------------------------------------------------------
\begin{figure}[!t]
 \centering
 \includegraphics[width=1.0\linewidth]{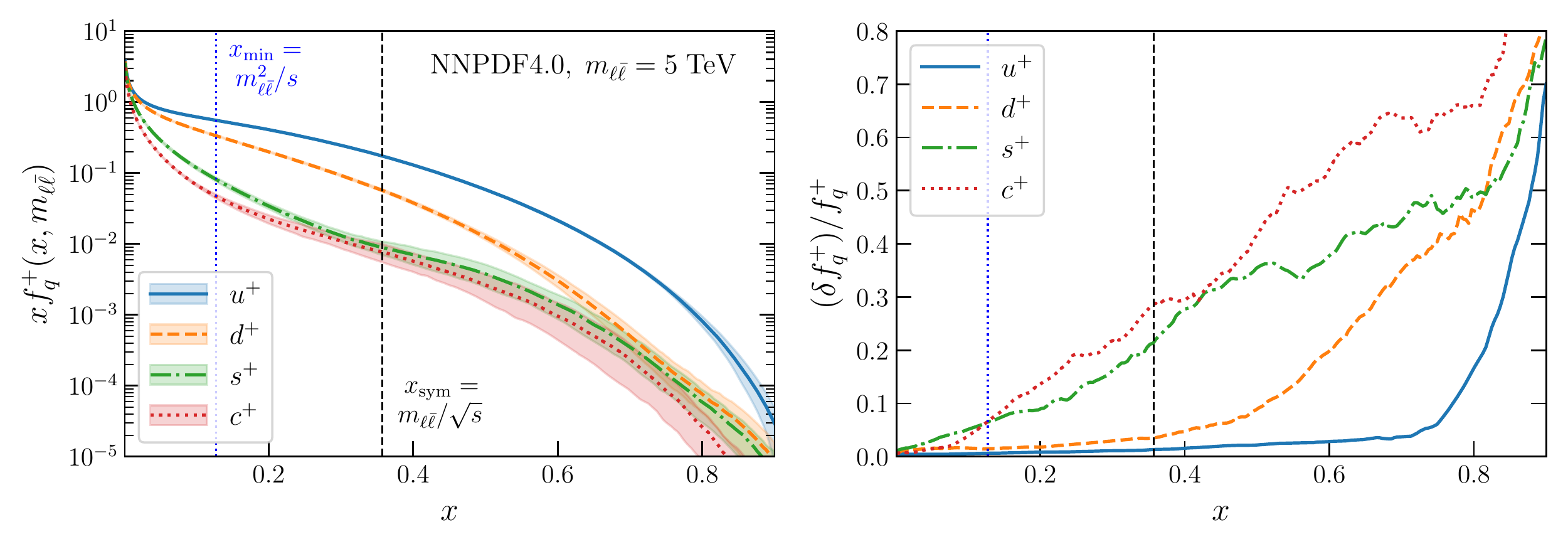}
 \includegraphics[width=1.0\linewidth]{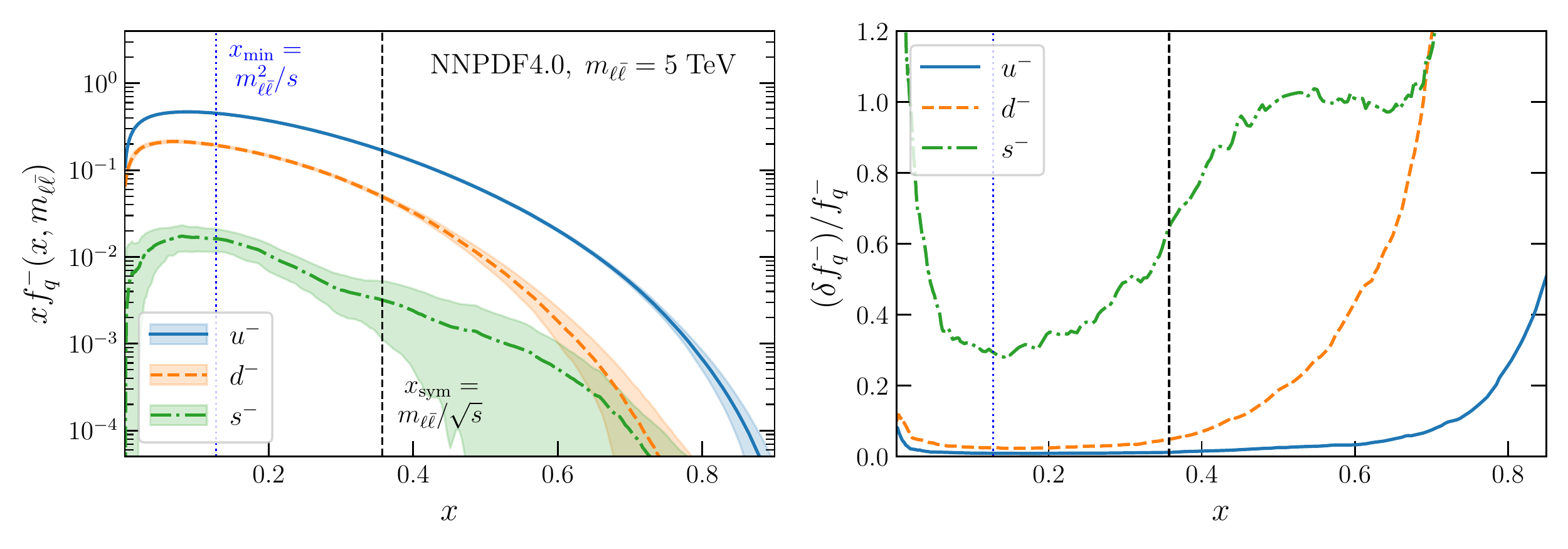}
 \caption{\small Comparison of the $xf^+_q$ (top) and $xf_q^-$ (bottom) quark
   PDF combinations for the up, down, strange, and charm quarks,
   evaluated at $\mll=5$~TeV for NNPDF4.0 NNLO.
   The right panels display the relative 68\% CL uncertainties.
   The two vertical lines indicate $x_{\rm min}=\mll^2/s$, the
   smallest allowed value of $x$ 
   for dilepton DY production for a collider
   CoM energy $\sqrt{s}=14$~TeV, and the value of $x$
   corresponding to a symmetric partonic collision $x_1=x_2$, namely
 $x_{\rm  sym}=\mll/\sqrt{s}$.
 }    
 \label{fig:pdfplot-abslargex}
\end{figure}
%-------------------------------------------------------------------------------

First, we provide a qualitative assessment of the relative size of the
PDFs corresponding to
individual quark flavors, both for the total and valence PDFs.
In Fig.~\ref{fig:pdfplot-abslargex} we
compare  the total $xf^+_q$ and valence $xf_q^-$  quark
   PDF combinations for the up, down, strange, and charm quarks,
   evaluated at $\mll=5$~TeV with the NNPDF4.0 NNLO PDF set.
   The right panels display the corresponding relative 68\% CL
   uncertainties. Note that, because of the way uncertainties are
   delivered by the various groups,  in this and all subsequent plots
   uncertainties 
   for NNPDF are given  are confidence levels (not necessarily Gaussian)
   determined from the Monte Carlo replica sample, and thus subject to
   point-to-point fluctuations, while for all other groups these are
   one-$\sigma$ Gaussian intervals determined from a Hessian PDF
   representation.

  The leftmost vertical line indicates $x_{\rm min}=\mll^2/s$, the
  smallest allowed value of $x$ 
   for dilepton DY production with invariant mass $\mll=5$~TeV for a collider
   CoM energy $\sqrt{s}=14$~TeV.
   The rightmost vertical line corresponds to
   the value of $x$ in a symmetric partonic collision where $x_1=x_2$, namely
   $x_{\rm  sym}\equiv\mll/\sqrt{s}$.

   From Fig.~\ref{fig:pdfplot-abslargex} one can observe that for
   $x\lesssim 0.3$ there is a clear hierarchy
$f_u^+>f_d^+ >f_s^+>f_c^+$, while for larger $x$ values the
   strange and charm PDFs become of comparable magnitude.
   The up and down quarks, both for $xf^+_q$ and $xf^-_q$, are significantly larger
   than the second-generation quark PDFs until $x\simeq 0.7$, and hence dominate the
   large-$\mll$ differential distributions in Drell-Yan production.
PDF uncertainties grow rapidly with $x$, reflecting the lack
of direct experimental constraints.
The same qualitative behavior of the lighter versus heavier flavor PDFs
is observed for other PDF sets.
Given the hierarchy $f_u^\pm, f_d^\pm \gg f_s^\pm, f_c^\pm $, in the following
we will discuss only the behavior of the first-generation quark
and antiquark PDFs which are those relevant for the interpretation
of neutral-current Drell-Yan production in the kinematic region used
for BSM searches.

%---------------------------------------------------------------------------
\begin{figure}[!t]
 \centering
 \includegraphics[width=1.00\linewidth]{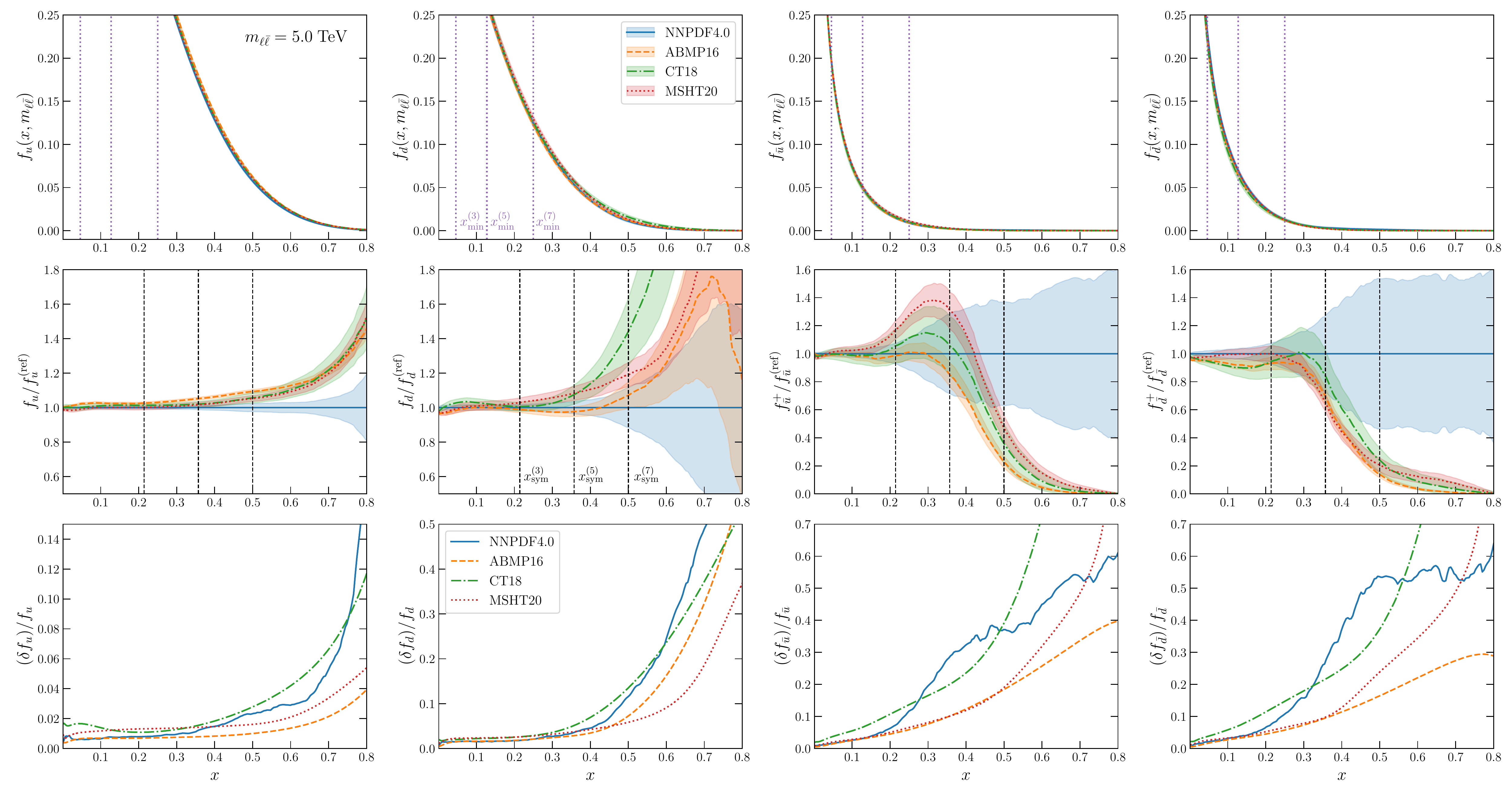}
 \caption{\small The up and down quark and antiquark PDFs evaluated at $\mll=5$ TeV
   for NNPDF4.0, CT18, MSHT20, and ABMP16 in the $x$ region relevant for
   high-mass Drell-Yan production. The upper panels display the absolute PDFs,
   the middle ones their ratio to the central NNPDF4.0 value, and the bottom panels
   the relative 68\% CL uncertainties.
   The vertical lines in the top
   row indicate the values of  $x_{\rm min}=\mll^2/s$ and in the central
   row those of $x_{\rm  sym}=\mll/\sqrt{s}$
   for three
   different values  $\mll=3,\,5,\,7$~TeV.
   Note that in the second row the
   range on the $y$ axis is not the same for quarks and antiquarks,
   and in the third row also for up and down quarks.
   Note also that the
   PDFs, their ratios and their uncertainties are essentially
   unchanged in the displayed large-$x$ region in the range $1~{\rm TeV}<\mll<7$~TeV.
}    
 \label{fig:mll_dep_pdfs}
\end{figure}
%---------------------------------------------------------------------------

We next compare the large-$x$ behavior 
of the four PDF sets ABMP16, CT18, MSHT20, and NNPDF4.0 in Fig.~\ref{fig:mll_dep_pdfs}
for $\mll=5$~TeV.
We display from top to bottom the absolute PDFs, their ratio to the central NNPDF4.0 value, and
their relative 68\% CL uncertainties.
As in the case of Fig.~\ref{fig:pdfplot-abslargex}, we indicate with two vertical lines
the values of $x_{\rm min}$ and $x_{\rm  sym}$,
both for $\mll=5$~TeV, and for a smaller and a larger value of $\mll$,
namely for $\mll=3$~TeV and  $\mll=7$~TeV.
For clarity, the values of $x_{\rm min}$ are only shown 
in the top row of plots, and the values of  $x_{\rm  sym}$ in the
central row. Note that the scale dependence
of the PDFs in this range of  $x$ and invariant mass is very
slight. Indeed, the PDFs shown in Fig.~\ref{fig:pdfplot-abslargex} are
essentially unchanged at $\mll=3$ TeV or  $\mll=7$ TeV;  only the
corresponding ranges of $x_1$, $x_2$ vary significantly.

Good agreement between all PDF
sets is found up to around $x\simeq 0.4$.
For $\mll=5$~TeV this corresponds to the value of $x_{\rm sym}$, i.e.\ central rapidity.
For larger values of $x\gsim 0.4$, the up  quark PDF $xf_u$ from the
NNPDF4.0 set is somewhat
suppressed in comparison to the other three sets, which in turn agree
among each other.
A rather stronger suppression of
NNPDF4.0  in comparison to  CT18 is observed for the down quark, with
MSHT20 and ABMP16 in a somewhat intermediate situation.
The opposite behavior is found in the same region $x\gsim 0.4$ for
antiquark PDFs  $xf_{\bar{u}}$ and $xf_{\bar{d}}$:
namely, the NNPDF4.0 PDF is significantly  larger than that of the other sets.
It follows that for a lower invariant mass value  $\mll = 3$~TeV, all
PDF sets are  in agreement in the $x$ range in which they are probed,
while for a higher value  $\mll = 7$~TeV the 
disagreement between NNPDF4.0 and the other PDF sets is present for
most of the $x\ge x_{\rm min}$ range.

It is interesting to observe that in the region with $0.4\lesssim x\lesssim 0.6$ the
PDFs are constrained by some fixed-target DIS structure functions and by
 forward $W$ and $Z$ production data from LHCb. Hence, at
the edge of the data region NNPDF4.0 starts disagreeing with the other
global PDF sets considered here, with the disagreement getting more
marked as $x$ grows outside the region covered by the
data.
Qualitatively, NNPDF4.0 is characterized by the fact that the
quark PDFs drop faster as a function of $x$, and the antiquark PDFs
drop less fast as $x$ grows towards $x=1$.
As we will show next, this feature will lead to significant differences
in the antisymmetric PDF luminosities $\mathcal{L}_{A,q}$ as the value of
the dilepton invariant mass $\mll$ is increased.

The relative PDFs uncertainties, shown
in the lower panels in
Fig.~\ref{fig:mll_dep_pdfs}  in all cases
grow with $x$ (see also  Fig.~\ref{fig:pdfplot-abslargex}).
The largest PDF uncertainties correspond to either CT18 or NNPDF4.0, 
depending on the $x$ range and the PDF flavor.
Specifically, the NNPDF4.0 uncertainties are largest for $f_d$ in the
region $x\gsim 0.6$ 
and for $f_{\bar{u}}$ and $f_{\bar{d}}$ when $0.3 \lsim x \lsim 0.5$.
The smallest PDF uncertainties are displayed by ABMP16 and  MSHT20.

\begin{figure}[!t]
 \centering
 \includegraphics[width=0.49\linewidth]{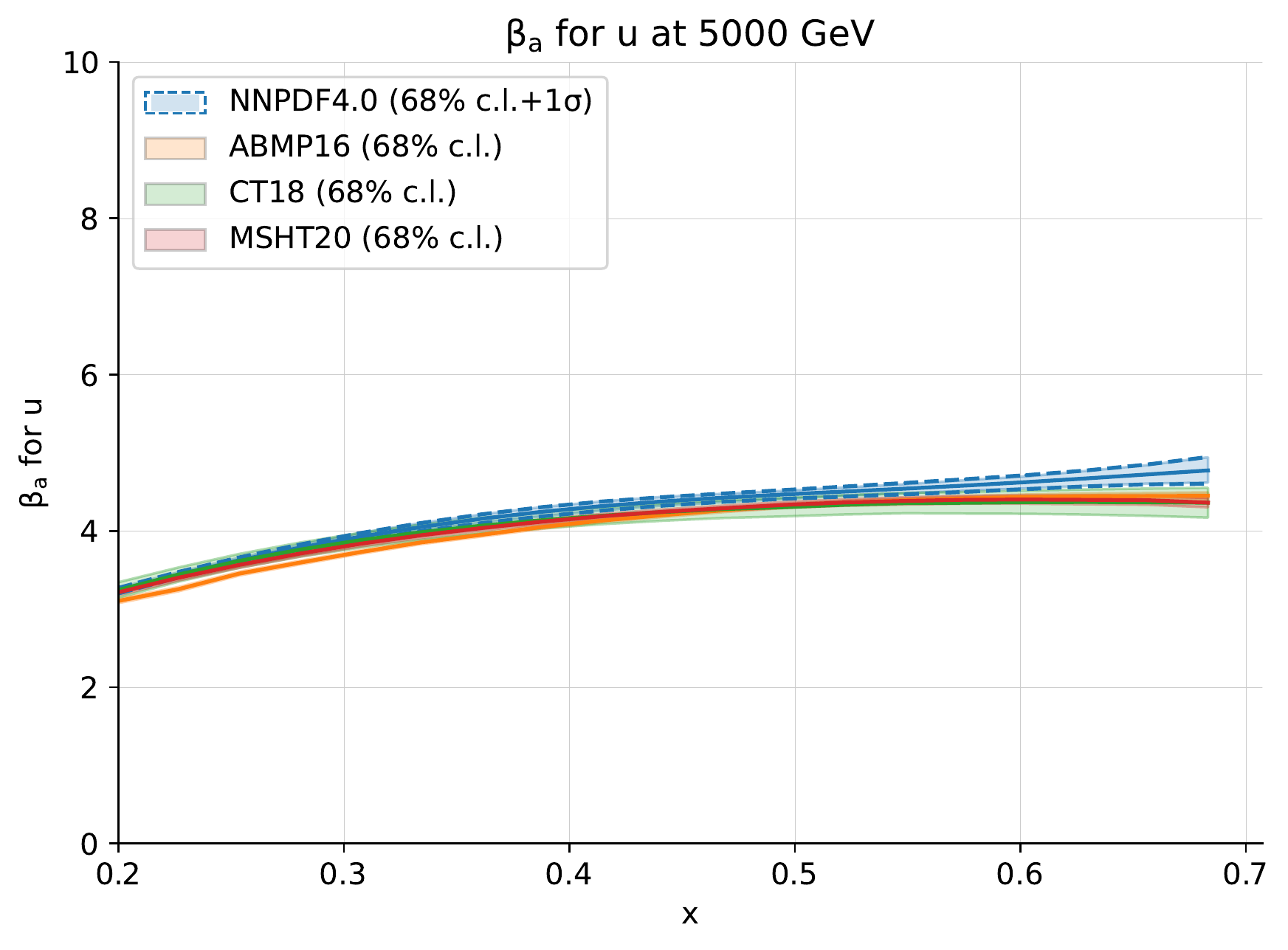}
 \includegraphics[width=0.49\linewidth]{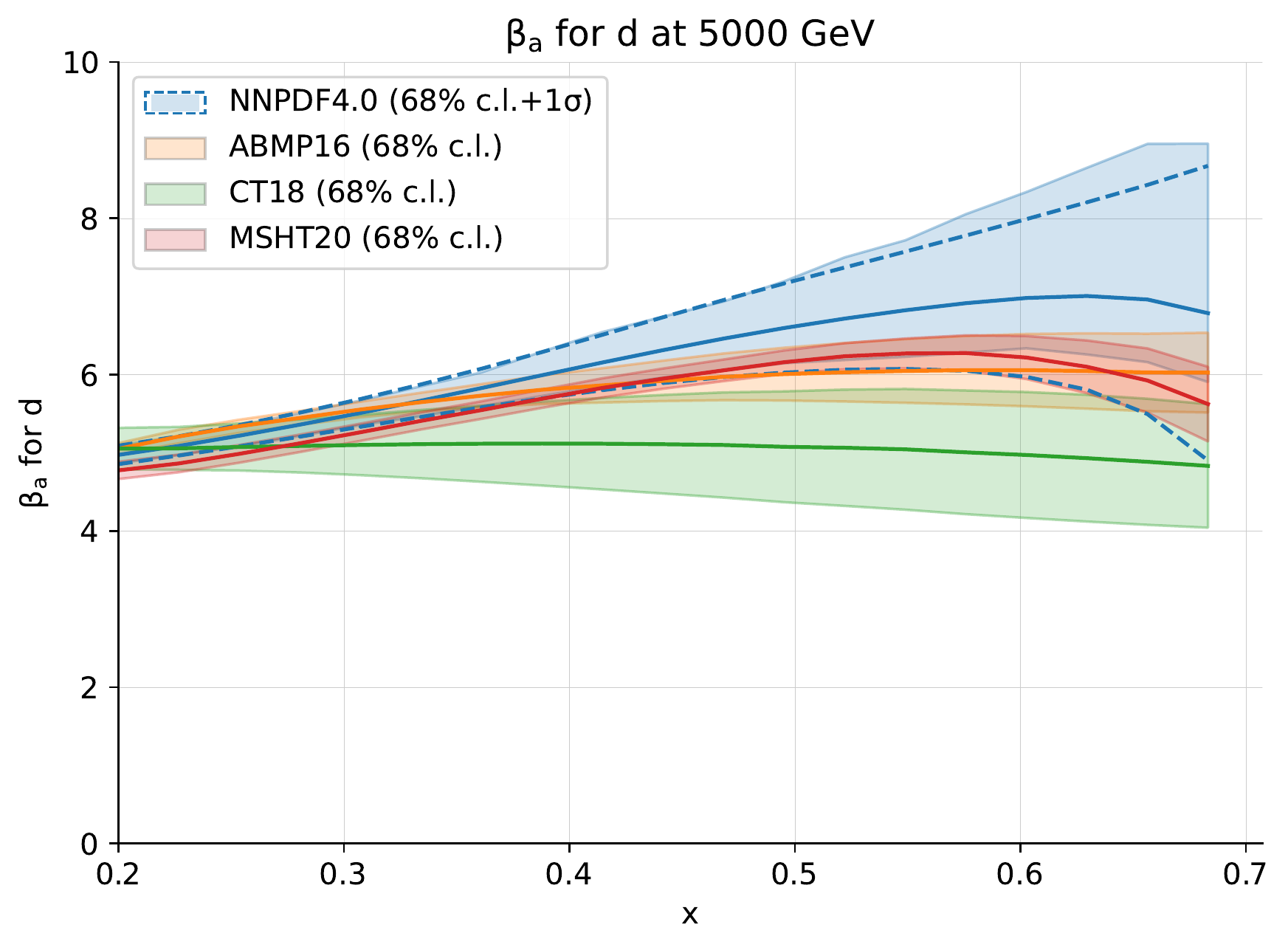}
 \includegraphics[width=0.49\linewidth]{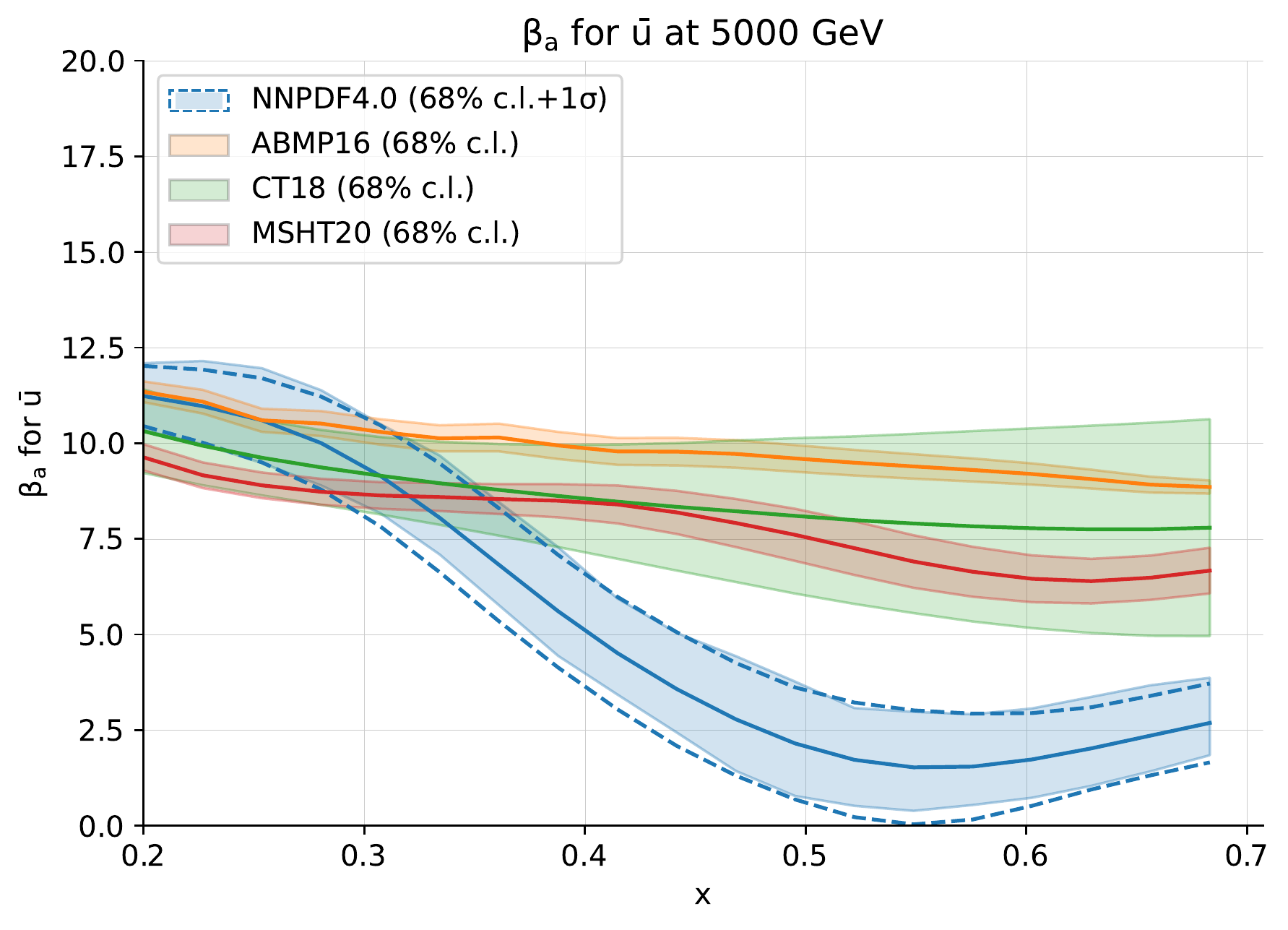}
 \includegraphics[width=0.49\linewidth]{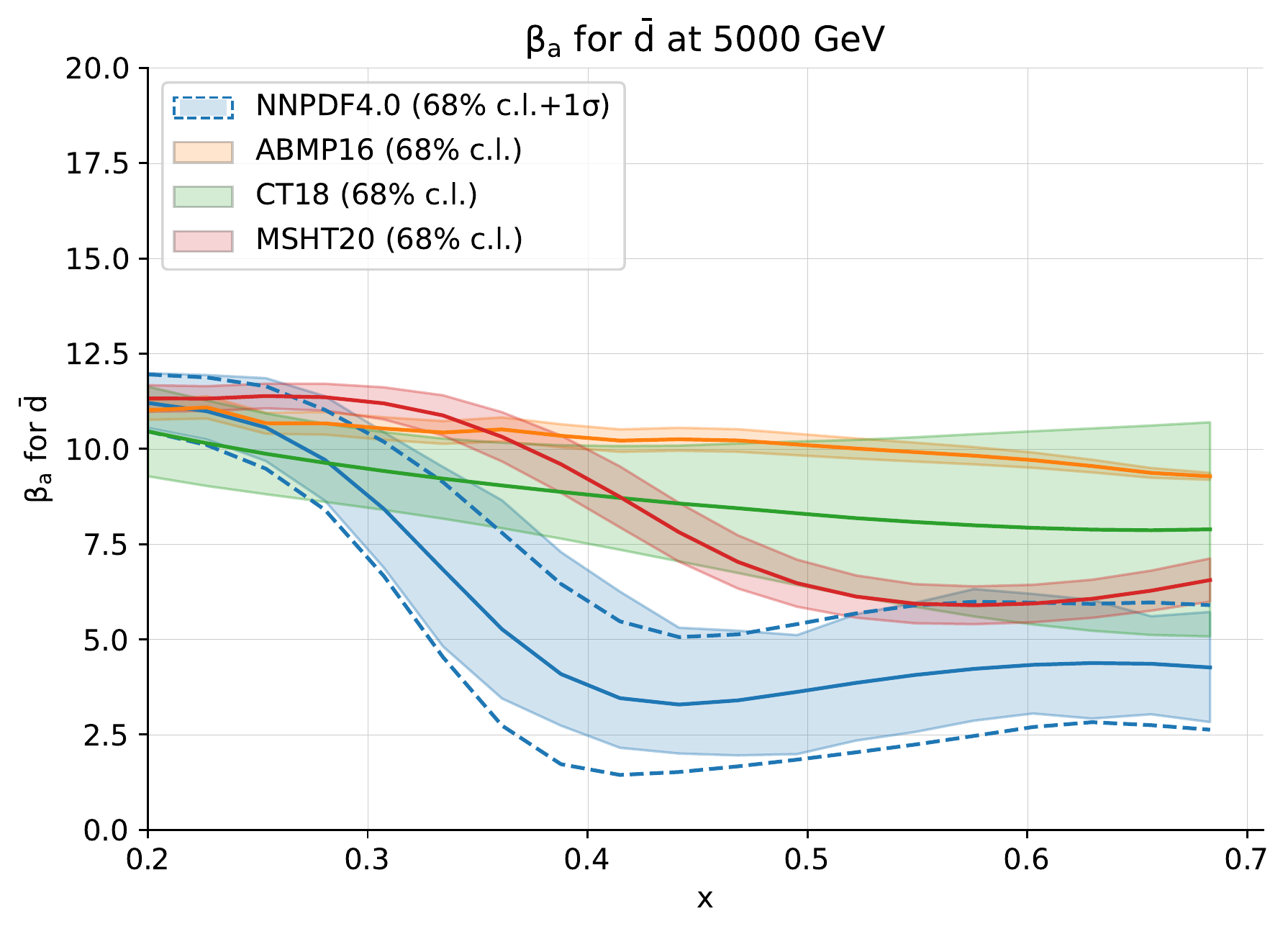}
 \caption{\small The large-$x$ asymptotic exponents $\beta_{a,q}(x,\mll)$, defined
   in Eq.~\eqref{eq:beta_asy},
   for ABMP16, CT18, MSHT20, and NNPDF4.0 evaluated at $\mll=5$~TeV
   for the up and down quark and antiquark PDFs.}    
 \label{fig:asy_exponents}
\end{figure}

The different behavior of the rate of decrease with $x$ of
PDFs  in the large $x$ region, specifically comparing NNPDF4.0
to other PDF sets, can  be seen most clearly from a comparison off
effective asymptotic exponents~\cite{Ball:2016spl}
\begin{equation}
  \beta_{a,q}(x,Q)\equiv\frac{\partial \ln|xf_q(x,Q)|}{\partial \ln(1-x)}\,,
  \label{eq:beta_asy}
\end{equation}
which of course for PDFs of the form of Eq.~(\ref{eq:toypdf}) just
coincide with the exponent $b$ up to $O(1-x)$ corrections.
In Fig.~\ref{fig:asy_exponents} we compare
the values of $\beta_{a,q}(x,\mll)$
for ABMP16, CT18, MSHT20, and NNPDF4.0 evaluated at $\mll=5$~TeV
for the up and down quark and antiquark PDFs in the  $x$ range of
Fig.~\ref{fig:pdfplot-abslargex}.

It is clear that while all PDF
sets have a similar effective asymptotic exponent for $x\lsim 0.35$, a
different behavior of NNPDF4.0 in comparison to other
determinations sets in for $x\gsim 0.4$.
Specifically, for quarks the NNPDF4.0 exponents are always larger,
and for antiquarks smaller than those found with other PDF
sets.
Interestingly, whereas for the up quark the
effective exponent $\beta_{a,u}$ is approximately constant for all
PDF sets when  $x\gsim 0.4$, with the NNPDF4.0 value being just
slightly higher and slowly increasing, for the down quark and all
antiquarks this approximately constant behavior is seen for other
PDF sets but not for NNPDF4.0.
Specifically, for the NNPDF4.0 down quark
the exponent slowly but markedly increases for $x\gsim 0.3$, together 
with its uncertainty.
In the case of NNPDF4.0 for both antiquarks the exponent
rapidly drops  in the region $0.3 \lsim x \lsim 0.4$.
This is consistent with the observation at the PDF level
(Fig.~\ref{fig:mll_dep_pdfs})  that for NNPDF4.0
at large-$x$, as compared to the other groups,
the up and especially the down  quark fall off more rapidly, while
the antiquark PDFs drop more slowly. Note in particular that for
the down PDF the antiquark effective exponent is significantly
smaller than the quark effective exponent for all $x\gsim0.4$.

%-------------------------------------------------------------------------------
%-------------------------------------------------------------------------------

The fact that a modification in behavior of the effective down quark
and especially antiquark PDFs is observed at the edge of the data
region for NNPDF4.0, but not for other PDF sets, suggests that this
might be related to the fact that  NNPDF4.0 generally adopts a more
flexible PDF parametrization in comparison to other
groups. Conversely, the fact that other groups display similar
behaviors suggests that this is related to their common choice of
parametrizing the large $x$ behavior of PDFs as $(1-x)^{\beta_i}$,
with the exponents $\beta_i$ fixed for each PDF flavor or combination
of flavors.
Also, the  uncertainties on the effective exponents
$\beta_{a,q}(x,\mll)$ tend to be larger for NNPDF4.0 (and also to a
lesser extent for CT18) in comparison to those of other groups.
Note however that the full PDF uncertainty contains also a
contribution from the overall 
magnitude, which is not captured by the effective exponents displayed here.

\subsection{Parton luminosities}
\label{subsec:partoniclumis}

\begin{figure}[!t]
 \centering
 \includegraphics[width=\linewidth]{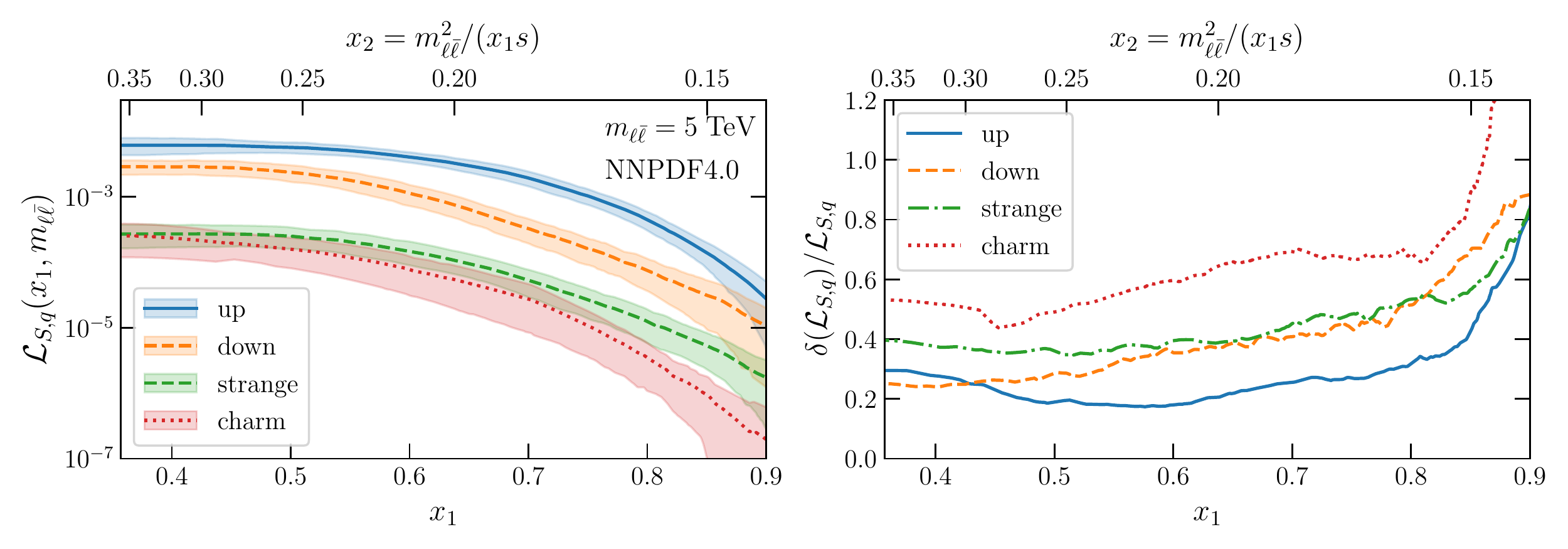}
 \includegraphics[width=\linewidth]{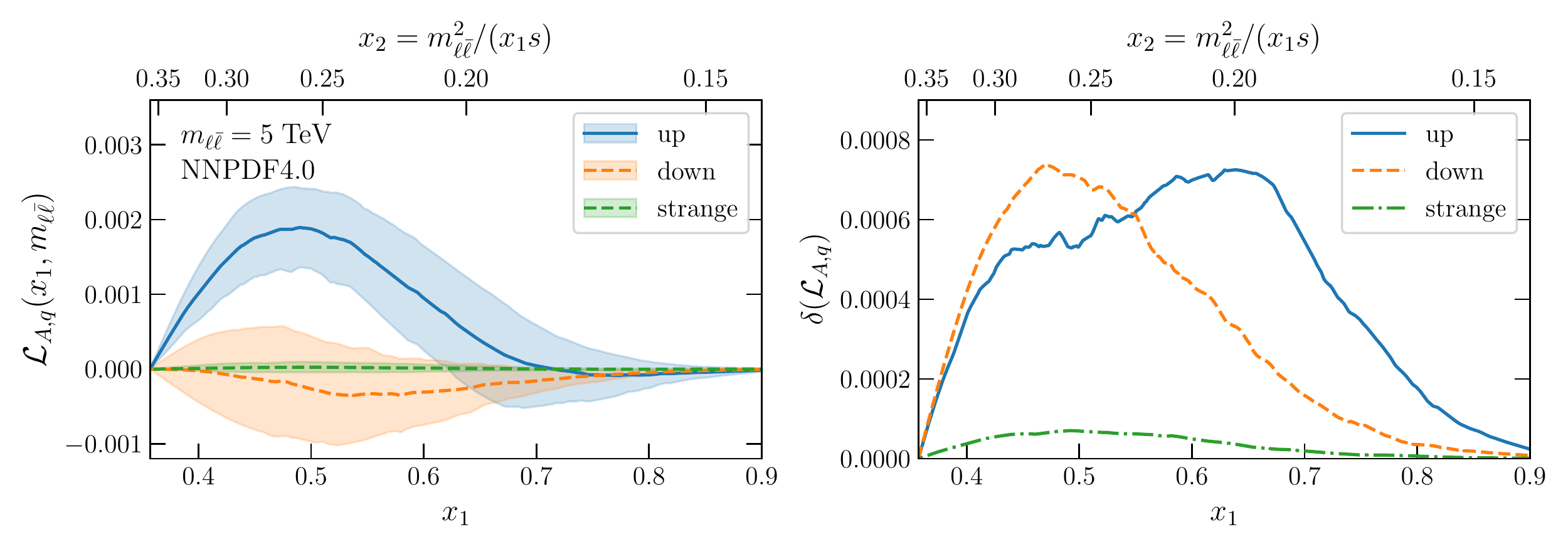}
 \caption{The symmetric $\mathcal{L}_{S,q}$ (top)
   and antisymmetric $\mathcal{L}_{A,q}$ (bottom)
   parton
   luminosities (left) and relative uncertainties (right) evaluated with
   NNPDF4.0 NNLO at $\mll=5$ TeV and $\sqrt{s}=14$ TeV.
The bottom  and top $x$-axes in each plot show respectively the values
of $x_1$ and $x_2$  at which the
luminosities are being evaluated, within the allowed range
$x \ge x_{\rm sym}=\mll/\sqrt{s}$, with the convention $x_1>x_2$.}    
 \label{fig:pdfplot-absDYlumis-plus-nnpdf40}
\end{figure}

We finally turn to the behavior of parton luminosities, with
particular regard for the antisymmetric combination which is relevant
for the forward-backward asymmetry.
As for PDFs, we first assess the qualitative features of the
luminosities corresponding to different quark flavors.
Specifically, the symmetric $\mathcal{L}_{S,q}$ and antisymmetric
$\mathcal{L}_{A,q}$ luminosities
Eq.~(\ref{eq:symm_asymm_lumis})  for individual flavors are
displayed in Fig.~\ref{fig:pdfplot-absDYlumis-plus-nnpdf40},
evaluated with NNPDF4.0 NNLO for $\mll=5$ TeV and $\sqrt{s}=14$ TeV.
The left panels display the absolute luminosities (in logarithmic and
linear scale respectively
for the $y$ and $x$ axes)
while the right panels show the corresponding PDF uncertainties (relative and absolute for
$\mathcal{L}_{S,q}$ and $\mathcal{L}_{A,q}$, respectively).
The bottom  and top $x$-axes in each plot show respectively the values
of $x_1$ and $x_2$  at which the
luminosities are being evaluated, within the allowed range
$x \ge x_{\rm sym}=\mll/\sqrt{s}$, with the convention $x_1>x_2$.

%-------------------------------------------------------------------------------
%---------------------------------------------------------------------------

The symmetric parton luminosities exhibit of course the same hierarchy
between flavors
as the corresponding PDF plots of Fig.~\ref{fig:pdfplot-abslargex}. 
The luminosity $\mathcal{L}_{S,q}$  drops rapidly for
$x_1\gsim 0.6$. PDF  uncertainties  depend weakly on  $x$
up to $x_1\gsim 0.8$, after which they blow up, and range between $\sim 20\%$
for the up quark luminosity to $\sim 60\%$ for the charm quark one,
with down and strange intermediate and of similar magnitude.

\begin{figure}[!t]
 \centering
 \includegraphics[width=1.00\linewidth]{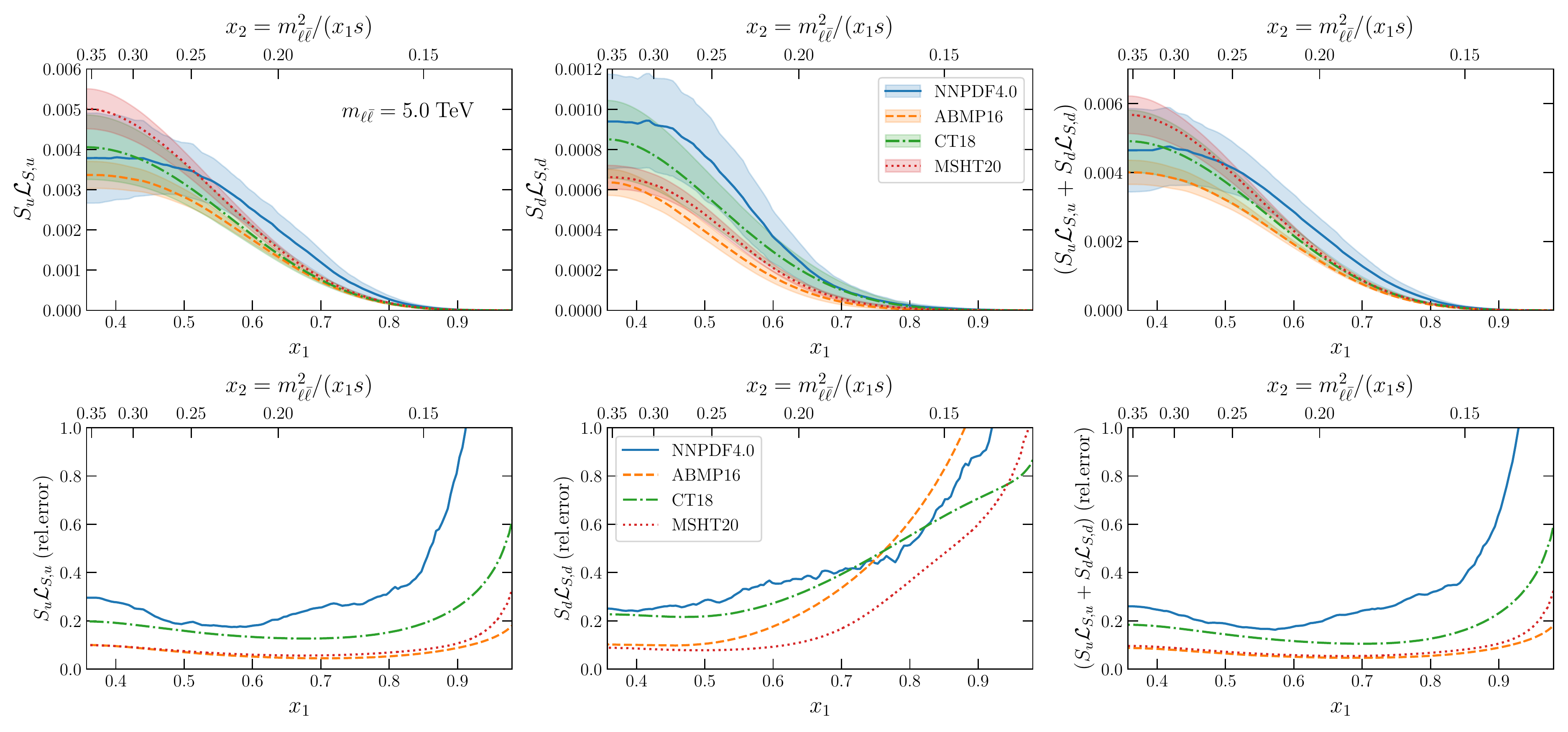}
  \caption{The symmetric 
   parton luminosities $\mathcal{L}_{S,q}(x_1,\mll)$ for the NNPDF4.0, ABMP16,
   CT18, and MSHT20 NNLO PDF sets for dilepton
   invariant masses of $\mll=5$ TeV.
   The luminosities are multiplied by the effective charges
   $S_q$ defined in Eq.~(\ref{eq:coup}).
   From left to right, we display $\mathcal{L}_{S,u}$,  $\mathcal{L}_{S,d}$,
   and their weighted sum that enters the  coefficient $g_{S,q}$ in Eq.~(\ref{eq:dsigma-dcos-v2}).
   The bottom panels display the relative 68\% CL PDF uncertainties.
    }    
 \label{fig:mll_dep_lumi_plus}
\end{figure}

As displayed in Fig.~\ref{fig:mll_dep_lumi_plus}, the light quark symmetric luminosities of other global PDF sets
are qualitatively similar.
We show $\mathcal{L}_{S,u}$,  $\mathcal{L}_{S,d}$,
and their weighted sum that enters the  enters the
symmetric coefficient $g_{S,q}$ in Eq.~(\ref{eq:dsigma-dcos-v2})
for the NNPDF4.0, ABMP16,
CT18, and MSHT20 at $\mll=5$ TeV.
The luminosities are multiplied by the effective charges
$S_q$ defined in Eq.~(\ref{eq:coup}),
and the bottom panels display the corresponding 68\% CL PDF uncertainties.
Good agreement between the four sets, with a similar shape
of $\mathcal{L}_{S,q}$, is observed.
The PDF luminosities for the dominant $\mathcal{L}_{S,u}$ contribution are the largest for NNPDF4.0.

%-------------------------------------------------------------------------------
%---------------------------------------------------------------------------

Turning to the antisymmetric PDF luminosities $\mathcal{L}_{A,q}$,
we note  that, for NNPDF4.0, while the up luminosity is
positive, the central value of the down luminosity is negative, though
the luminosity is compatible with zero at the one sigma level.
Recalling
from Fig.~\ref{fig:pdfplot-abslargex} that $xf_{d}^-$ itself is
positive for all values of $x$,  this provides an explicit example in
which the condition Eq.~(\ref{eq:signas}) is satisfied without the valence
combination being negative.
We conclude that for NNPDF4.0, the faster
drop of the quark distribution and slower drop of the antiquark
distribution that was displayed by the effective exponents of
Fig.~\ref{fig:asy_exponents} leads to a negative antisymmetric
luminosity, in agreement with Eq.~(\ref{eq:signas}).
The absolute PDF uncertainties are of a similar size for
$\mathcal{L}_{A,u}$ and $\mathcal{L}_{A,d}$, with a different shape
reflecting the underlying central values.

\begin{figure}[!t]
 \centering
 \includegraphics[width=1.00\linewidth]{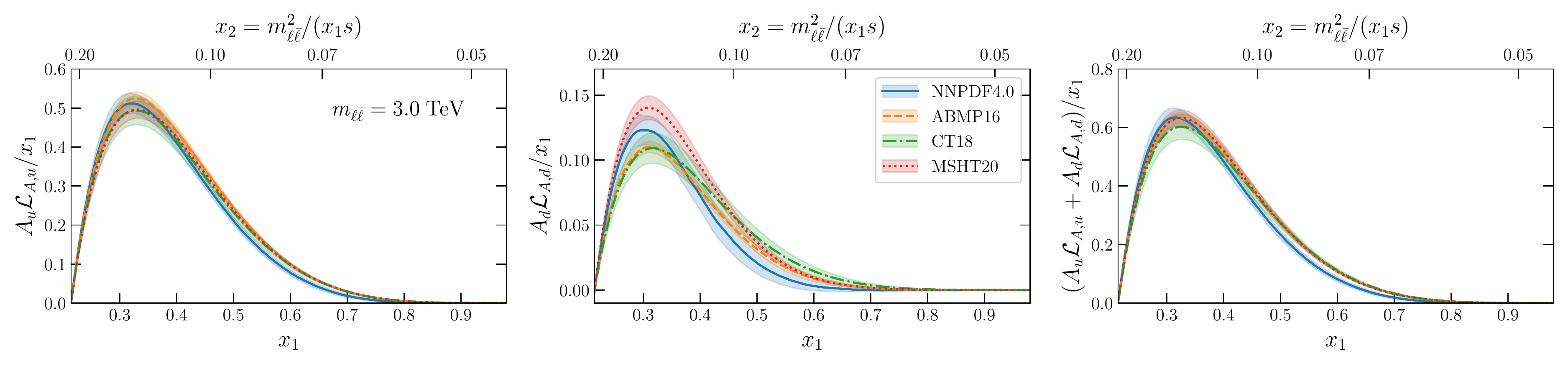}
 \includegraphics[width=1.00\linewidth]{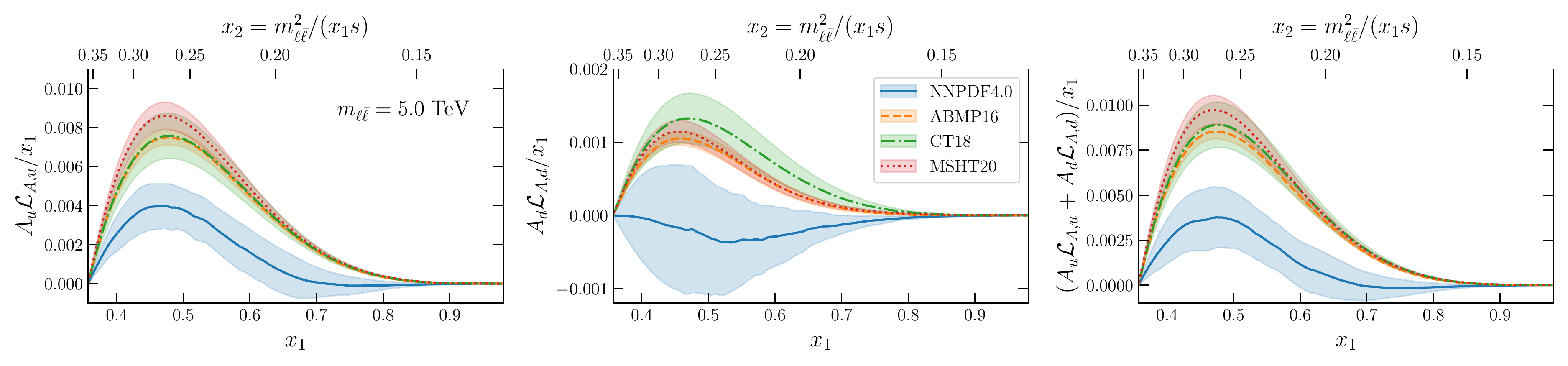}
 \caption{The antisymmetric 
   parton luminosities $\mathcal{L}_{A,q}(x_1,\mll)$ for the NNPDF4.0, ABMP16,
   CT18, and MSHT20 NNLO PDF sets for dilepton
   invariant masses of
   $\mll=3$ TeV (top) and $\mll=5$ TeV (bottom).
   The luminosities are multiplied by the effective charges
   $A_q$ defined in Eq.~(\ref{eq:coup}).
   From left to right, we display $\mathcal{L}_{A,u}$,  $\mathcal{L}_{A,d}$,
   and their weighted sum that enters the  coefficient $g_{A,q}$ Eq.~(\ref{eq:gAq_integrated_1}).
    }    
 \label{fig:mll_dep_lumi_minus}
\end{figure}

\begin{figure}[!t]
 \centering
 \includegraphics[width=1.00\linewidth]{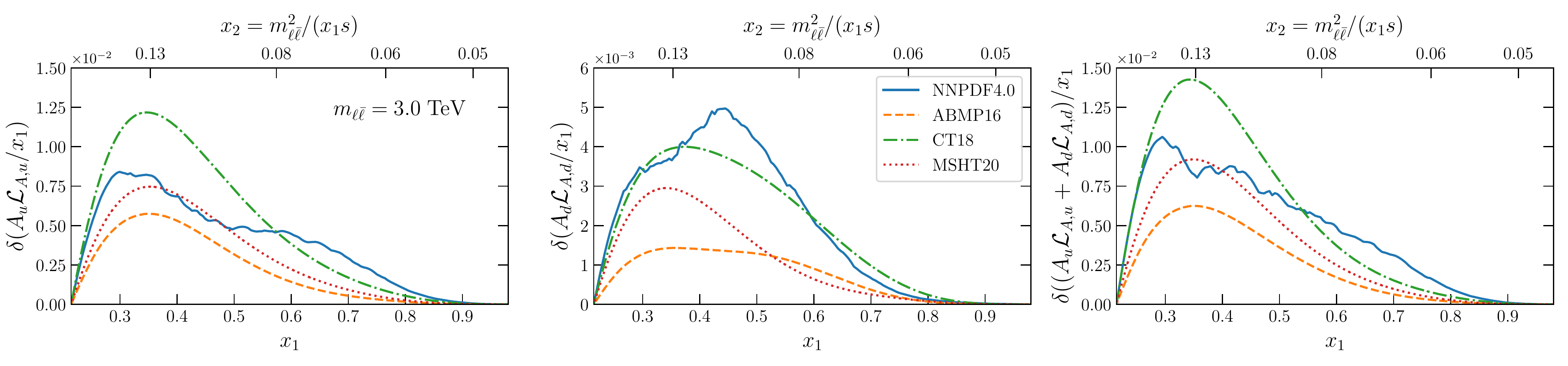}
 \includegraphics[width=1.00\linewidth]{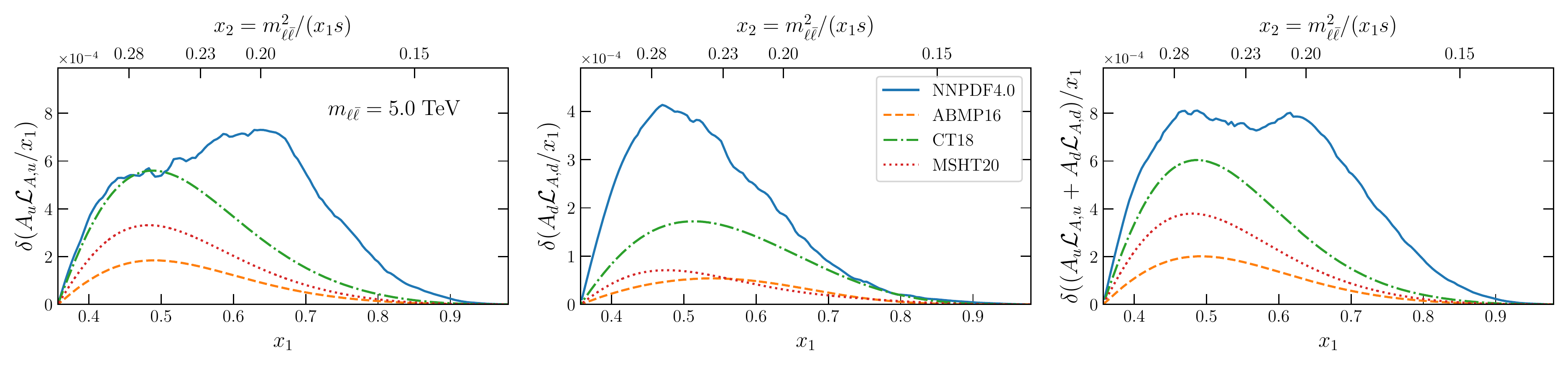}
 \caption{Same as Fig.~\ref{fig:mll_dep_lumi_minus} now for the absolute PDF uncertainties.
    }    
 \label{fig:mll_dep_lumi_minus_pdferrs}
\end{figure}

We compare
in Fig.~\ref{fig:mll_dep_lumi_minus}
the behavior of the antisymmetric luminosities for all PDF
sets for $\mll=3$ TeV (top) and $\mll=5$ TeV (bottom).
In order to facilitate the understanding of the way the PDF behavior
determines that of the asymmetry, we show both the contribution of
individual flavors and the total contribution
to the antisymmetric coefficient $g_{A,q}$ of
Eq.~(\ref{eq:gAq_integrated_1}). Namely, in
Fig.~\ref{fig:mll_dep_lumi_minus} the luminosities corresponding to
individual flavors are multiplied by the corresponding flavor-dependent
effective charges $A_q$ defined in Eq.~(\ref{eq:coup}):
from left to right we display $\mathcal{L}_{A,u}$,  $\mathcal{L}_{A,d}$,
and their weighted sum
 which determines
the sign and magnitude of the total forward-backward asymmetry.
The corresponding absolute PDF uncertainties for each of the four PDF sets
are displayed in  Fig.~\ref{fig:mll_dep_lumi_minus_pdferrs}.

%-------------------------------------------------------------------------------
%---------------------------------------------------------------------------

%-------------------------------------------------------------------------------
%---------------------------------------------------------------------------

Fig.~\ref{fig:mll_dep_lumi_minus} shows that for ABMP16, CT18, and
MSHT20 the antisymmetric
parton luminosities depend only mildly on $\mll$, whereas for NNPDF4.0
they exhibit a strong $\mll$ dependence.
Indeed, for dilepton invariant masses of $\mll=3$ TeV there is good
agreement between the three groups, but
for $\mll=5$ TeV the NNPDF4.0 up quark luminosity, while preserving a
similar valence-like shape, is suppressed
by a factor 2 in comparison  to other groups, and the down quark luminosity becomes compatible with zero with a negative
central value,  as already noted. 
For all PDF sets and  $\mll$ values the weighted sum is dominated by the up quark contribution.
The strong scale dependence of $\mathcal{L}_{A,q}$ in NNPDF4.0
reflects the underlying PDF behavior seen in  Fig.~\ref{fig:mll_dep_pdfs}
and highlighted by the effective exponents Fig.~\ref{fig:asy_exponents}.
As the scale $\mll$ increases, a range of increasingly large $x$ values is probed,
for which, in the case of
NNPDF4.0, the quark effective exponent slowly increases and the
antiquark exponent rapidly drops.
This leads to a negative asymmetry, 
following  Eq.~(\ref{eq:signas}). 

A comparison of the corresponding PDF uncertainties, displayed in
Fig.~\ref{fig:mll_dep_lumi_minus_pdferrs}, clearly shows the transition
from the data region to the extrapolation region.
For
$\mll=3$~TeV the uncertainty $\delta \mathcal{L}_{A,u}$ is generally
small for all sets, with CT18
showing a somewhat larger uncertainty for the up quark, and comparable
uncertainties for the down quark for all PDF sets.
As the scale  increases to $\mll=5$~TeV, where the large-$x$ region is
probed,  the uncertainty 
increases, though more markedly for NNPDF4.0.
For all PDF sets but
NNPDF4.0, the
uncertainty is approximately unchanged when the scale is further increased,
while for NNPDF4.0 it grows markedly.

\begin{figure}[!t]
 \centering
 \includegraphics[width=0.60\linewidth]{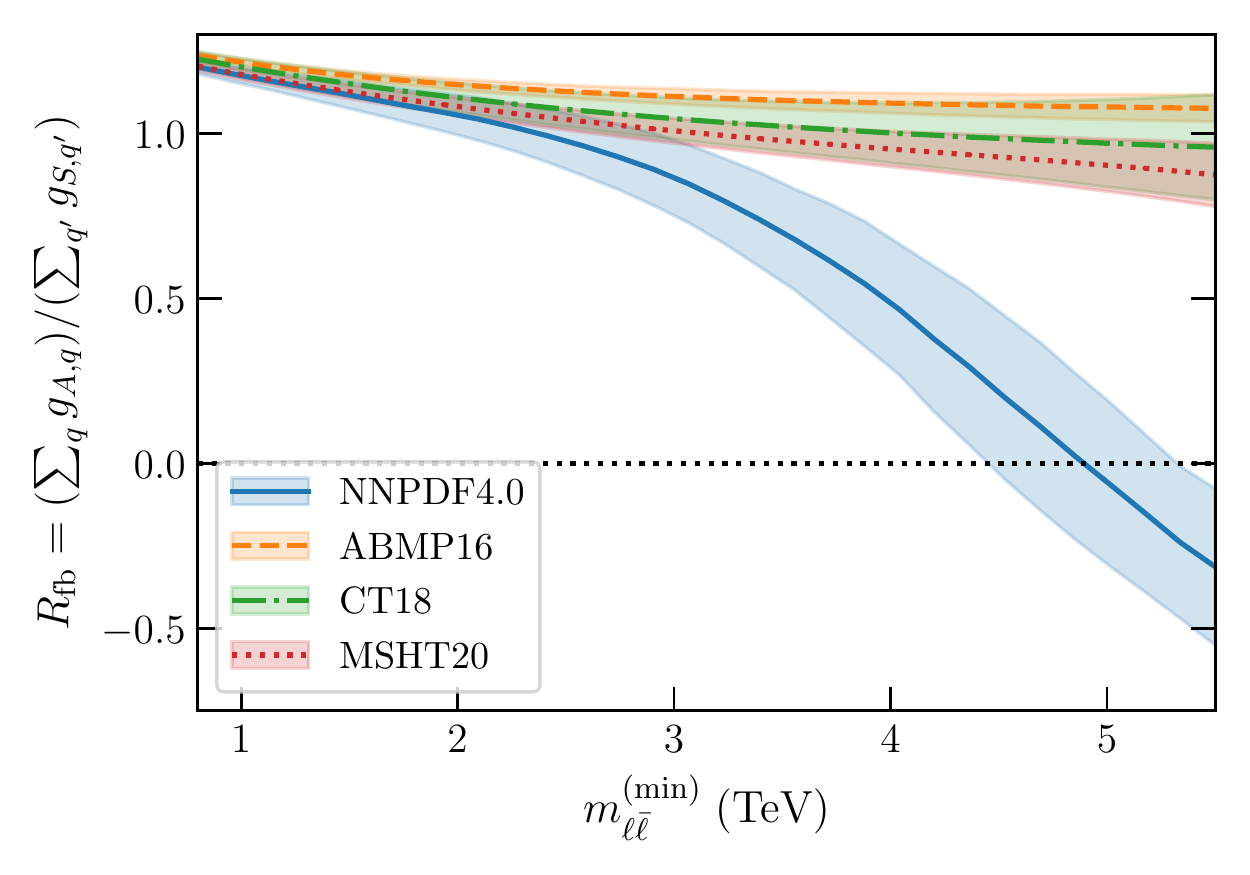}
 \caption{The coupling ratio $R_{\rm fb}$,
   Eq.~(\ref{eq:coupling_ratio}),
   that enters the forward-backward asymmetry $A_{\rm
     fb}(\cos\theta^*)$ at LO,  Eq.~(\ref{eq:afb_lo}), for different PDF
   sets, as  a function of the lower cut in the dilepton
   invariant mass $\mll^{\rm min}$.
 }    
 \label{fig:asym_coeff_mlldep}
\end{figure}

Finally, in Fig.~\ref{fig:asym_coeff_mlldep} we display for all
PDF sets the
ratio of antisymmetric to symmetric couplings
\be
\label{eq:coupling_ratio}
R_{\rm fb}\equiv \frac{\sum_q g_{A,q}}{\sum_{q'} g_{S,{q'}}} \, ,
\ee
that, according to
Eq.~(\ref{eq:afb_lo}), determines at leading order
the sign and magnitude
of the forward-backward asymmetry distribution $A_{\rm fb}(\cos\theta^*)$.
The symmetric and antisymmetric coefficients are obtained by integrating
the corresponding symmetric $\mathcal{L}_{S,q}$ and antisymmetric
$\mathcal{L}_{A,q}$ partonic luminosities according to
Eq.~(\ref{eq:gAq_integrated_1}), and the result is shown as a function of the lower integration cut $\mll^{\rm min}$.
In all cases the correlation between PDF uncertainties in the numerator and
the denominator are kept into account.

%-------------------------------------------------------------------------------
%---------------------------------------------------------------------------

Fig.~\ref{fig:asym_coeff_mlldep} shows that, consistently
with the behavior of the luminosity of
Fig.~\ref{fig:mll_dep_lumi_minus},  for $\mll^{{\rm
    min}}\lsim 3$~TeV results agree within uncertainties for all PDF
sets.
The situation is different for higher dilepton invariant masses $\mll^{{\rm min}}\gsim 3$~TeV:
the ratio $R_{\rm fb}$ starts to decrease for NNPDF4.0, while it
remains approximately  constant 
for the other  PDF sets. In particular, for NNPDF4.0 the coupling ratio
vanishes around $\mll^{{\rm min}}\sim5$~TeV, and it becomes negative
for yet larger   $\mll^{{\rm min}}$ values.
It follows that the forward-backward
asymmetry in high-mass Drell-Yan production should decrease  and
eventually vanish (and possibly even turn negative)
in NNPDF4.0 as the $\mll^{{\rm min}}$ cut is increased,
while for CT18, MSHT20, and ABMP16 it should remain positive
with a similar magnitude irrespective of the cut  $\mll^{{\rm min}}$ adopted.

%-------------------------------------------------------------------------------
\begin{figure}[!t]
 \centering
 \includegraphics[width=0.49\linewidth]{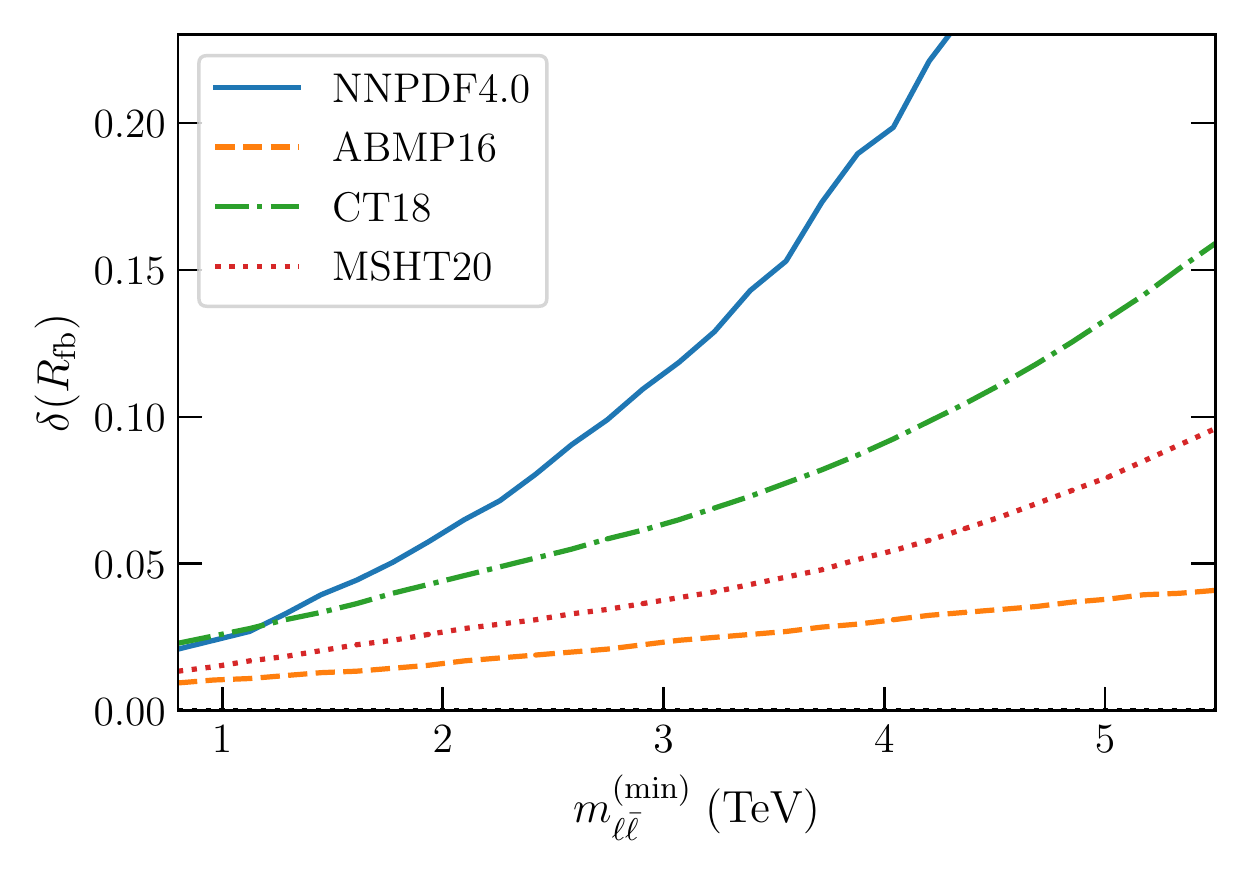}
 \includegraphics[width=0.49\linewidth]{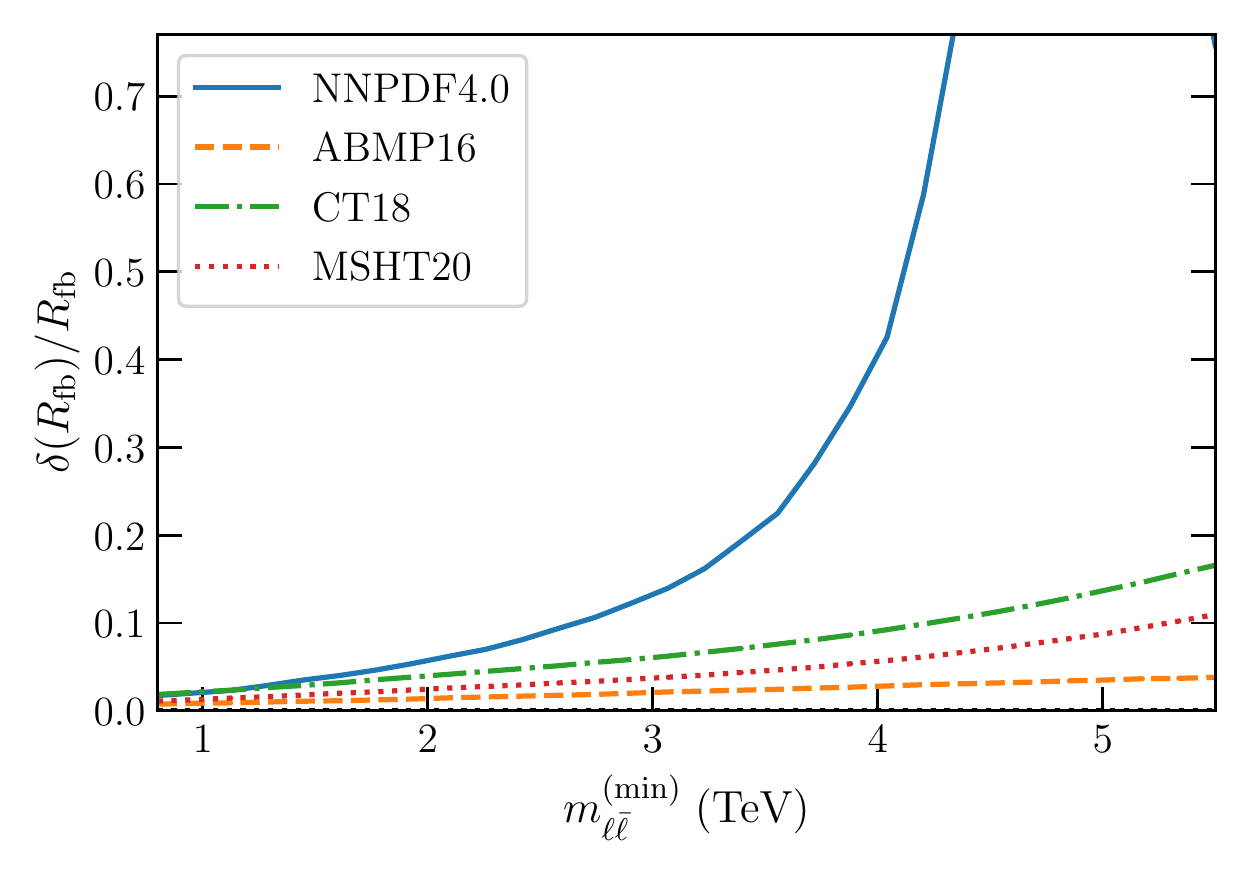}
 \caption{The absolute (left) and relative (right panel) uncertainties
   in the coupling ratio $R_{\rm fb}$ shown in Fig.~\ref{fig:asym_coeff_mlldep}.
 }    
 \label{fig:asym_coeff_mlldep_err}
\end{figure}
%---------------------------------------------------------------------------

Fig.~\ref{fig:asym_coeff_mlldep_err} displays the 
absolute  and relative  uncertainties
associated to the coupling ratio $R_{\rm fb}$.
We observe that NNPDF4.0 shows
the most marked increase of the uncertainties in $R_{\rm fb}$
as $\mll^{\rm min}$ grows.
For instance, for  $\mll^{\rm min}\gtrsim4$~TeV
the absolute PDF uncertainty in NNPDF4.0
is about twice as large as that found using CT18 
four times as large as MSHT20,
and about one order of magnitude larger than ABMP16.
This trend is magnified for the relative uncertainties
due to the decrease in the central value of $R_{\rm fb}$
as $\mll^{\rm min}$ increases.

\section{The Drell-Yan forward-backward asymmetry at the LHC}
\label{sec:afb}

After the qualitative discussion of the previous sections, here we
present results for the $\cos\theta^*$
distributions Eq.~(\ref{eq:dsigma-dcos}) and the
forward-backward asymmetry Eq.~(\ref{eq:forward-backward-asymmetry}), with
NLO QCD and electroweak corrections included and
with realistic selection and acceptance cuts for the LHC at $\sqrt{s} = 14$~TeV
and different values of the invariant mass $\mll$ relevant for SM
studies and BSM searches.

Computations are performed using {\sc\small MadGraph5\_aMC@NLO}~\cite{Alwall:2014hca},
interfaced to {\sc\small PineAPPL}~\cite{Carrazza:2020gss,christopher_schwan_2022_7023438} to generate
fast interpolation grids.
In order to account for realistic detector acceptances,
we impose phase-space cuts on the transverse momentum and the pseudo-rapidity of the two
leading leptons,
\be
p_T^{\ell} > 10~{\rm GeV}  \, ,\qquad |\eta_{\ell}| < 2.4 \,.
\ee
We then consider various regions of dilepton invariant mass $\mll$:
either close to the $Z$-boson peak ($60$~GeV $< \mll < 120$~GeV),
relevant for precision SM studies, or the
high-mass region relevant for BSM searches, with  various choices of a
lower mass invariant cutoff ($\mll > 3,4,5,6$~TeV).
In all cases, in order to facilitate the interpretation of
hadron-level results  and the connection to
the discussion of the PDF features from Sect.~\ref{sec:largexpdfs},
we also provide results for the two partonic channels that give the
largest contribution to the cross-section.
As in Sect.~\ref{sec:largexpdfs}, we compare results obtained using
the  ABMP16, CT18, MSHT20, and NNPDF4.0 PDF sets.
In all cases, we
use the  NNLO sets corresponding to the value $\alpha_s(m_Z)=0.118$
of the strong coupling.
Results obtained using the NNPDF3.1 PDF set are reported in App.~\ref{app:nnpdf31}.

%-------------------------------------------------------------------------------
\begin{figure}[!t]
 \centering
 \includegraphics[width=0.49\linewidth]{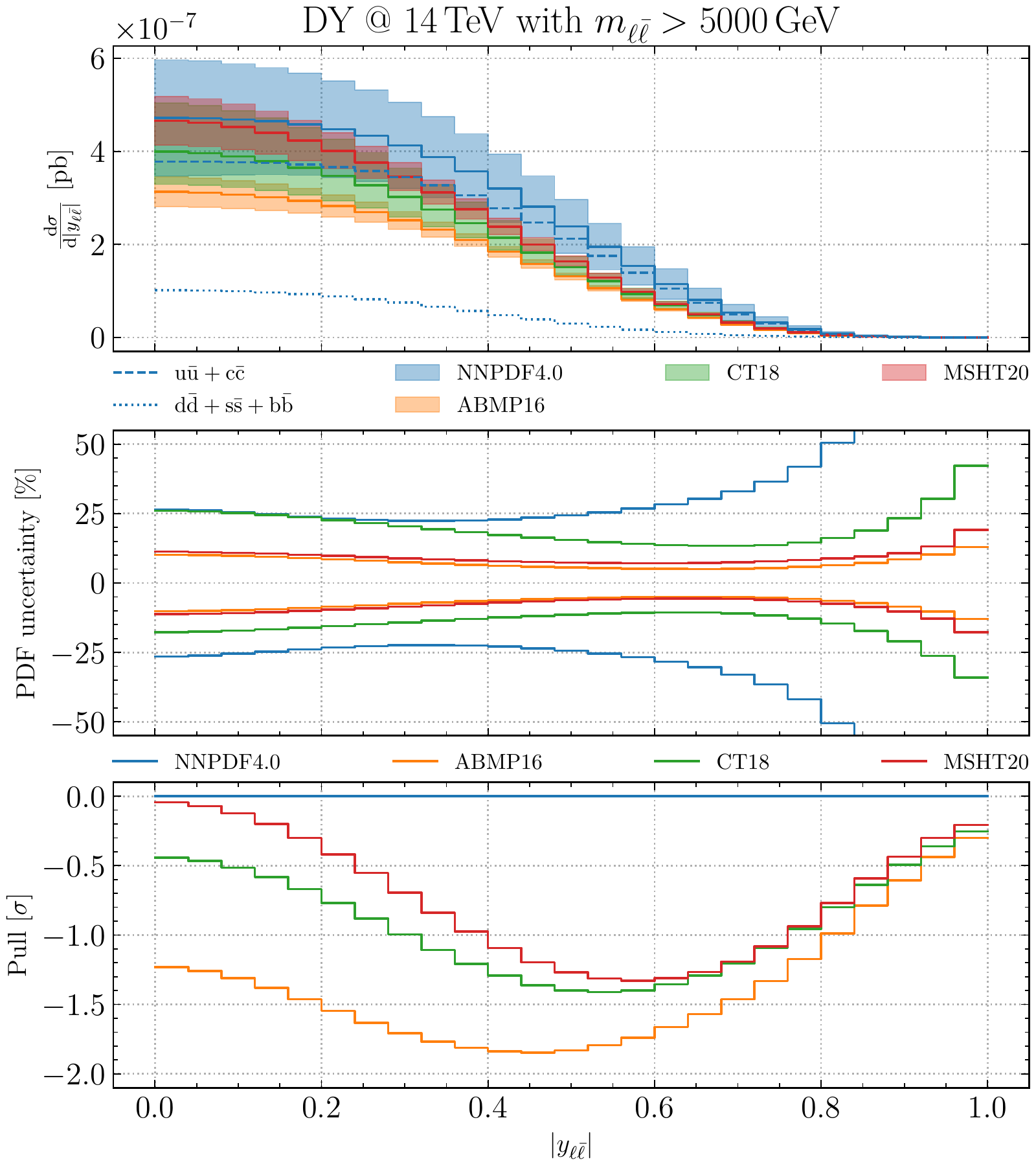}
 \caption{The differential distribution in absolute dilepton rapidity $|\yll|$, given in
Eq.~(\ref{eq:dsigma-dyll}),
for dilepton invariant masses of $\mll > 5$~TeV
for neutral current Drell-Yan production at the
LHC 14 TeV,
obtained using ABMP16, CT18, MSHT20, and NNPDF4.0 NNLO PDFs with $\alpha_s(m_Z)=0.118$.
All
uncertainties shown are 68\% CL PDF uncertainties, computed at NLO QCD with
realistic cuts (see text).
We show the absolute distributions (top), relative uncertainties (normalized
to the central curve of each set, middle) and the pull with respect to the
NNPDF4.0 result, Eq.~(\ref{eq:pulldef_xsec}) (bottom).
For the central NNPDF4.0 prediction
the contributions of the $u\bar{u}+c\bar{c}$ and $d\bar{d}+s\bar{s}+b\bar{b}$
parton subchannels are also shown.
 \label{fig:CMS_DY_14TEV_MLL_5000_rap}}
\end{figure}
%-------------------------------------------------------------------------------

Before considering the angular distributions, in
Fig.~\ref{fig:CMS_DY_14TEV_MLL_5000_rap} we display the 
differential distribution in absolute dilepton rapidity $|\yll|$,
defined in Eq.~(\ref{eq:dsigma-dyll}),
for a dilepton invariant mass of $\mll > 5$~TeV.
This is the kinematic region relevant for searches of high-mass resonances
in the dilepton channel at the LHC, e.g.~\cite{ATLAS:2019erb,Khachatryan:2016zqb}.
We display the absolute differential distributions with the 68\% CL PDF uncertainties
(top), the relative PDF uncertainty (center) normalized for each PDF
set to the corresponding central prediction, and the pull between the
NNPDF4.0 result, taken as a reference, and other sets (bottom).
This pull is    defined as
    \be
\label{eq:pulldef_xsec}
{\rm Pull}_i= \frac{ \sigma^{(0)}_{2,i} -\sigma^{(0)}_{1,i} }{
  \sqrt{ \lp  \delta \sigma_{2,i}\rp^2+\lp  \delta \sigma_{1,i}\rp^2 }} \, , \qquad i=1,\ldots,n_{\rm bin} \, ,
\ee
where  $\sigma^{(0)}_{1,i}$ and $\sigma^{(0)}_{2,i}$ are the central values of the
theory prediction in the $i$-th bin of the distribution and $\delta \sigma_{1,i}$, $\delta \sigma_{2,i}$ are
the corresponding PDF uncertainties.
For the central NNPDF4.0 prediction in the upper panel we also display the
contributions from the dominant parton subchannels, namely
$u\bar{u}+c\bar{c}$ and $d\bar{d}+s\bar{s}+b\bar{b}$.

As discussed in Sect.~\ref{sec:HMDY}, the  $|y_{\ell\bar{\ell}}|$  distribution
depends  on the symmetric partonic luminosities $\mathcal{L}_{S,q}$, Eq.~(\ref{eq:lumiss_qpm}),
which in turn are driven by the total PDFs $xf^+_q$.
The $|y_{\ell\bar{\ell}}|$  distribution
is dominated by the $u\bar{u}$
contribution and its qualitative behavior is found to be similar for the four PDF sets considered.
PDF uncertainties are the largest in NNPDF4.0, ranging between 25\% and 50\%,
and the pull between NNPDF4.0 and  CT18 and MSHT20 is at most at the
$1.5\sigma$ level, and slightly larger  
with ABMP16.
The dependence of the $|y_{\ell\bar{\ell}}|$  distribution on the dilepton mass $\mll$
is moderate, and the same qualitative features are
obtained if $\mll$ is lowered down to the $Z$-peak region, or
raised to yet higher values.
Hence, for the absolute rapidity distribution there is a
reasonable agreement between all  PDF sets for all scales considered.

%-------------------------------------------------------------------------------------
\begin{figure}[t]
\centering
\includegraphics[width=0.49\linewidth]{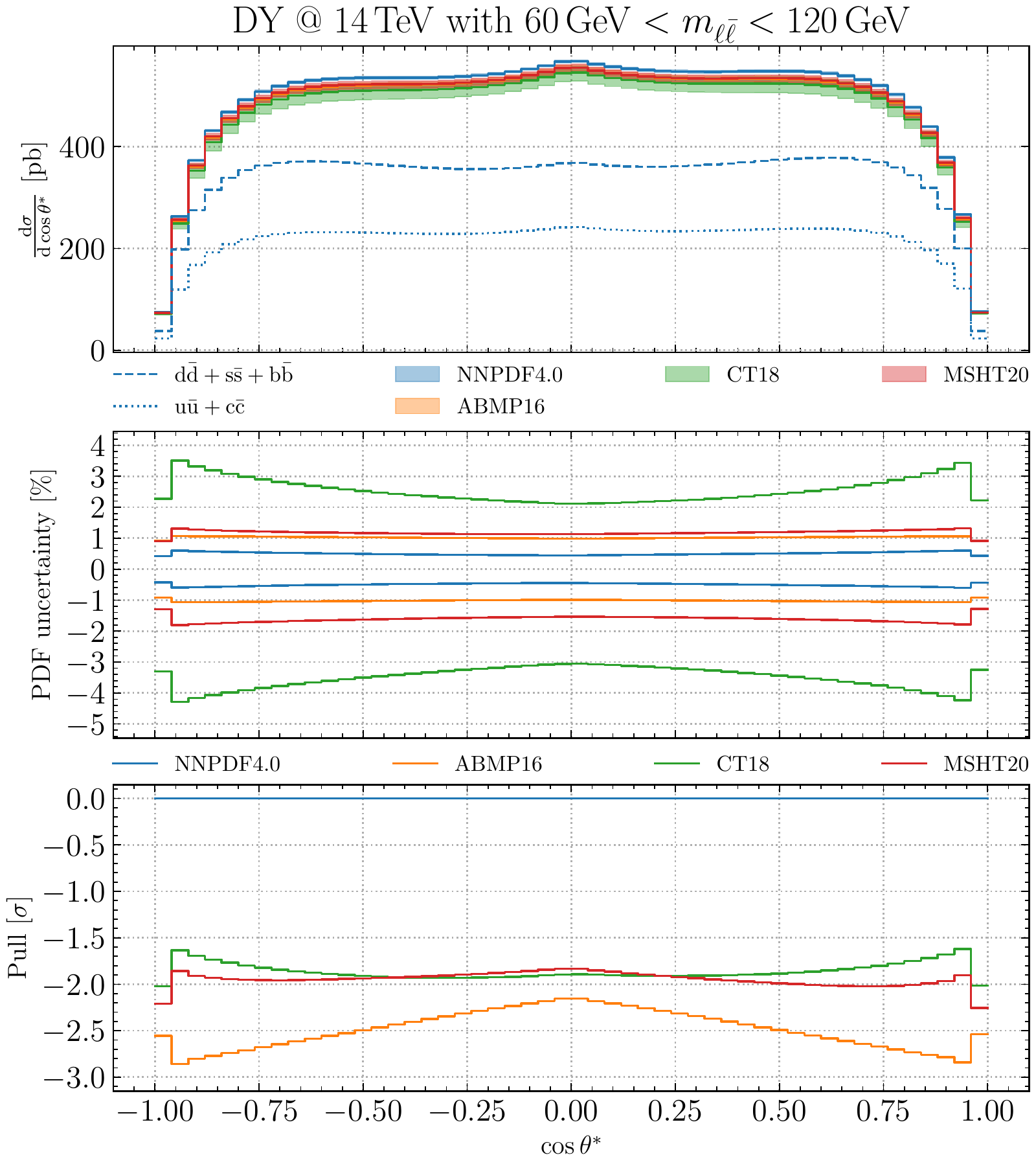}
\includegraphics[width=0.49\linewidth]{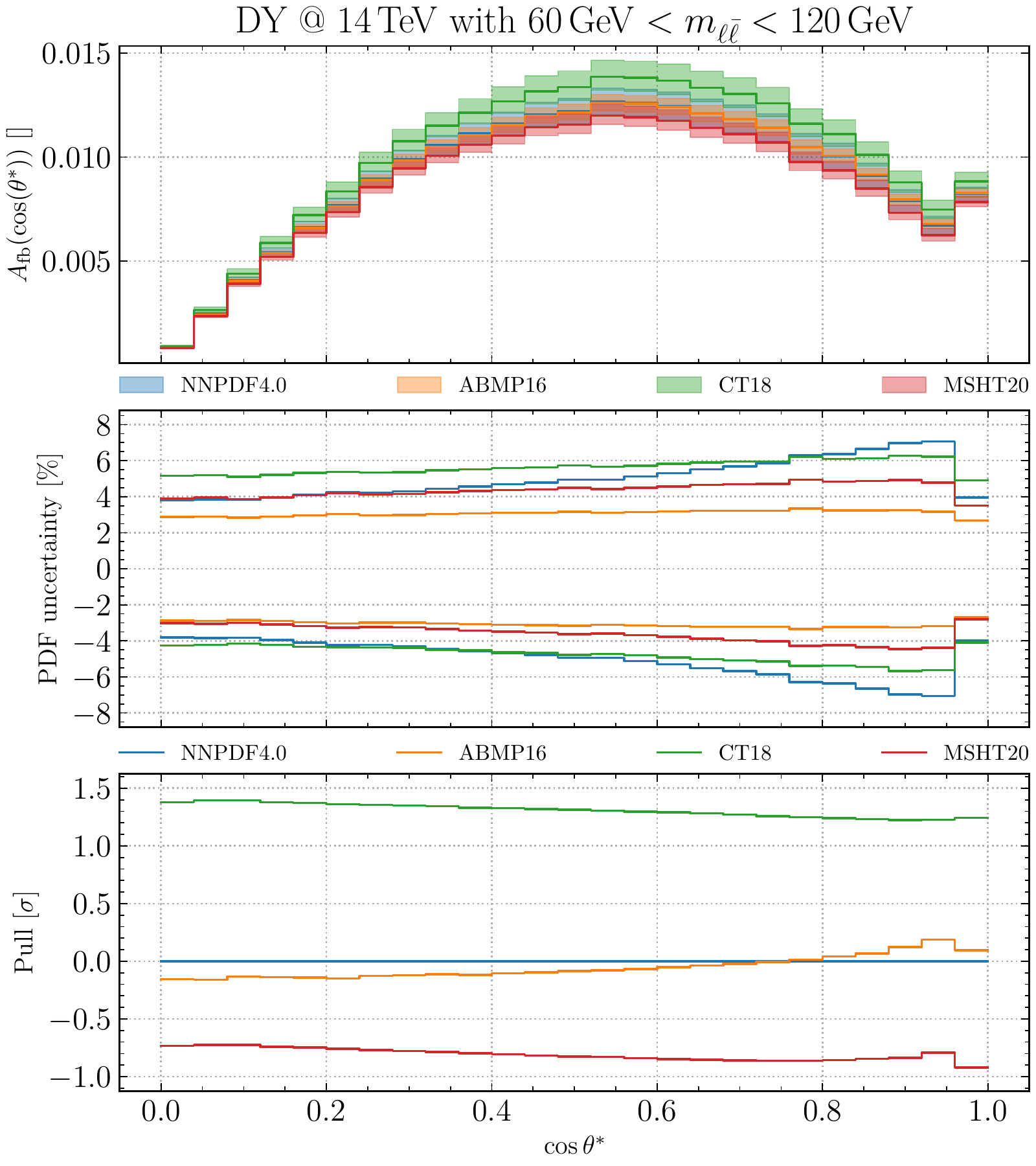}
\caption{Same as Fig.~\ref{fig:CMS_DY_14TEV_MLL_5000_rap}, now for the differential distribution in
  $\cos\theta^*$ (left)
  and the corresponding forward-backward asymmetry
  $A_{\rm fb}(\cos\theta^*)$ (right), in the $Z$-peak region defined by $60~{\rm GeV} < \mll < 120$ GeV.}
\label{fig:CMS_DY_14TEV_MLL_zpeak}
\end{figure}
%-------------------------------------------------------------------------------------

We now turn to the differential distribution in
  $\cos\theta^*$ 
  and the corresponding forward-backward asymmetry $A_{\rm
    fb}(\cos\theta^*)$.
We first consider the $Z$-peak region, $60~{\rm GeV} < \mll <
120$~GeV, in Fig.~\ref{fig:CMS_DY_14TEV_MLL_zpeak}.
The $\cos\theta^*$ 
distribution exhibits a small but non-negligible asymmetry,
and uncertainties are  smallest for NNPDF4.0.
The four PDF sets predict a similar behavior and magnitude
of the asymmetry $A_\mathrm{fb}$.
PDF uncertainties in the asymmetry
are  comparable for  all PDF sets when $\cos\theta^* \approx0$,
and actually  largest for NNPDF4.0 when $\cos\theta^* \approx 1$.
In all cases the predictions are compatible within $2 \sigma$,
with ABMP16 showing larger differences of up to $2.8 \sigma$ for the $\cos\theta^*$
distribution.
Note that the
sharp drop-off at the edges $|\cos\theta^*| \approx  1$, appearing in
all plots in this section, is a consequence of the phase-space cuts which
limit the phase-space volume.
Indeed, using  LO kinematics
\begin{equation}
| \cos\theta^* | = \tanh \left| \frac{\eta_\ell - \eta_{\bar{\ell}}}{2} \right| = \sqrt{1 - \frac{4 (p_T^\ell)^2}{\mll^2}} \text{,}
\end{equation}
so $| \cos\theta^* | \approx 1$ requires a lepton pair with either
a large rapidity separation, or a very large invariant mass and small
transverse momenta. 

\begin{figure}[t!]
 \centering
 \includegraphics[width=0.49\linewidth]{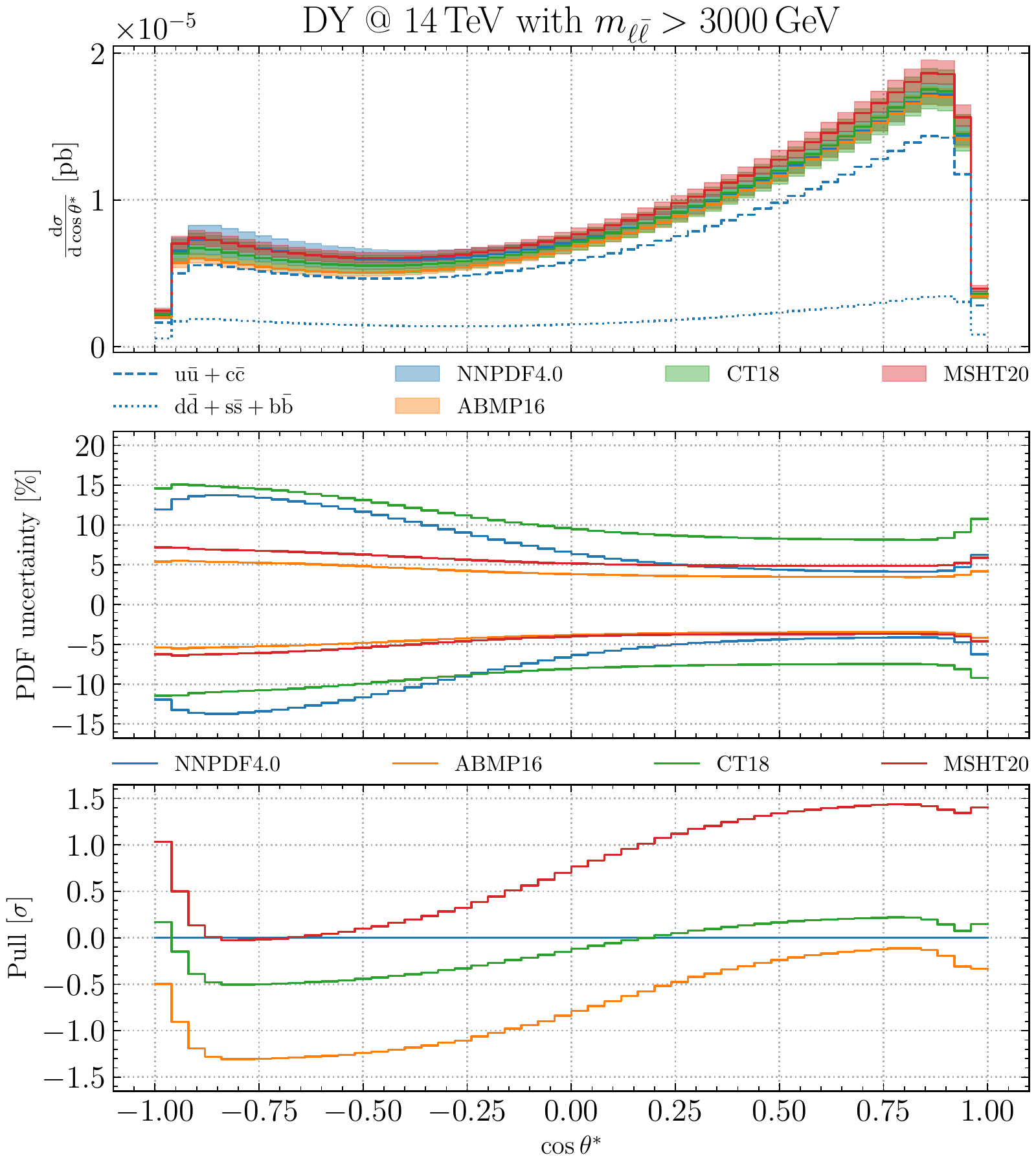}
 \includegraphics[width=0.49\linewidth]{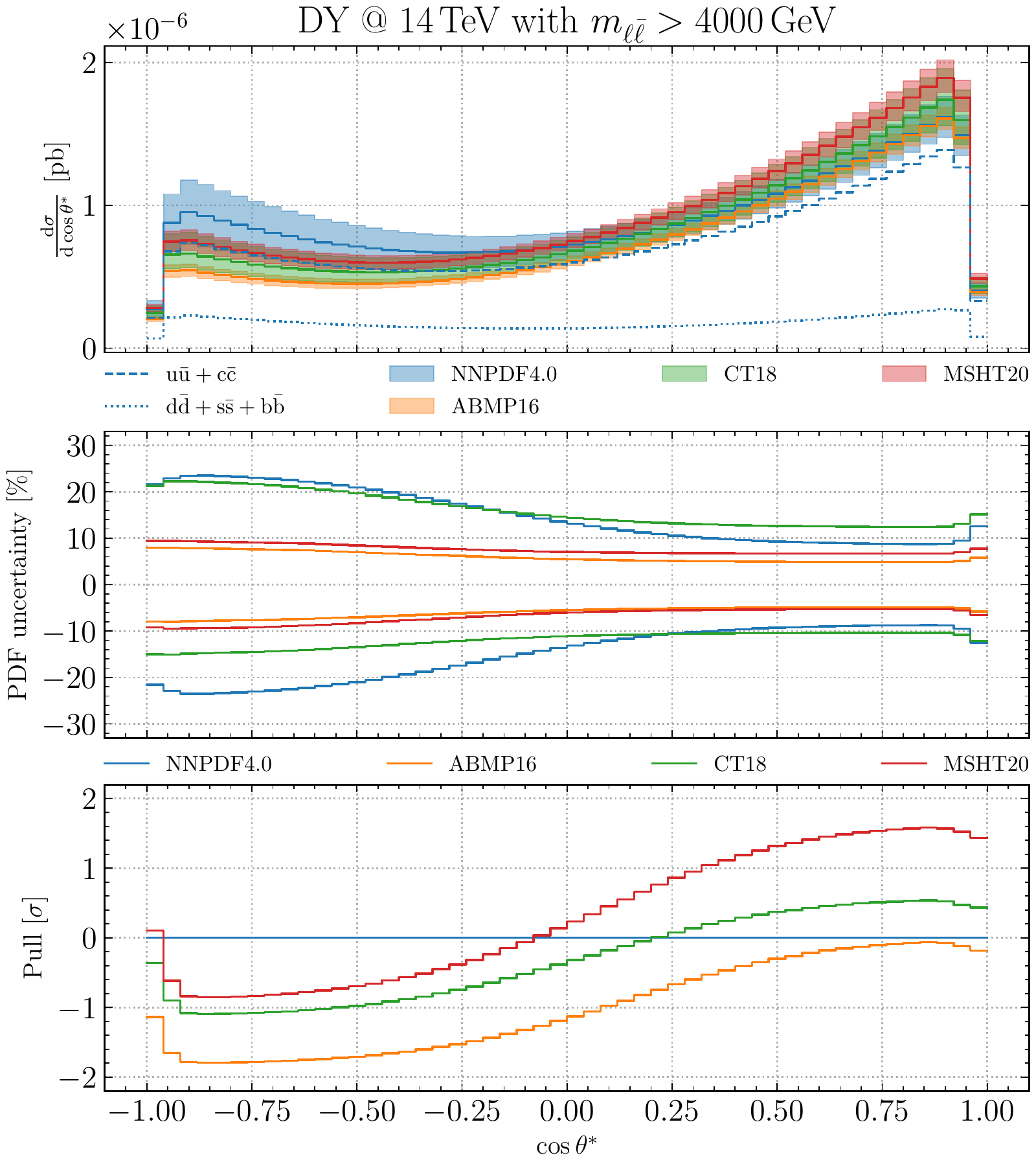}
 \includegraphics[width=0.49\linewidth]{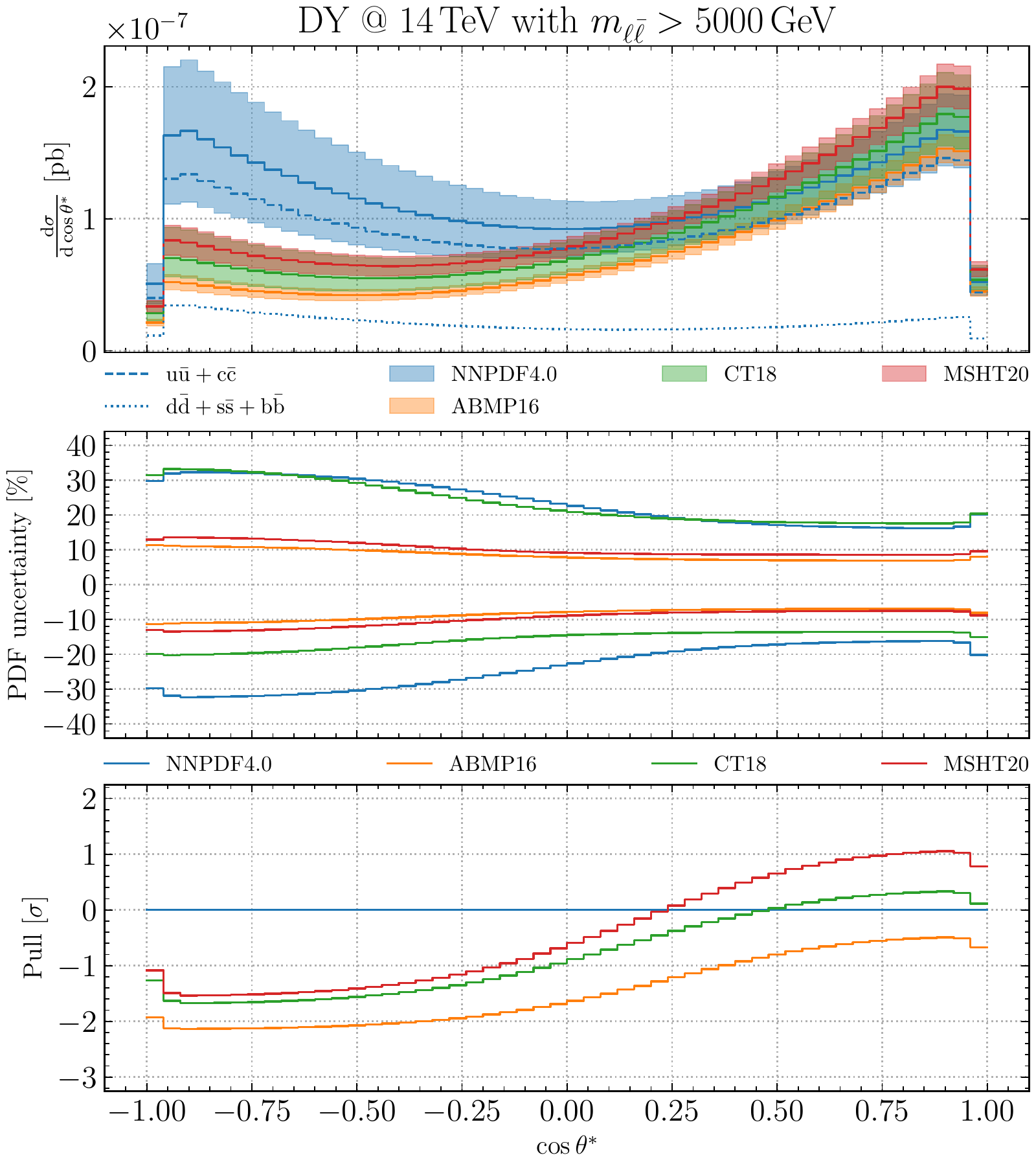}
 \includegraphics[width=0.49\linewidth]{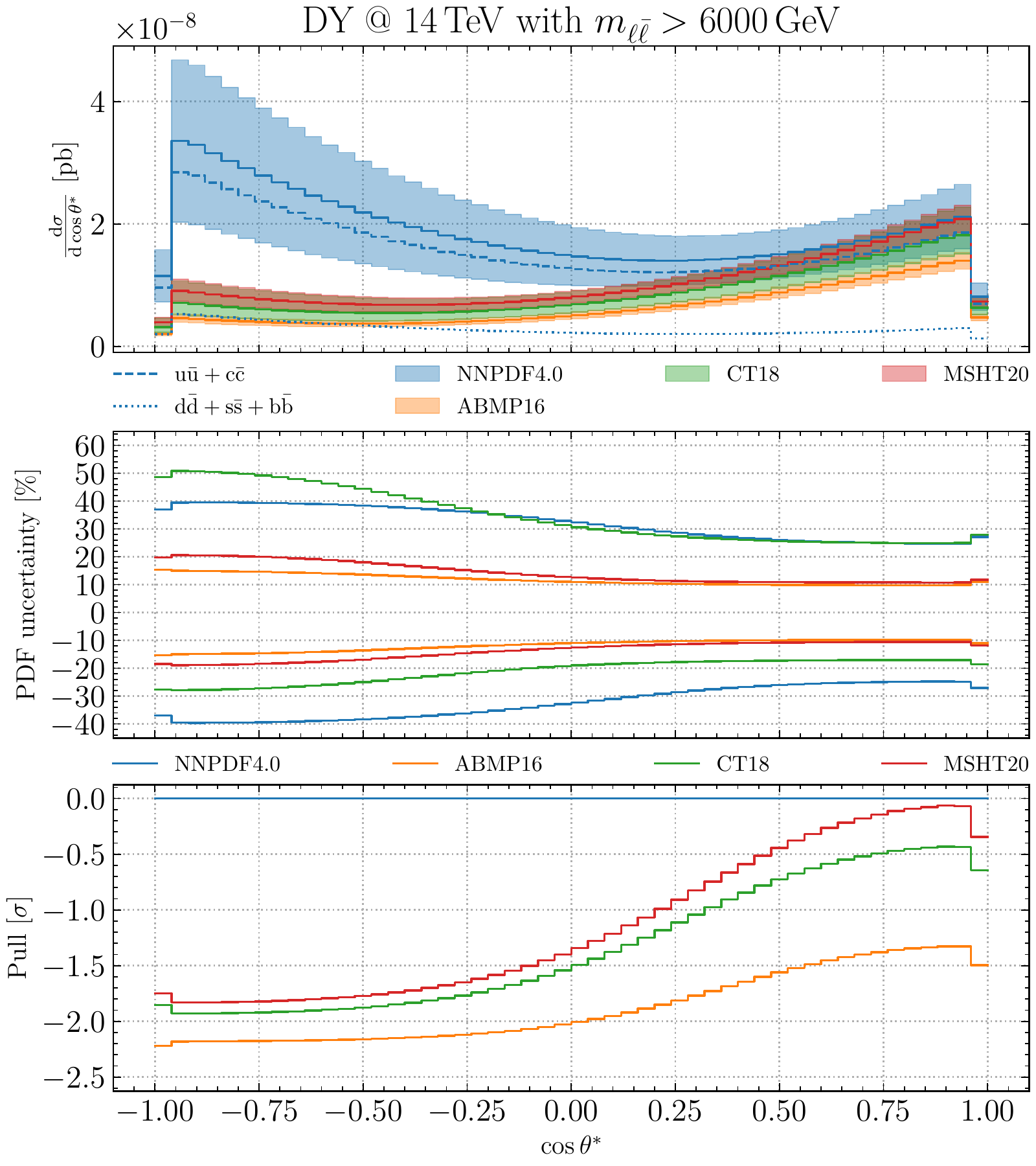}
 \caption{Same as Fig.~\ref{fig:CMS_DY_14TEV_MLL_zpeak} (left)
   for different values of the  lower cut in the dilepton
   invariant mass: $\mll \ge 3, 4,5,$ and 6 TeV respectively.
  }    
 \label{fig:CMS_DY_14TEV_MLL_others}
\end{figure}

\begin{figure}[t!]
 \centering
 \includegraphics[width=0.49\linewidth]{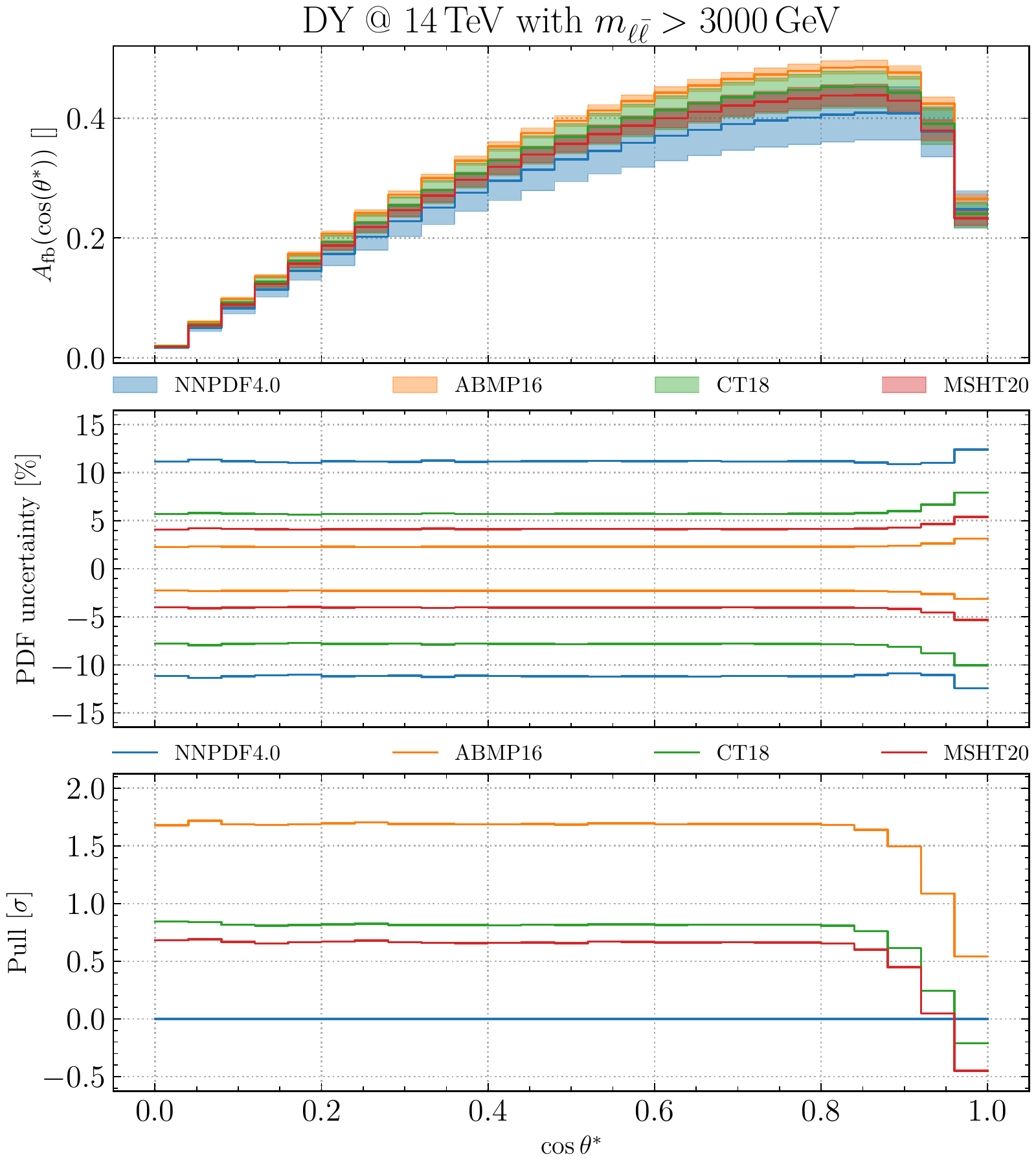}
 \includegraphics[width=0.49\linewidth]{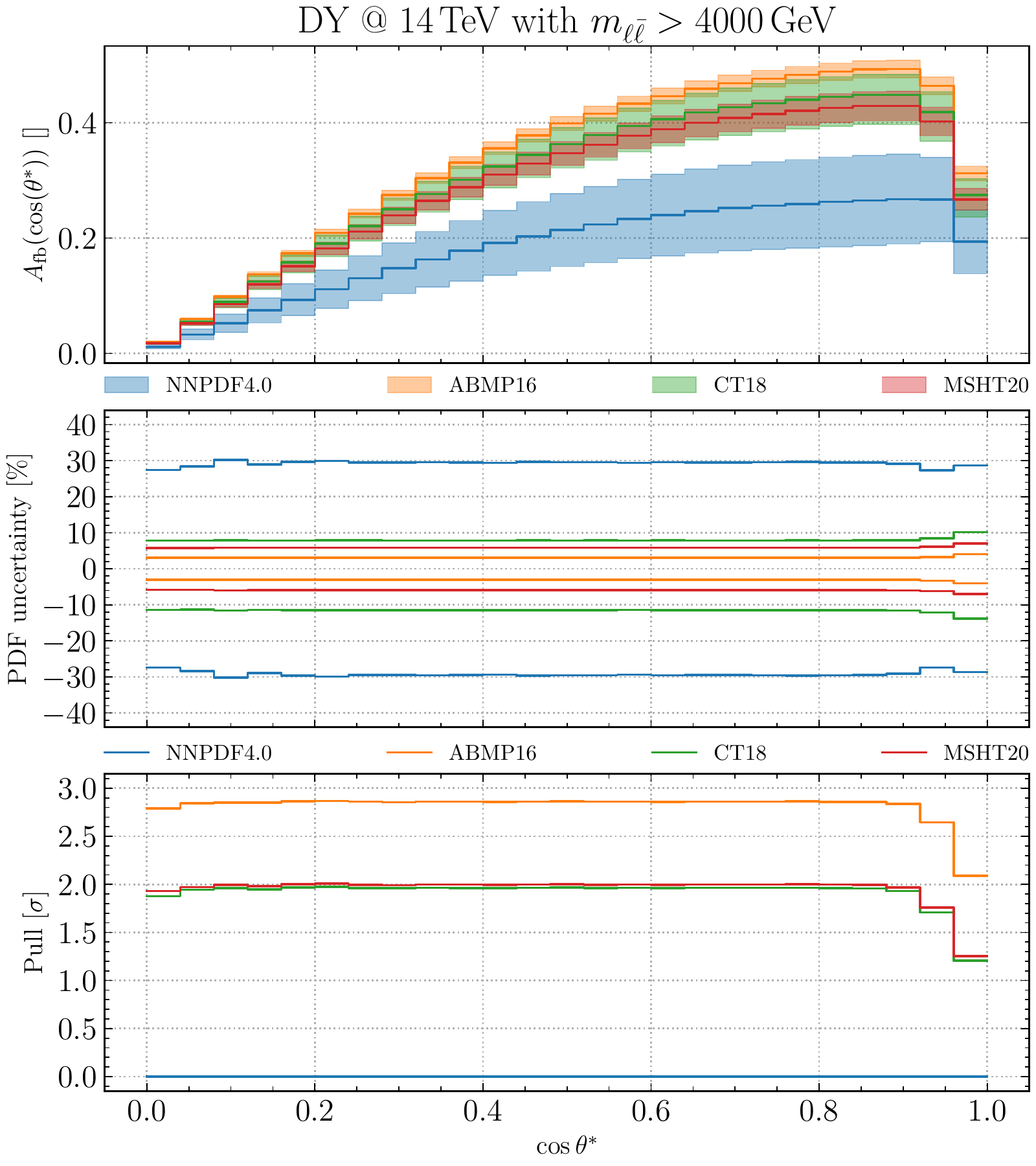}
 \includegraphics[width=0.49\linewidth]{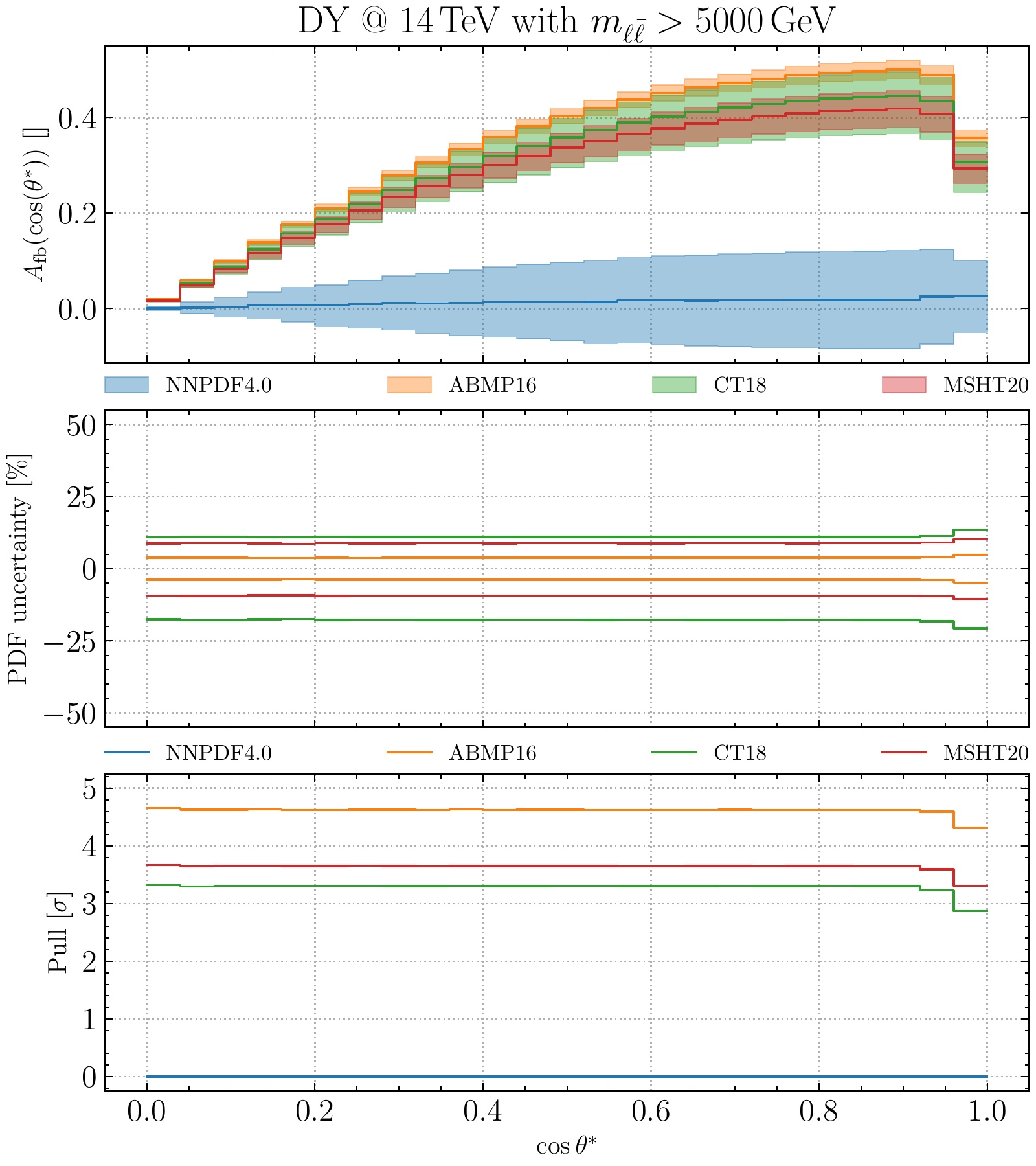}
 \includegraphics[width=0.49\linewidth]{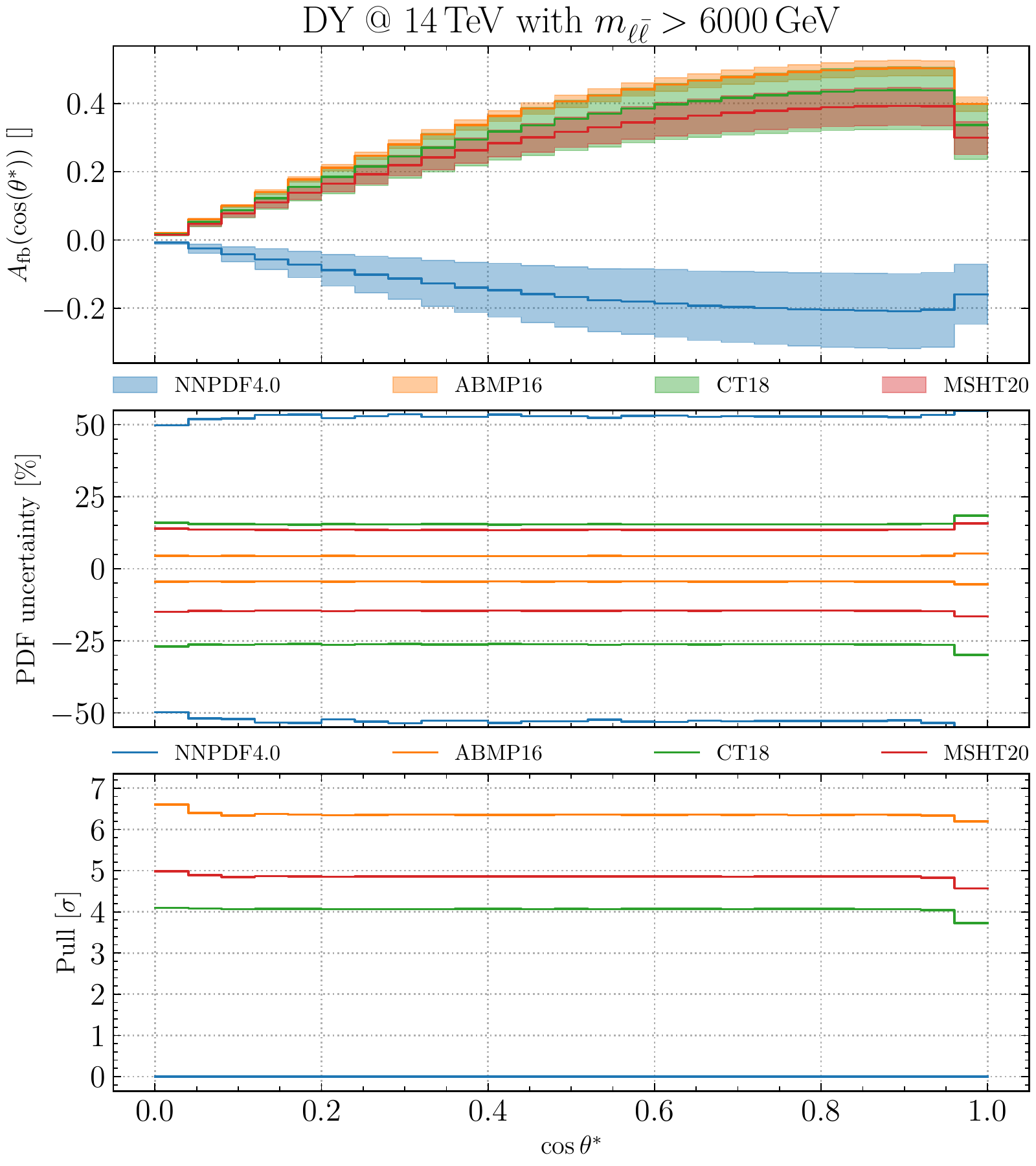}
 \caption{Same as Fig.~\ref{fig:CMS_DY_14TEV_MLL_zpeak} (right)
   for different values of the  lower cut in the dilepton
   invariant mass: $\mll^{\rm min}=3, 4,5,$ and 6 TeV.
  }    
 \label{fig:CMS_DY_14TEV_MLL_others_asy}
\end{figure}

As expected from the antisymmetric partonic luminosities studied in
Sect.~\ref{subsec:partoniclumis}, the situation is quite different when
considering distributions with a higher dilepton invariant mass range.
The angular distribution and forward-backward asymmetry
in the high-mass region, for different values of the  lower cut in the dilepton
 invariant mass, namely $\mll^{\rm min}=3, 4,5$ and 6 TeV, are
 respectively
 shown in
Fig.~\ref{fig:CMS_DY_14TEV_MLL_others} and Fig.~\ref{fig:CMS_DY_14TEV_MLL_others_asy}.

%-------------------------------------------------------------------------------
%-------------------------------------------------------------------------------

 Consistent with the underlying parton luminosities, the $\cos\theta^*$ distribution
 is dominated by $u\bar{u}$ scattering, while  $d\bar{d}$ provides
 a subdominant contribution.
 When the lower cut  is $\mll^{(\rm min)}=3$ TeV is used, the four PDF
 sets are in agreement at the $1\sigma$ level: they all
 display a 
 positive forward-backward asymmetry, and exhibit PDF uncertainties ranging between 10\% and 15\%.
 As the invariant mass cut is raised, the qualitative behavior of the
 angular distribution and
 asymmetry change substantially for NNPDF4.0, while they remain
 approximately the same for all other PDF sets, consistent with the
 behavior of the PDFs and luminosities discussed in
 Sect.~\ref{sec:subsec-largexPDFs}-\ref{subsec:partoniclumis}.
 Specifically,
 raising the cut to
 $\mll \ge 4$ TeV, for NNPDF4.0
 the backwards cross-section starts increasing, though the asymmetry remains
positive.

For $\mll\ge 5$ TeV the central value of the NNPDF4.0 $\cos\theta^*$
 distribution  becomes symmetric, though the  PDF uncertainty band is
 rather asymmetric. Also, PDF uncertainties
 are now the largest for NNPDF4.0, reaching up to 30\%.
 Finally, for $\mll\ge 6$ TeV  the central value of 
 forward-backward asymmetry for NNPDF4.0 becomes negative, with the
 PDF uncertainties increasing further so the asymmetry remains compatible
 with zero at about the 1.1~$\sigma$ level.
 For all other PDF sets there is little change in the shape of the distribution as the
 dilepton invariant mass cut is increased.
 Because of the large uncertainty on the NNPDF4.0 result for the $\cos\theta^*$
distribution, even with
the highest value of the  $\mll^{(\rm  min)}$ cut, where NNPDF4.0 finds a
symmetric distributions while all other PDF sets find an asymmetry,
the pull is always below $2 \sigma$.

%-------------------------------------------------------------------------------
%-------------------------------------------------------------------------------

\begin{figure}[t!]
 \centering
 \includegraphics[width=0.49\linewidth]{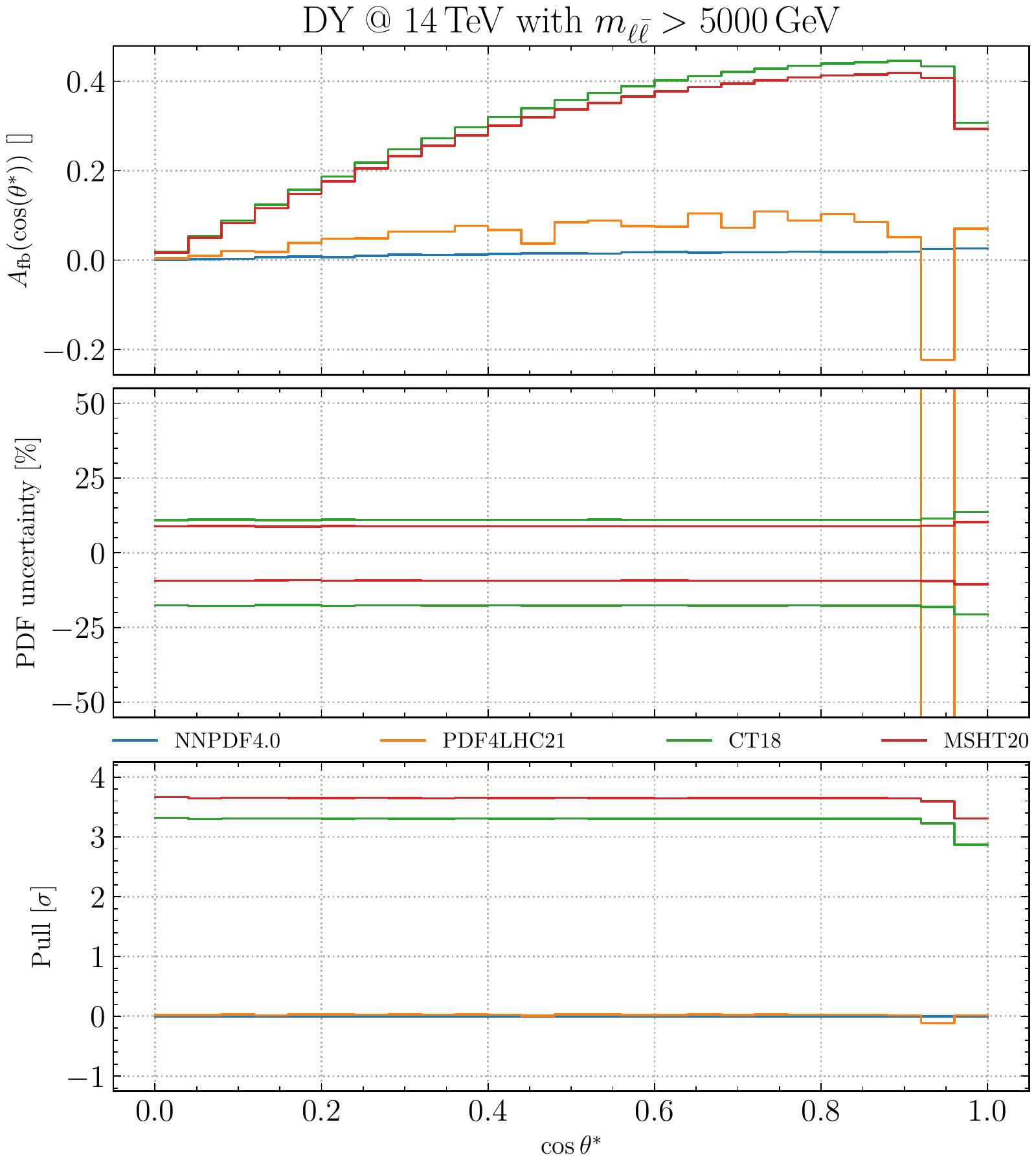}
 \includegraphics[width=0.49\linewidth]{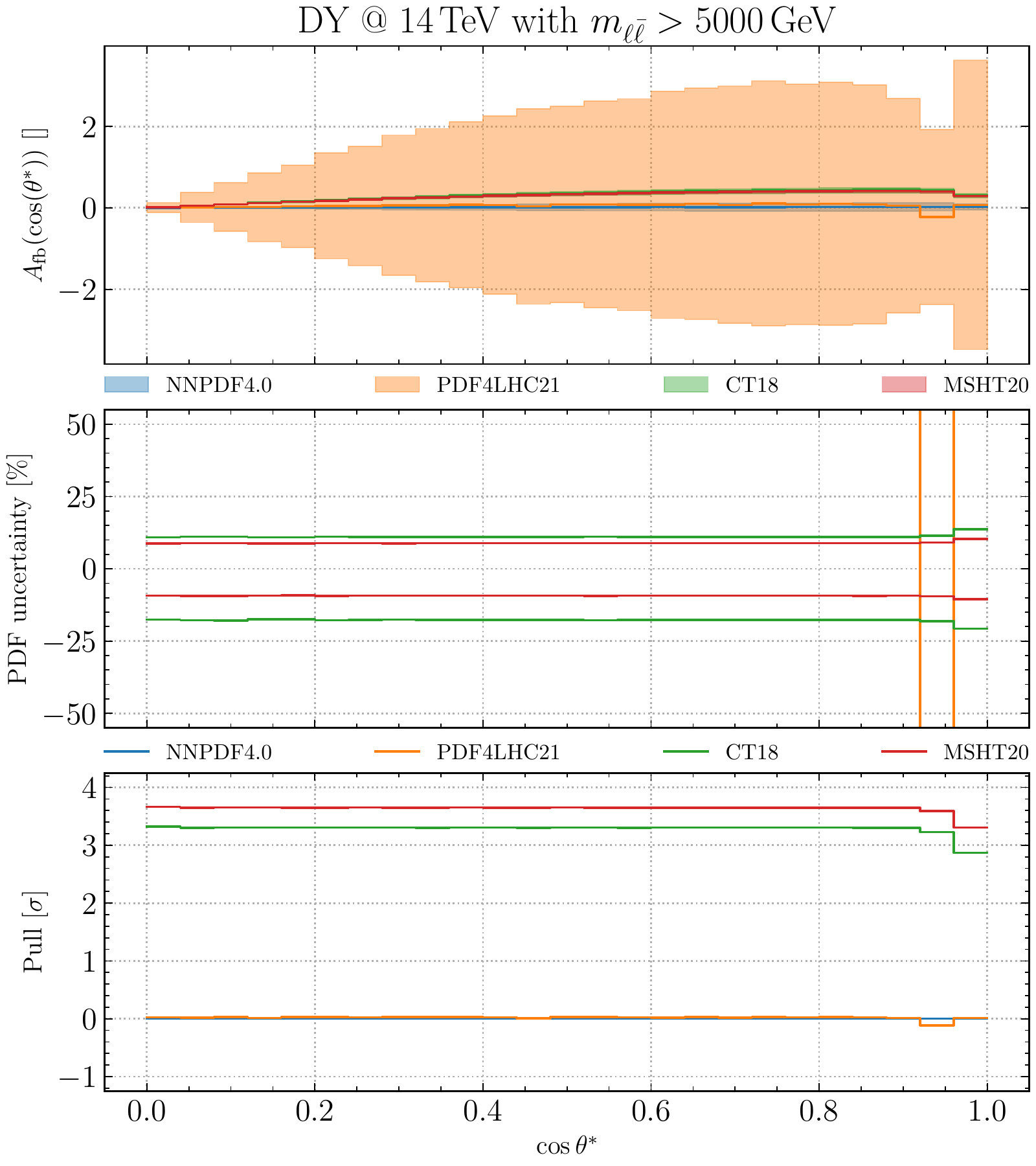}
 \caption{Same as Fig.~\ref{fig:CMS_DY_14TEV_MLL_others_asy} (bottom left), now
   also including the PDF4LHC21 prediction. In the left plot the PDF
   uncertainty is not shown, in the right plot the scale on the $y$
   axis is suitably expanded. Note  that the PDF4LHC percentage PDF
   uncertainty does not show as it falls outside the plot. 
  }    
 \label{fig:CMS_DY_14TEV_MLL_pdf4lhc}
\end{figure}

In Fig.~\ref{fig:CMS_DY_14TEV_MLL_pdf4lhc} the forward-backward asymmetry
with  $\mll^{(\rm min)}=5$~TeV shown in
Fig.~\ref{fig:CMS_DY_14TEV_MLL_others_asy} (bottom left) is shown again,
now also including a prediction obtained using the PDF4LHC21
combination of parton
distributions~\cite{PDF4LHCWorkingGroup:2022cjn}, specifically its
compressed Monte Carlo representation~\cite{Carrazza:2015hva}, based on
the CT18, MSHT and NNPDF3.1 PDF sets.
Because the PDF uncertainties
for NNPDF3.1  are
generally, and in particular at large $x$, rather larger than those on
NNPDF4.0 (see also App.~\ref{app:nnpdf31}) the uncertainty on the
$A_{\rm fb}$ distribution found using the PDF4LHC21 combination is extremely large,
and no signal for the asymmetry can be seen.
PDF4LHC recommends~\cite{PDF4LHCWorkingGroup:2022cjn} usage of the
combination for BSM searches, and that of individual PDF sets for
comparison between data and theory for SM
measurements.
The results presented here suggest that
the uncertainty estimate of  NNPDF4.0 in the extrapolation
region, which is rather more conservative than that  of the other PDF
sets shown here,  might be 
desirable and lead to more robust predictions for the
forward-backward asymmetry in the high-mass region which is relevant
for new physics searches.

%-------------------------------------------------------------------------------
%-------------------------------------------------------------------------------

\section{Summary and outlook}
\label{sec:summary}

In this work we have scrutinised the PDF dependence of  neutral current
Drell-Yan production
at large dilepton invariant masses $\mll$, focusing on the behavior
of the forward-backward asymmetry $A_{\rm fb}$
in the Collins-Soper angle $\cos\theta^*$, an observable frequently
considered in the context of searches for new physics beyond the SM.
We have demonstrated that while theoretical
predictions for the sign and magnitude of $A_{\rm fb}$ are very
similar for all PDF sets in the
$Z$ peak region, they
depend markedly on the choice of PDF set for  large values of $\mll$. 
We have traced this behavior to that of the PDFs, which agree in the
data region, but differ in the large-$x$
region, where PDFs are mostly unconstrained by data.

We have specifically shown that the uncertainty on the asymmetry
differs substantially between PDF sets, with NNPDF4.0 displaying
a more marked increase as  $\mll$ grows, leading to
an absolute uncertainty that e.g.\ for $\mll^{\rm min}\gtrsim4$~TeV
is about twice as large as that found using CT18, 
four times as large as MSHT20,
and about one order of magnitude larger than ABMP16.
Also, whereas other PDF sets predict a shape of the asymmetry
which is unchanged when  $\mll$ increases from the $Z$-peak region to
the TeV range, namely a 
positive
asymmetry implying a larger cross-section  for $\cos\theta^*\ge 0$, NNPDF4.0 finds that as
$\mll$ increases, the asymmetry is reduced, and the $\cos\theta^*$ distribution
becomes symmetric when $\mll^{\rm min}\sim5$~TeV.

We have traced this behavior to that of the underlying
PDFs in the large-$x$ region, where PDFs are mostly unconstrained by
data.
Specifically we have seen that in this region NNPDF4.0 has
generally wider uncertainties.
Also, while for  all PDF  sets the
quark and antiquark distributions vanish as
a power of $(1-x)$ as $x\to 1$, for all groups but NNPDF4.0 this power is
constant for light quarks to the right of the valence peak, while for
NNPDF4.0 it changes as $x$ increases, slowly for up quarks, more rapidly
for down quarks and even more rapidly for antiquarks.
All this suggests that the different behavior of
NNPDF4.0 is due to its more flexible PDF parametrization.

Our general conclusion is that the 
behavior of the forward-backward asymmetry
observed at lower invariant masses is not necessarily reproduced at
large masses if flexible enough
PDFs are used:  the characteristic positive asymmetry observed
for low $\mll$ values
can be washed out in the high-mass region.
Hence, deviations from the traditional expectation of a positive forward-backward
asymmetry in high-mass Drell-Yan cannot be taken as an indication of 
BSM physics,
at least based on  our current understanding of proton structure in the large-$x$ region.

Turning the argument around, future measurements of the $\cos\theta^*$
distribution and the associated forward-backward asymmetry 
$A_{\rm fb}$ when included in PDF determinations could help in
constraining PDFs at large $x$.
For instance, Fig.~\ref{fig:CMS_DY_14TEV_MLL_others} indicates that for
$\mll^{\rm min}=5$ TeV and $\sqrt{s}=14$ TeV the
asymmetry $A_{\rm fb}$ can be as large as 50\% for ABMP16
while it vanishes (within large uncertainties) in the case of NNPDF4.0.
By rebinning the $\cos\theta^*$ distribution, for an integrated
luminosity of $\mathcal{L}=6$ ab$^{-1}$, corresponding to the
combination at ATLAS and CMS 
at the end of the HL-LHC data-taking period, $\mathcal{O}(10)$ events are expected in the backward region,
with an statistical uncertainty of $\delta_{\rm stat}\sim 30\%$ which could be sufficient to
discriminate between these two limiting scenarios at the $2\sigma$ level.

Higher event counts are expected if the $\mll$ cut is loosened, though one is
then less sensitive to the large-$x$ region where differences between PDF sets and their
uncertainties are the largest.
Ultimately, the constraining power of high-mass Drell-Yan in general and of the forward-backward
asymmetry in particular can only be addressed by means of a dedicated projections
based on binned pseudo-data such as those carried
out for the HL-LHC and the Electron Ion Collider in e.g.~\cite{AbdulKhalek:2018rok,Khalek:2021ulf}.
While we leave this exercise for a future study, the investigations
presented in this work indicate that $A_{\rm fb}$
at high-invariant masses represents a promising and mostly
unexplored channel to pin down large-$x$ light
quark and antiquark PDFs at the HL-LHC.

While in this work
we have focused on the forward-backward asymmetry in neutral-current Drell-Yan production,
similar considerations apply for other processes relevant
for BSM searches at high mass at the LHC.
Indeed, the HL-LHC will be sensitive to a broad range of hypothetical
new massive particles, from resonances in the $m_{jj}$ dijet invariant mass distribution up to 11 TeV,
heavy vector triplet resonances decaying into a diboson $VV'$ pair up to 5 TeV,
and gluinos with masses up to $m_{\tilde{g}}=3$ TeV in the minimal
supersymmetric standard model (MSSM) with a massless lightest SUSY
particle~\cite{CidVidal:2018eel}.

For all these channels, a robust understanding of PDFs
and their uncertainties at large $x$, including the role of
methodological and model assumptions, will be necessary to fully exploit
the HL-LHC discovery potential for BSM signatures.
Conversely, once BSM phenomena have been excluded in some high-energy channel,
the corresponding search can be unfolded into a measurement to provide direct
constraints on the PDFs in this key large-$x$ region, which in turn will
enhance the reach of other searches. It would be very interesting to
perform a detailed study, in the same vein as
Ref.~\cite{AbdulKhalek:2018rok}, of the impact of future HL-LHC data
on large-$x$ PDFs and the prospect of asymmetry measurements in
searches for new physics, but this will be left for future work.

\subsection*{Acknowledgments}

We are grateful to Dimitri Bourilkov, Alexander Grohsjean, Meng Lu, and Jan Schulte for raising with us the issue of the
PDF dependence of $A_{\rm fb}$ at high invariant masses and for the subsequent discussions.
A.~C., S.~F., and F.~H.\ are supported by
the European Research Council under 
the European Union's Horizon 2020 research and innovation Programme
(grant agreement n.740006).
R.~D.~B.\ is supported by the U.K.\
Science and Technology Facility Council (STFC) grant ST/P000630/1.
E.~R.~N.\ is supported by the Italian Ministry of University and Research (MUR)
through the ``Rita Levi-Montalcini'' Program.
J.~R.\ is partially supported by NWO (Dutch Research Council).
C.~S.\ is supported by the German Research Foundation (DFG) under
reference number DE~623/6-2.

\appendix
\section{\texorpdfstring{$A_{\rm fb}$}{Afb} in NNPDF3.1}
\label{app:nnpdf31}

In this appendix we compare partonic luminosities
and LHC differential distributions obtained with NNPDF4.0 in
Sects.~\ref{sec:largexpdfs}
and~\ref{sec:afb} with those based 
on its predecessor
NNPDF3.1, as well as with a variant of NNPDF4.0
where positivity is imposed at the level of observable cross-sections but
not at the PDF level, as was the case in  NNPDF3.1, which we will denote
NNPDF4.0(3.1pos).

\begin{figure}[!t]
 \centering
 \includegraphics[width=0.99\linewidth]{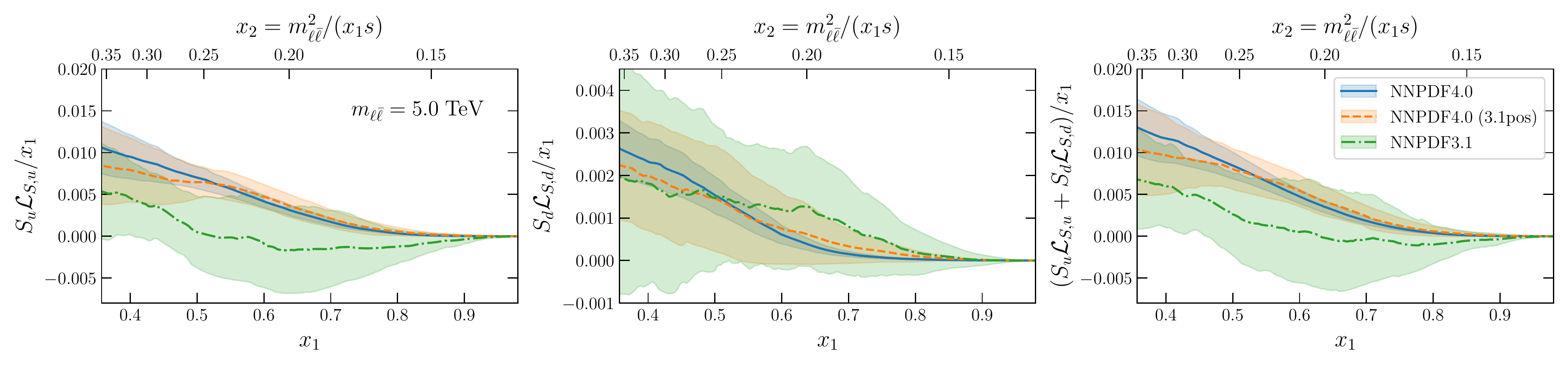}
 \caption{Same as Fig.~\ref{fig:mll_dep_lumi_plus} (upper panels) comparing
NNPDF4.0, NNPDF4.0(3.1pos), and NNPDF3.1.
 }    
 \label{fig:pdfplot-absDYlumis-pdfsets-plus-q5tev-nnpdf31}
\end{figure}

Fig.~\ref{fig:pdfplot-absDYlumis-pdfsets-plus-q5tev-nnpdf31}
compares the 
symmetric partonic luminosities $\mathcal{L}_{S,q}$
evaluated for $\mll=5$~TeV.
The three sets are found to agree within uncertainties,
with NNPDF4.0 having the smallest uncertainties.
This increase in precision
arises only marginally due to the more restrictive positivity constraints,
since predictions with the NNPDF4.0(3.1pos) variant 
are close to the baseline NNPDF4.0, especially 
for the $u\bar{u}$ contribution, for both central values and uncertainties.
The comparison in Fig.~\ref{fig:pdfplot-absDYlumis-pdfsets-plus-q5tev-nnpdf31}
indicates that phenomenological predictions for high-mass Drell-Yan
production based on NNPDF3.1 are expected
to be consistent within errors with those of NNPDF4.0 for the contributions
symmetric in $\cos\theta^*$, such as the $|\yll|$ distribution.

%-------------------------------------------------------------------------------
%---------------------------------------------------------------------------

\begin{figure}[!t]
 \centering
 \includegraphics[width=0.99\linewidth]{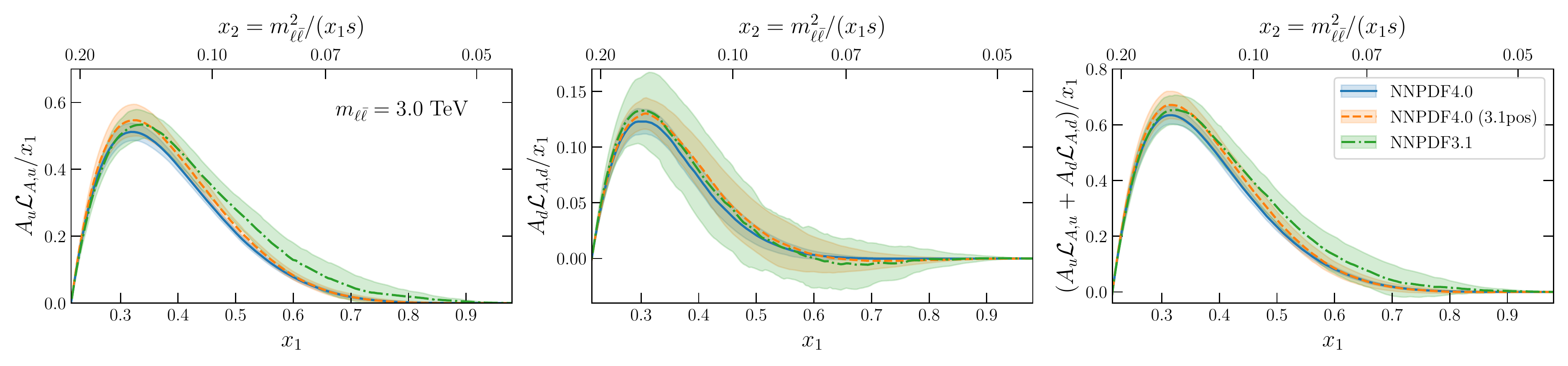}
 \includegraphics[width=0.99\linewidth]{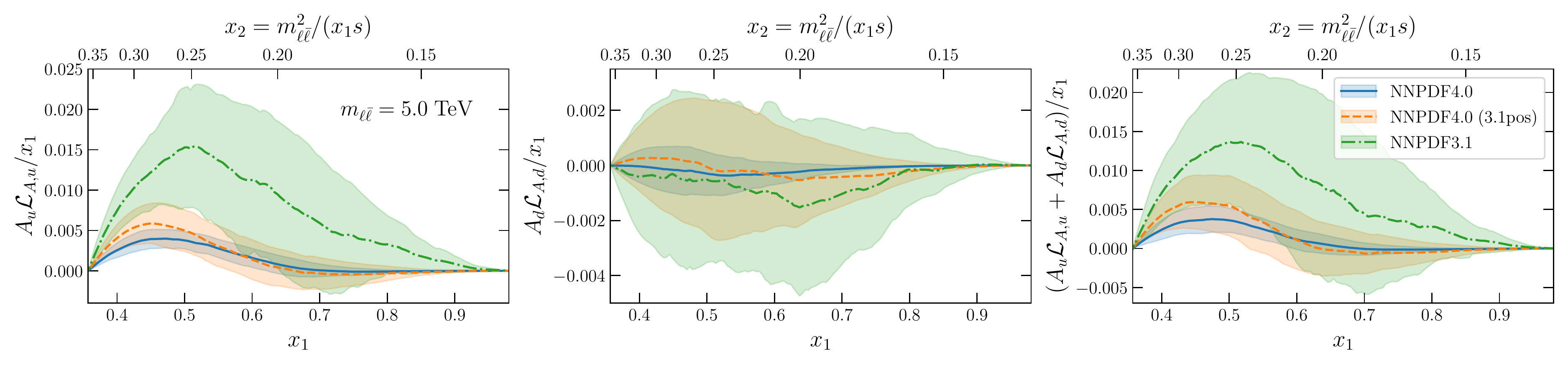}
 \caption{Same as Fig.~\ref{fig:mll_dep_lumi_minus} for the antisymmetric partonic luminosities $\mathcal{L}_{A,q}$,
   comparing NNPDF4.0, NNPDF4.0(3.1pos), and NNPDF3.1.
 }    
 \label{fig:pdfplot-absDYlumis-pdfsets-minus-q5tev-nnpdf31}
\end{figure}

The antisymmetric luminosities $\mathcal{L}_{A,q}$, relevant for the
forward-backward asymmetry, are displayed in Fig.~\ref{fig:pdfplot-absDYlumis-pdfsets-minus-q5tev-nnpdf31}
for $\mll = 3$ and 5 TeV respectively.
Their qualitative behavior is similar for all  PDF sets,
with a marked decrease of PDF uncertainties first from NNPDF3.1
to  NNPDF4.0(3.1pos)  then
to NNPDF4.0.
Specifically, the qualitative $\mll$ dependence
of $\mathcal{L}_{A,q}$ remains unchanged. Namely, the positive $A_{\rm fb}$
found for $\mll= 3$ TeV decreases 
as the dilepton invariant mass is increased.
Hence also for the component of the Drell-Yan cross-section which is odd
in $\cos\theta^*$ we expect LHC predictions based on NNPDF3.1 to be consistent
with those obtained from NNPDF4.0.

%-------------------------------------------------------------------------------
%---------------------------------------------------------------------------

\begin{figure}[!t]
 \centering
 \includegraphics[width=0.49\linewidth]{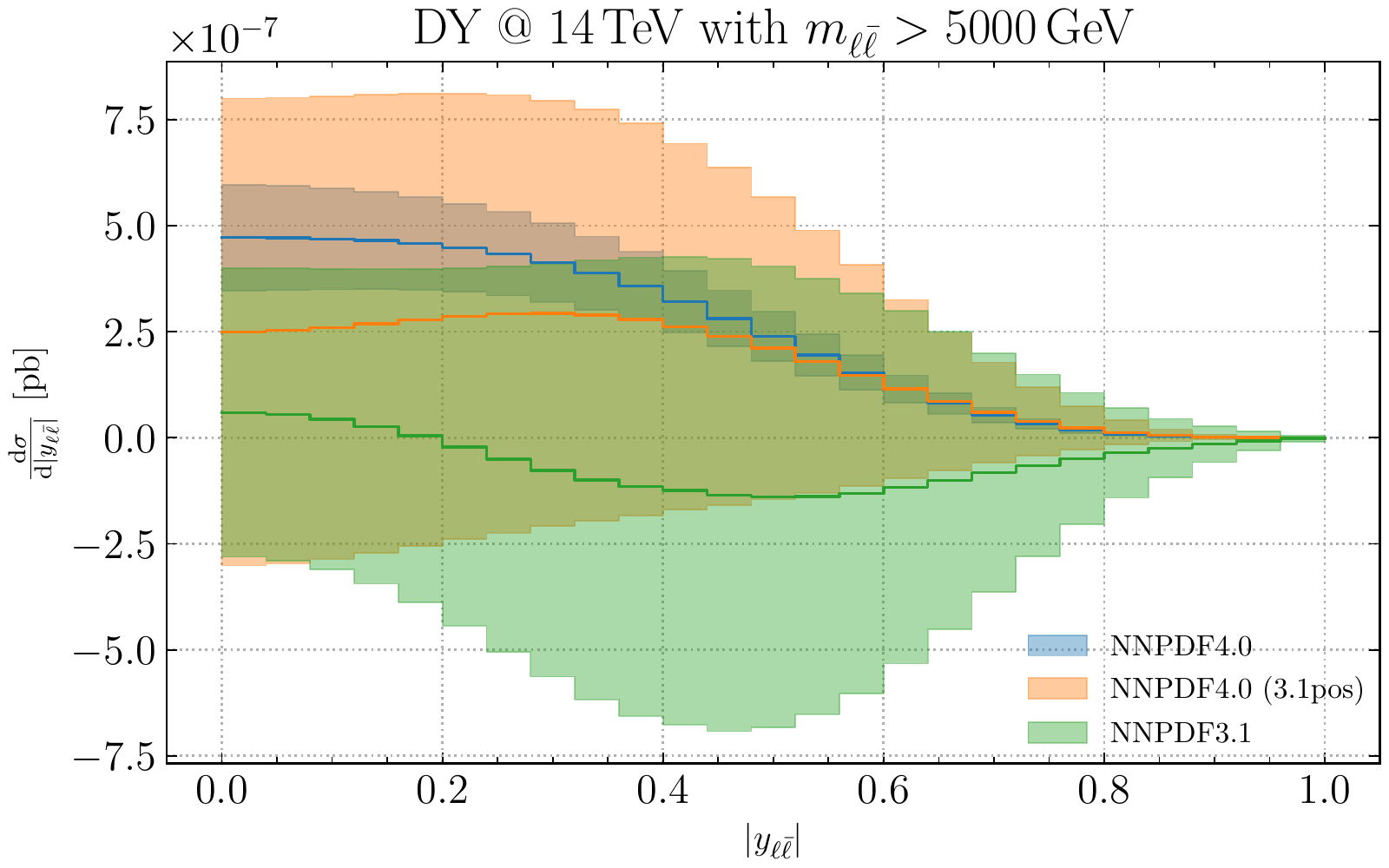}
 \includegraphics[width=0.49\linewidth]{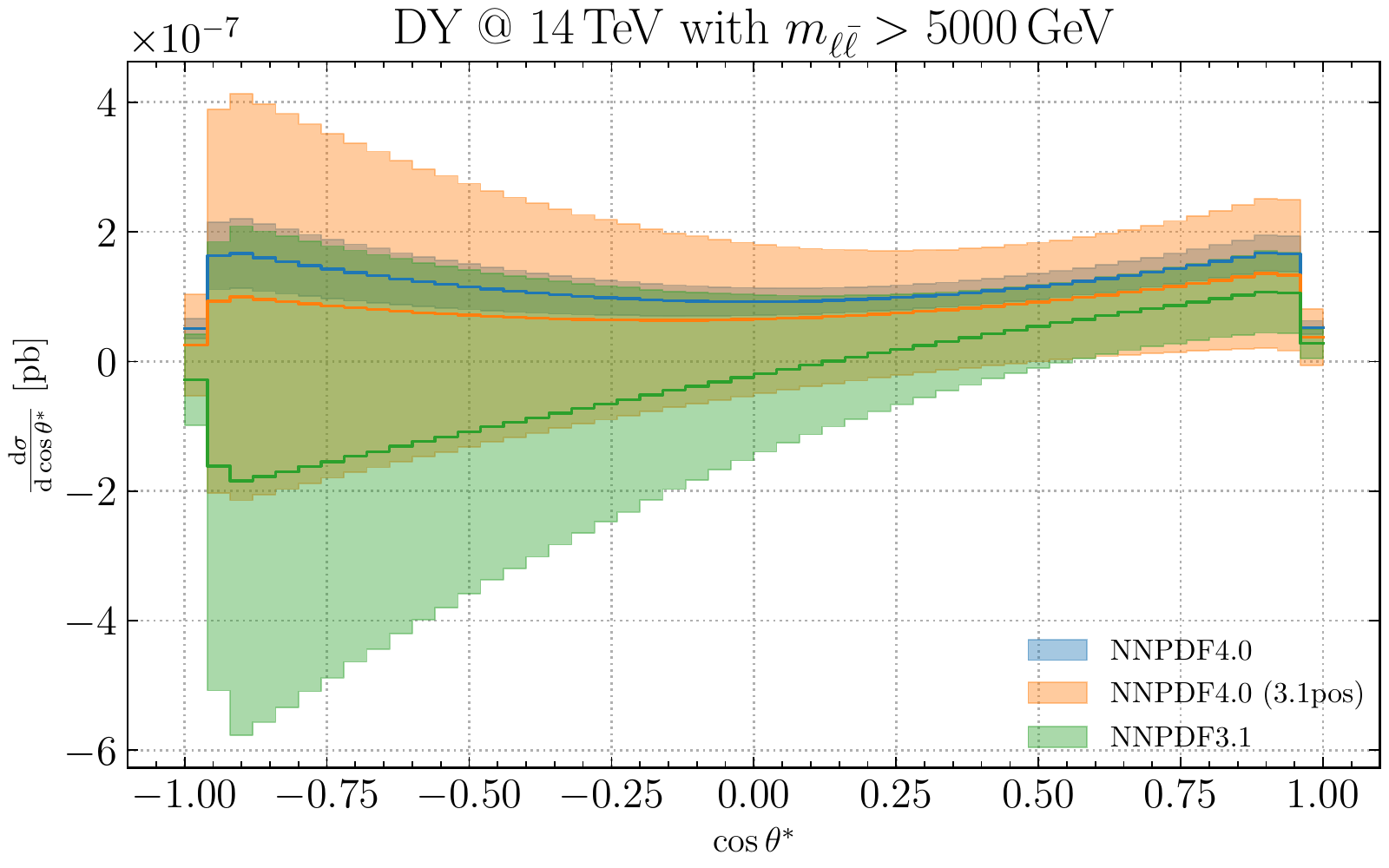}
 \caption{Same as Figs.~\ref{fig:CMS_DY_14TEV_MLL_5000_rap}
and~\ref{fig:CMS_DY_14TEV_MLL_others}
for the absolute dilepton rapidity $|y_{\ell\bar{\ell}}|$ (left)
and the $\cos \theta^*$  (right) distributions
for dilepton invariant masses of $m_{\ell\bar{\ell}}\ge 5$ TeV
 comparing
NNPDF4.0, NNPDF4.0(3.1pos), and NNPDF3.1.
 }    
 \label{fig:CMS_DY_14TEV_COSTH_5000_YLL40-vs-31}
\end{figure}

These expectations are confirmed by
Fig.~\ref{fig:CMS_DY_14TEV_COSTH_5000_YLL40-vs-31}, which shows the
dilepton rapidity $|y_{\ell\bar{\ell}}|$ 
and the Collins-Soper angle $\cos\theta^*$ distributions for neutral-current DY production
at the LHC 14 TeV for dilepton invariant masses of $m_{\ell\bar{\ell}}\ge 5$ TeV,
comparing the baseline NNPDF4.0 predictions with those from NNPDF3.1
and NNPDF4.0(3.1pos).
Indeed, 
good agreement within the three PDF sets is observed with a significant reduction
of PDF uncertainties between NNPDF3.1 and NNPDF4.0, consistent
with the behaviour exhibited by the corresponding partonic luminosities.

%-------------------------------------------------------------------------------
%---------------------------------------------------------------------------

%\bibliography{bsm_afb}

\begin{thebibliography}{10}

\bibitem{CidVidal:2018eel}
X.~Cid~Vidal et~al., {\it {Report from Working Group 3}: {Beyond the Standard
  Model physics at the HL-LHC and HE-LHC}},  {\em CERN Yellow Rep. Monogr.}
  {\bf 7} (2019) 585--865, [\href{http://arxiv.org/abs/1812.07831}{{\tt
  arXiv:1812.07831}}].

\bibitem{Gao:2017yyd}
J.~Gao, L.~Harland-Lang, and J.~Rojo, {\it {The Structure of the Proton in the
  LHC Precision Era}},  {\em Phys. Rept.} {\bf 742} (2018) 1--121,
  [\href{http://arxiv.org/abs/1709.04922}{{\tt arXiv:1709.04922}}].

\bibitem{Kovarik:2019xvh}
K.~Kova\v{r}\'\i{}k, P.~M. Nadolsky, and D.~E. Soper, {\it {Hadronic structure
  in high-energy collisions}},  {\em Rev. Mod. Phys.} {\bf 92} (2020), no.~4
  045003, [\href{http://arxiv.org/abs/1905.06957}{{\tt arXiv:1905.06957}}].

\bibitem{CDF:1996yow}
{\bf CDF} Collaboration, F.~Abe et~al., {\it {Inclusive jet cross section in
  ${\bar p p}$ collisions at $\sqrt{s}=1.8$ TeV}},  {\em Phys. Rev. Lett.} {\bf
  77} (1996) 438--443, [\href{http://arxiv.org/abs/hep-ex/9601008}{{\tt
  hep-ex/9601008}}].

\bibitem{Lai:1996mg}
H.~L. Lai, J.~Huston, S.~Kuhlmann, F.~I. Olness, J.~F. Owens, D.~E. Soper,
  W.~K. Tung, and H.~Weerts, {\it {Improved parton distributions from global
  analysis of recent deep inelastic scattering and inclusive jet data}},  {\em
  Phys. Rev. D} {\bf 55} (1997) 1280--1296,
  [\href{http://arxiv.org/abs/hep-ph/9606399}{{\tt hep-ph/9606399}}].

\bibitem{Beenakker:2015rna}
W.~Beenakker, C.~Borschensky, M.~Krämer, A.~Kulesza, E.~Laenen, S.~Marzani,
  and J.~Rojo, {\it {NLO+NLL squark and gluino production cross-sections with
  threshold-improved parton distributions}},  {\em Eur. Phys. J.} {\bf C76}
  (2016), no.~2 53, [\href{http://arxiv.org/abs/1510.00375}{{\tt
  arXiv:1510.00375}}].

\bibitem{Ethier:2021bye}
{\bf SMEFiT} Collaboration, J.~J. Ethier, G.~Magni, F.~Maltoni, L.~Mantani,
  E.~R. Nocera, J.~Rojo, E.~Slade, E.~Vryonidou, and C.~Zhang, {\it {Combined
  SMEFT interpretation of Higgs, diboson, and top quark data from the LHC}},
  {\em JHEP} {\bf 11} (2021) 089, [\href{http://arxiv.org/abs/2105.00006}{{\tt
  arXiv:2105.00006}}].

\bibitem{Dawson:2018dxp}
S.~Dawson, P.~P. Giardino, and A.~Ismail, {\it {Standard model EFT and the
  Drell-Yan process at high energy}},  {\em Phys. Rev. D} {\bf 99} (2019),
  no.~3 035044, [\href{http://arxiv.org/abs/1811.12260}{{\tt
  arXiv:1811.12260}}].

\bibitem{Ellis:2020unq}
J.~Ellis, M.~Madigan, K.~Mimasu, V.~Sanz, and T.~You, {\it {Top, Higgs, Diboson
  and Electroweak Fit to the Standard Model Effective Field Theory}},  {\em
  JHEP} {\bf 04} (2021) 279, [\href{http://arxiv.org/abs/2012.02779}{{\tt
  arXiv:2012.02779}}].

\bibitem{Greljo:2021kvv}
A.~Greljo, S.~Iranipour, Z.~Kassabov, M.~Madigan, J.~Moore, J.~Rojo, M.~Ubiali,
  and C.~Voisey, {\it {Parton distributions in the SMEFT from high-energy
  Drell-Yan tails}},  {\em JHEP} {\bf 07} (2021) 122,
  [\href{http://arxiv.org/abs/2104.02723}{{\tt arXiv:2104.02723}}].

\bibitem{ATLAS:2014gys}
{\bf ATLAS} Collaboration, G.~Aad et~al., {\it {Search for contact interactions
  and large extra dimensions in the dilepton channel using proton-proton
  collisions at $\sqrt{s}$ = 8 TeV with the ATLAS detector}},  {\em Eur. Phys.
  J. C} {\bf 74} (2014), no.~12 3134,
  [\href{http://arxiv.org/abs/1407.2410}{{\tt arXiv:1407.2410}}].

\bibitem{ATLAS:2020yat}
{\bf ATLAS} Collaboration, G.~Aad et~al., {\it {Search for new non-resonant
  phenomena in high-mass dilepton final states with the ATLAS detector}},  {\em
  JHEP} {\bf 11} (2020) 005, [\href{http://arxiv.org/abs/2006.12946}{{\tt
  arXiv:2006.12946}}]. [Erratum: JHEP 04, 142 (2021)].

\bibitem{ATLAS:2019erb}
{\bf ATLAS} Collaboration, G.~Aad et~al., {\it {Search for high-mass dilepton
  resonances using 139 fb$^{-1}$ of $pp$ collision data collected at
  $\sqrt{s}=$13 TeV with the ATLAS detector}},  {\em Phys. Lett. B} {\bf 796}
  (2019) 68--87, [\href{http://arxiv.org/abs/1903.06248}{{\tt
  arXiv:1903.06248}}].

\bibitem{CMS:2021ctt}
{\bf CMS} Collaboration, A.~M. Sirunyan et~al., {\it {Search for resonant and
  nonresonant new phenomena in high-mass dilepton final states at $ \sqrt{s} $
  = 13 TeV}},  {\em JHEP} {\bf 07} (2021) 208,
  [\href{http://arxiv.org/abs/2103.02708}{{\tt arXiv:2103.02708}}].

\bibitem{ATLAS:2021mla}
{\bf ATLAS} Collaboration, G.~Aad et~al., {\it {Search for New Phenomena in
  Final States with Two Leptons and One or No $b$-Tagged Jets at $\sqrt{s} =
  13$ TeV Using the ATLAS Detector}},  {\em Phys. Rev. Lett.} {\bf 127} (2021),
  no.~14 141801, [\href{http://arxiv.org/abs/2105.13847}{{\tt
  arXiv:2105.13847}}].

\bibitem{CMS:2018nlk}
{\bf CMS} Collaboration, A.~M. Sirunyan et~al., {\it {Search for contact
  interactions and large extra dimensions in the dilepton mass spectra from
  proton-proton collisions at $\sqrt{s} =$ 13 TeV}},  {\em JHEP} {\bf 04}
  (2019) 114, [\href{http://arxiv.org/abs/1812.10443}{{\tt arXiv:1812.10443}}].

\bibitem{Duhr:2021vwj}
C.~Duhr and B.~Mistlberger, {\it {Lepton-pair production at hadron colliders at
  N$^{3}$LO in QCD}},  {\em JHEP} {\bf 03} (2022) 116,
  [\href{http://arxiv.org/abs/2111.10379}{{\tt arXiv:2111.10379}}].

\bibitem{Buccioni:2020cfi}
F.~Buccioni, F.~Caola, M.~Delto, M.~Jaquier, K.~Melnikov, and R.~R\"ontsch,
  {\it {Mixed QCD-electroweak corrections to on-shell Z production at the
  LHC}},  {\em Phys. Lett. B} {\bf 811} (2020) 135969,
  [\href{http://arxiv.org/abs/2005.10221}{{\tt arXiv:2005.10221}}].

\bibitem{Buccioni:2022kgy}
F.~Buccioni, F.~Caola, H.~A. Chawdhry, F.~Devoto, M.~Heller, A.~von Manteuffel,
  K.~Melnikov, R.~R\"ontsch, and C.~Signorile-Signorile, {\it {Mixed
  QCD-electroweak corrections to dilepton production at the LHC in the high
  invariant mass region}},  {\em JHEP} {\bf 06} (2022) 022,
  [\href{http://arxiv.org/abs/2203.11237}{{\tt arXiv:2203.11237}}].

\bibitem{Bonciani:2020tvf}
R.~Bonciani, F.~Buccioni, N.~Rana, and A.~Vicini, {\it {Next-to-Next-to-Leading
  Order Mixed QCD-Electroweak Corrections to on-Shell Z Production}},  {\em
  Phys. Rev. Lett.} {\bf 125} (2020), no.~23 232004,
  [\href{http://arxiv.org/abs/2007.06518}{{\tt arXiv:2007.06518}}].

\bibitem{Bonciani:2021zzf}
R.~Bonciani, L.~Buonocore, M.~Grazzini, S.~Kallweit, N.~Rana, F.~Tramontano,
  and A.~Vicini, {\it {Mixed Strong-Electroweak Corrections to the Drell-Yan
  Process}},  {\em Phys. Rev. Lett.} {\bf 128} (2022), no.~1 012002,
  [\href{http://arxiv.org/abs/2106.11953}{{\tt arXiv:2106.11953}}].

\bibitem{Armadillo:2022bgm}
T.~Armadillo, R.~Bonciani, S.~Devoto, N.~Rana, and A.~Vicini, {\it {Two-loop
  mixed QCD-EW corrections to neutral current Drell-Yan}},  {\em JHEP} {\bf 05}
  (2022) 072, [\href{http://arxiv.org/abs/2201.01754}{{\tt arXiv:2201.01754}}].

\bibitem{ATLAS:2017rue}
{\bf ATLAS} Collaboration, M.~Aaboud et~al., {\it {Measurement of the Drell-Yan
  triple-differential cross section in $pp$ collisions at $\sqrt{s} = 8$ TeV}},
   {\em JHEP} {\bf 12} (2017) 059, [\href{http://arxiv.org/abs/1710.05167}{{\tt
  arXiv:1710.05167}}].

\bibitem{CMS:2022uul}
{\bf CMS} Collaboration, A.~Tumasyan et~al., {\it {Measurement of the Drell-Yan
  forward-backward asymmetry at high dilepton masses in proton-proton
  collisions at $\sqrt{s}$ = 13 TeV}},
  \href{http://arxiv.org/abs/2202.12327}{{\tt arXiv:2202.12327}}.

\bibitem{Fiaschi:2021sin}
J.~Fiaschi, F.~Giuli, F.~Hautmann, and S.~Moretti, {\it {Enhancing the Large
  Hadron Collider sensitivity to charged and neutral broad resonances of new
  gauge sectors}},  {\em JHEP} {\bf 02} (2022) 179,
  [\href{http://arxiv.org/abs/2111.09698}{{\tt arXiv:2111.09698}}].

\bibitem{Fiaschi:2021okg}
J.~Fiaschi, F.~Giuli, F.~Hautmann, and S.~Moretti, {\it {Lepton-Charge and
  Forward-Backward Asymmetries in Drell-Yan Processes for Precision Electroweak
  Measurements and New Physics Searches}},  {\em Nucl. Phys. B} {\bf 968}
  (2021) 115444, [\href{http://arxiv.org/abs/2103.10224}{{\tt
  arXiv:2103.10224}}].

\bibitem{Accomando:2019vqt}
E.~Accomando et~al., {\it {PDF Profiling Using the Forward-Backward Asymmetry
  in Neutral Current Drell-Yan Production}},  {\em JHEP} {\bf 10} (2019) 176,
  [\href{http://arxiv.org/abs/1907.07727}{{\tt arXiv:1907.07727}}].

\bibitem{Accomando:2018nig}
E.~Accomando, J.~Fiaschi, F.~Hautmann, and S.~Moretti, {\it {Neutral current
  forward\textendash{}backward asymmetry: from $\theta _W$ to PDF
  determinations}},  {\em Eur. Phys. J. C} {\bf 78} (2018), no.~8 663,
  [\href{http://arxiv.org/abs/1805.09239}{{\tt arXiv:1805.09239}}]. [Erratum:
  Eur.Phys.J.C 79, 453 (2019)].

\bibitem{Fiaschi:2022wgl}
J.~Fiaschi, F.~Giuli, F.~Hautmann, S.~Moch, and S.~Moretti, {\it
  {$\mathbf{Z^\prime}$-boson dilepton searches and the high-$\mathbf{x}$ quark
  density}},  \href{http://arxiv.org/abs/2211.06188}{{\tt arXiv:2211.06188}}.

\bibitem{CMS:2018ktx}
{\bf CMS} Collaboration, A.~M. Sirunyan et~al., {\it {Measurement of the weak
  mixing angle using the forward-backward asymmetry of Drell-Yan events in pp
  collisions at 8 TeV}},  {\em Eur. Phys. J. C} {\bf 78} (2018), no.~9 701,
  [\href{http://arxiv.org/abs/1806.00863}{{\tt arXiv:1806.00863}}].

\bibitem{Ball:2021leu}
{\bf NNPDF} Collaboration, R.~D. Ball et~al., {\it {The path to proton
  structure at 1\% accuracy}},  {\em Eur. Phys. J. C} {\bf 82} (2022), no.~5
  428, [\href{http://arxiv.org/abs/2109.02653}{{\tt arXiv:2109.02653}}].

\bibitem{Alekhin:2017kpj}
S.~Alekhin, J.~Blümlein, S.~Moch, and R.~Placakyte, {\it {Parton distribution
  functions, $\alpha_s$, and heavy-quark masses for LHC Run II}},  {\em Phys.
  Rev.} {\bf D96} (2017), no.~1 014011,
  [\href{http://arxiv.org/abs/1701.05838}{{\tt arXiv:1701.05838}}].

\bibitem{Hou:2019efy}
T.-J. Hou et~al., {\it {New CTEQ global analysis of quantum chromodynamics with
  high-precision data from the LHC}},  {\em Phys. Rev. D} {\bf 103} (2021),
  no.~1 014013, [\href{http://arxiv.org/abs/1912.10053}{{\tt
  arXiv:1912.10053}}].

\bibitem{Bailey:2020ooq}
S.~Bailey, T.~Cridge, L.~A. Harland-Lang, A.~D. Martin, and R.~S. Thorne, {\it
  {Parton distributions from LHC, HERA, Tevatron and fixed target data: MSHT20
  PDFs}},  {\em Eur. Phys. J. C} {\bf 81} (2021), no.~4 341,
  [\href{http://arxiv.org/abs/2012.04684}{{\tt arXiv:2012.04684}}].

\bibitem{Alwall:2014hca}
J.~Alwall, R.~Frederix, S.~Frixione, V.~Hirschi, F.~Maltoni, et~al., {\it {The
  automated computation of tree-level and next-to-leading order differential
  cross sections, and their matching to parton shower simulations}},  {\em
  JHEP} {\bf 1407} (2014) 079, [\href{http://arxiv.org/abs/1405.0301}{{\tt
  arXiv:1405.0301}}].

\bibitem{Carrazza:2020gss}
S.~Carrazza, E.~R. Nocera, C.~Schwan, and M.~Zaro, {\it {PineAPPL: combining EW
  and QCD corrections for fast evaluation of LHC processes}},  {\em JHEP} {\bf
  12} (2020) 108, [\href{http://arxiv.org/abs/2008.12789}{{\tt
  arXiv:2008.12789}}].

\bibitem{christopher_schwan_2022_7023438}
C.~Schwan, A.~Candido, F.~Hekhorn, and S.~Carrazza, {\em N3PDF/pineappl:
  v0.5.5}, Aug., 2022.

\bibitem{Collins:1977iv}
J.~C. Collins and D.~E. Soper, {\it {Angular Distribution of Dileptons in
  High-Energy Hadron Collisions}},  {\em Phys. Rev. D} {\bf 16} (1977) 2219.

\bibitem{Peskin:1995ev}
M.~E. Peskin and D.~V. Schroeder, {\em {An Introduction to quantum field
  theory}}.
\newblock Addison-Wesley, Reading, USA, 1995.

\bibitem{Candido:2020yat}
A.~Candido, S.~Forte, and F.~Hekhorn, {\it {Can $ \overline{\mathrm{MS}} $
  parton distributions be negative?}},  {\em JHEP} {\bf 11} (2020) 129,
  [\href{http://arxiv.org/abs/2006.07377}{{\tt arXiv:2006.07377}}].

\bibitem{Brodsky:1973kr}
S.~J. Brodsky and G.~R. Farrar, {\it {Scaling Laws at Large Transverse
  Momentum}},  {\em Phys. Rev. Lett.} {\bf 31} (1973) 1153--1156.

\bibitem{Brodsky:1974vy}
S.~J. Brodsky and G.~R. Farrar, {\it {Scaling Laws for Large Momentum Transfer
  Processes}},  {\em Phys. Rev. D} {\bf 11} (1975) 1309.

\bibitem{Ball:2017nwa}
{\bf NNPDF} Collaboration, R.~D. Ball et~al., {\it {Parton distributions from
  high-precision collider data}},  {\em Eur. Phys. J.} {\bf C77} (2017), no.~10
  663, [\href{http://arxiv.org/abs/1706.00428}{{\tt arXiv:1706.00428}}].

\bibitem{Ball:2016spl}
R.~D. Ball, E.~R. Nocera, and J.~Rojo, {\it {The asymptotic behaviour of parton
  distributions at small and large $x$}},  {\em Eur. Phys. J.} {\bf C76}
  (2016), no.~7 383, [\href{http://arxiv.org/abs/1604.00024}{{\tt
  arXiv:1604.00024}}].

\bibitem{Khachatryan:2016zqb}
{\bf CMS} Collaboration, V.~Khachatryan et~al., {\it {Search for narrow
  resonances in dilepton mass spectra in proton-proton collisions at $\sqrt{s}$
  = 13 TeV and combination with 8 TeV data}},  {\em Phys. Lett.} {\bf B768}
  (2017) 57--80, [\href{http://arxiv.org/abs/1609.05391}{{\tt
  arXiv:1609.05391}}].

\bibitem{PDF4LHCWorkingGroup:2022cjn}
{\bf PDF4LHC Working Group} Collaboration, R.~D. Ball et~al., {\it {The
  PDF4LHC21 combination of global PDF fits for the LHC Run III}},  {\em J.
  Phys. G} {\bf 49} (2022), no.~8 080501,
  [\href{http://arxiv.org/abs/2203.05506}{{\tt arXiv:2203.05506}}].

\bibitem{Carrazza:2015hva}
S.~Carrazza, J.~I. Latorre, J.~Rojo, and G.~Watt, {\it {A compression algorithm
  for the combination of PDF sets}},  {\em Eur. Phys. J.} {\bf C75} (2015) 474,
  [\href{http://arxiv.org/abs/1504.06469}{{\tt arXiv:1504.06469}}].

\bibitem{AbdulKhalek:2018rok}
R.~Abdul~Khalek, S.~Bailey, J.~Gao, L.~Harland-Lang, and J.~Rojo, {\it {Towards
  Ultimate Parton Distributions at the High-Luminosity LHC}},  {\em Eur. Phys.
  J. C} {\bf 78} (2018), no.~11 962,
  [\href{http://arxiv.org/abs/1810.03639}{{\tt arXiv:1810.03639}}].

\bibitem{Khalek:2021ulf}
R.~A. Khalek, J.~J. Ethier, E.~R. Nocera, and J.~Rojo, {\it {Self-consistent
  determination of proton and nuclear PDFs at the Electron Ion Collider}},
  {\em Phys. Rev. D} {\bf 103} (2021), no.~9 096005,
  [\href{http://arxiv.org/abs/2102.00018}{{\tt arXiv:2102.00018}}].

\end{thebibliography}
\providecommand{\href}[2]{#2}\begingroup\raggedright\endgroup

\end{document}